\documentclass[twocolumn]{revtex4}
\usepackage[]{graphicx} 
\usepackage{latexsym,amsmath,amsfonts,amssymb,bm,empheq}  
\usepackage{multirow}
\usepackage{threeparttable}
\usepackage{here}

\begin{document}
\title{Electric-field-induced oscillations in ionic fluids: a unified formulation of modified Poisson-Nernst-Planck models and its relevance to correlation function analysis}

\author{Hiroshi Frusawa}
\email{frusawa.hiroshi@kochi-tech.ac.jp}
\affiliation{Laboratory of Statistical Physics, Kochi University of Technology, Tosa-Yamada, Kochi 782-8502, Japan.}
\date{\today}
\begin{abstract}
We theoretically investigate an electric-field-driven system of charged spheres as a primitive model of concentrated electrolytes under an applied electric field.
First, we provide a unified formulation for the stochastic charge and density dynamics of the electric-field-driven primitive model using the stochastic density functional theory (DFT).
The stochastic DFT integrates various frameworks of the equilibrium and dynamic DFTs, the liquid state theory, and the field-theoretic approach, which allows us to justify in a unified manner various modifications previously made for the Poisson-Nernst-Planck model.
Next, we consider stationary density-density and charge-charge correlation functions of the primitive model with a static electric field.
We focus on an electric-field-induced synchronization between the emergence of density and charge oscillations, or the crossover from monotonic to oscillatory decay of density-density and charge-charge correlations.
The correlation function analysis demonstrates the appearance of stripe states formed by segregation bands perpendicular to the external field.
We also predict the following: (i) the electric-field-induced crossover occurs prior to the conventional Kirkwood crossover without an applied electric field, and (ii) the ion concentration dependence of the decay lengths at the electric-field-induced crossovers bears a similarity to the underscreening behavior found by simulation and theoretical studies on the oscillatory decay length in equilibrium.
\end{abstract}

%\pacs{}
\maketitle
%%%%%%%%%%%%%%%%%%%%%%%%
%%%%%%%%%%%%%%%%%%%%%%%%
\section{Introduction}
{\itshape Primitive model}.---
Ionic fluids cover a wide range of charged materials, including solvent-in-salt electrolytes, room-temperature ionic liquids (RTILs), various colloidal dispersions, and polyelectrolytes \cite{review bio, review fluid}.
Recently, ionic fluids are increasingly attracting much attention, due to their diverse applications not only in chemistry and biology \cite{review bio} but also in renewable energy devices such as batteries, supercapacitors, and separation media \cite{review fluid, rtil review, review energy}.
Here we consider electrolytes, RTILs, and mixtures of oppositely charged colloids under applied electric fields, as examples of symmetric ionic fluids driven by external electric fields.
These appear in biological ion channels, micro/nanofluidic devices for environmental and biomedical applications, and electrolyte-immersed porous electrodes for electrochemical applications \cite{review bio, review fluid, rtil review, review energy}.

Among models of the ionic fluids on target is a primitive model, or a symmetric collection of charged spheres whose cationic and anionic species have equal size and equal but opposite charge.
The primitive model under a static electric field, with which we are concerned, has been widely used to explain the structural and dynamical properties of concentrated electrolytes driven by external electric fields in confined geometries \cite{review fluid,bazant dynamics,eisenberg,mpnp channel}.

%%%%%%%%%%%%%%%%%%%%%%%
{\itshape Poisson-Nernst-Planck (PNP) model}.---
The Poisson-Nernst-Planck (PNP) model is the standard approach to describe the primitive model with a static electric field applied in the first approximation \cite{review fluid,bazant dynamics,eisenberg,mpnp channel}.
The Nernst-Planck equation, also known as the drift-diffusion equation, treats ionic currents arising from the combination of Fick's law of diffusion due to a concentration gradient and Ohm's law for drift of ions in a gradient of Coulomb potential.
The Nernst-Planck equation represents a conservation law:
\begin{flalign}
\label{conservation}%%%%%%%%%
\partial_tn_l({\bm r},t)=-\nabla\cdot\bm{J}_l(\bm{r},t),
\end{flalign}
where $n_l(\bm{r},t)$ ($l=1,\,2$) denotes an instantaneous number density of either cations ($l=1$) or anions ($l=2$) and $\bm{J}_l(\bm{r},t)$ a current vector of the $l$-th ion density.
The PNP model considers the coupled set of the Poisson and NP equations by relating $n_l(\bm{r},t)$ to a Coulomb potential via the Poisson equation.

%%%%%%%%%%%%%%%%%%%%%%%
{\itshape Deficiencies of the PNP model}.---
The PNP model provides a basic description of linear response dynamics of dilute electrolytes perturbed from equilibrium.
However, the original PNP model takes no account of (i) {\itshape steric interactions}, (ii) {\itshape ion-ion Coulomb correlations}, and (iii) {\itshape dielectric boundary effects}.
Hence, the conventional PNP model is insufficient to predict various electrokinetic phenomena when ions are crowded and/or when the ion distributions are spatially inhomogeneous \cite{bazant dynamics,eisenberg}.

For example, the PNP model is not relevant to the interfacial electrokinetic phenomena which are found not only in biological ion channels but also in advanced devices for micro/nanofluidic and electrochemical applications \cite{bazant dynamics,eisenberg}.
Steric effects become significant in either RTILs or thin electric double layers formed at large applied voltages, which, however, are not included in the PNP model \cite{bazant dynamics,eisenberg,mpnp channel, yochelis,mpnp rtil,mpnp self,witt,ddft,demery,gole,andelman,donev}.

There are also bulk properties for which the PNP model is not valid:
inhomogeneous steady states have been reported by theoretical, experimental, and simulation studies in the bulk region of either electrolytes or oppositely charged colloidal mixtures driven by electric fields \cite{demery,lowen,band,under exp,under simu}.
Theoretically, on the one hand, stationary correlation functions of electric-field-driven electrolytes were calculated, suggesting a tendency to form chains of cations and anions in the external field direction at larger electric fields \cite{demery}.
On the other hand, experimental and simulation studies have provided dynamic phase diagrams of steady states including laned, jammed or clogged, and mixed states of oppositely charged particles under a DC or AC electric field \cite{lowen,band}.
It is well known that lane formation of like-charge particles occurs at a high enough field strength along the applied field \cite{lowen,band}.
At the same time, previous studies have also observed that bands of like-charge particles are aligned in a direction non-parallel to the applied field direction when the electric-field-driven colloidal mixtures are in jammed or mixed states \cite{lowen,band}.

%%%%%%%%%%%%%%%%%%%%%%%
{\itshape Modified PNP models: deterministic case}.---
A variety of modified PNP (mPNP) models have thus been proposed so far; these arise either from semi-phenomenological methods \cite{bazant dynamics,eisenberg,mpnp channel, yochelis,mpnp rtil,mpnp self} or from deterministic density functional theory (DFT) \cite{witt,ddft}.
The modifications have aimed to overcome the above shortcomings given in (i) to (iii) as follows:
(i) {\itshape Steric effects} are included by adding a density current, or its associated chemical potential due to non-Coulombic short-range interactions. (ii) {\itshape Ion-ion Coulomb correlations} are taken into account by modified Poisson equations such as higher-order Poisson equation \cite{bsk,bazant bsk,high pb,frusawa review}. (iii) {\itshape Dielectric boundary effects} are investigated according to a generalized Born theory evaluating solvation energy from an ionic self-energy \cite{static self}.

Some of the results achieved by the deterministic mPNP models are as follows \cite{bazant dynamics,eisenberg,mpnp channel, yochelis,mpnp rtil,mpnp self}:
First, for biological ion channels, the numerical results have been found to agree with experimental or simulation data on the ion channel characteristics of selectivity and rectification.
Next, for micro/nanofluidics, it has been demonstrated that a coupled set of the mPNP and Navier-Stokes equations is a good descriptor of the interfacial electrokinetic phenomena. These include electro-osmotic flow, streaming current, and ionic conductance in porous media or nanochannels filled with RTILs or concentrated electrolytes of high valence.
Then, in terms of renewable energy technologies, modified PNP models have successfully explained
differential capacitance and non-monotonic oscillatory decay of electric double layers at solid-liquid interfaces with large voltages applied to RTILs or concentrated electrolytes.

%%%%%%%%%%%%%%%%%%%%%%%
{\itshape Modified PNP models: stochastic case}.---
An alternative approach to extend the PNP dynamics to a stochastic process is the stochastic DFT (SDFT) \cite{witt,sdft general}.
In the SDFT, we use the Dean-Kawasaki model \cite{witt,sdft general} that contains multiplicative noise by adding a stochastic current in eqn (\ref{conservation}).
The Dean-Kawasaki equation can be linearized for fluctuating density field around a reference density \cite{demery,gole,andelman,sdft linear, frusawa sdft}.
The linearized Dean-Kawasaki equation has proved relevant to describe various dynamics.
It is an outstanding feature of the linearized SDFT to justify the inclusion of stochastic processes into the PNP model.
The stochastic nature allows us to compute correlation functions for density and charge fluctuations around uniform states.

Recently, the stochastic mPNP models based on the linearized SDFT have provided the following results \cite{demery,gole,andelman,donev,sdft linear, frusawa sdft}:
First, the linear Dean-Kawasaki equation has formulated ion concentration-dependent electrical conductivity.
The obtained expression for conductivity reproduces the Debye-H\"uckel-Onsager theory and also explains the experimental results on concentrated electrolytes where the Debye-H\"uckel-Onsager theory breaks down \cite{demery,andelman,donev}.
It should be noted that we need to use a regularized interaction potential \cite{andelman} and to introduce hydrodynamic interactions \cite{andelman,donev} for explaining the high-density results \cite{andelman}; while this paper will justify the use of regularized form from the first principle, consideration of hydrodynamic interactions is beyond our scope.
Furthermore, it is found from the analysis of correlation functions that density-density and charge-charge correlations are long-range correlated even in the steady state.
The asymptotic decay of the correlation functions exhibits a power-law behavior with a dipolar character, thereby giving rise to a long-range fluctuation-induced force acting on uncharged confining plates \cite{gole}.

Yet more modifications of the stochastic mPNP models need to be made, following the deterministic mPNP models.
Namely, the above three issues (i.e., (i) to (iii) described above) have yet to be fully addressed by the stochastic mPNP models.
We would also like to note that the stochastic formulation focuses on linear response dynamics from a uniform density distribution and that there have been few systematic studies on an inhomogeneous density distribution in a steady state \cite{frusawa sdft}.
To pave the way for a more elaborate mPNP model, the stochastic mPNP models need to capture the benefits of the deterministic mPNP models.

%%%%%%%%% table1 %%%%%%%%%%%%%%
\begin{table*}
\begin{threeparttable}
\caption{Summary table of our formulation in comparison with previous theories on the mPNP models.}
\centering
\footnotesize
\begin{tabular}{clcccc}
\multicolumn{2}{c}{\shortstack{{\bf Equation}\\{\bf Type}}}$\quad$&
\hspace{2pt}\shortstack{{\bf Modification of}\\{\bf the Poisson Equation}}$\>$&
\hspace{2pt}\shortstack{{\bf Self-energy}\\{\bf Contribution}}$\>$&
\hspace{8pt}\shortstack{{\bf Stochastic}\\{\bf Density Current}}$\>$&
\hspace{2pt}\shortstack{{\bf Correlation}\\{\bf Functions}}\\
%\hspace{2pt}\shortstack{{\bf The Kirkwood}\\{\bf Crossover}}\\
\hline\\
\multirow{2}{*}{\shortstack{{\bf Self-energy-modified}\\{\bf PNP equations}}}&\hspace{2pt}Ref. \cite{mpnp self}$\quad$&\hspace{2pt}Eqn (\ref{high poisson})\tnote{b}$\>$&\hspace{2pt}Eqn (\ref{self energy}), (\ref{poisson}) and (\ref{previous gdh})$\>$&\hspace{8pt}---$\quad$&\hspace{2pt}---\\
%&\hspace{2pt}---\\
&\hspace{2pt}Ours$\quad$&\hspace{2pt}Eqn (\ref{finite spread})\tnote{a}
\hspace{3pt}or (\ref{high poisson})\tnote{b}$\>$&\hspace{2pt}Eqn (\ref{self energy}), (\ref{g0 poisson}) and (\ref{gdh})&\hspace{8pt}Eqn (\ref{pnp current}) and (\ref{mu})$\quad$&\hspace{2pt}TBD\\
&&&&&\\
\multirow{2}{*}{\shortstack{{\bf Linear}\\{\bf mPNP equations}}}&\hspace{2pt}Ref. \cite{demery,gole,andelman}$\quad$&\hspace{2pt}Poisson or Eqn (\ref{finite spread})\tnote{a}$\>$&\hspace{2pt}---$\>$&\hspace{8pt}Eqn (\ref{mat current rhoq})$\quad$&\hspace{2pt}Eqn (\ref{C solution2}) to (\ref{ans E0})\\
%&\hspace{2pt}---\\
&\hspace{2pt}Ours$\quad$&\hspace{2pt}Eqn (\ref{finite spread})\tnote{a}$\>$&\hspace{2pt}---$\>$&\hspace{8pt}Eqn (\ref{mat current rhoq})$\quad$&\hspace{2pt}Eqn (\ref{C solution2}) to (\ref{ans E0})%%%%%%%%%%)\\
\end{tabular}
\begin{tablenotes}
\item[a] Finite-spread Poisson equation \cite{demery,andelman,frusawa review,frydel review,finite gaussian,finite pb}.
\item[b] Higher-order Poisson equation \cite{bazant dynamics,eisenberg,mpnp channel, yochelis,mpnp rtil,mpnp self,bsk,bazant bsk,high pb,frusawa review}.
\end{tablenotes}
\end{threeparttable}
\end{table*}
%%%%%%%%%%%%%%%%
%%%%%%%%%%%%%%%%%%%%%%%%
%%%%%%%%%%%%%%%%%%%%%%%%

%%%%%%%%%%%%%%%%%%%%%%%
{\itshape The aim of this paper}.---
To summarize, the deterministic and stochastic mPNP models proposed so far are beneficial in the following respects:
while the deterministic mPNP models have provided elaborate and tractable methods to include short-range correlations and interactions of Coulombic and non-Coulombic origins, the stochastic mPNP models have demonstrated the relevance to correlation function analysis on fluctuation phenomena in uniform and steady states.
Integration of these modifications should lead to a deeper understanding of the electrokinetic phenomena in concentrated electrolytes.

Thus, this paper serves two purposes.
The first aim is to provide a unified formulation that combines the above results of the deterministic and stochastic mPNP models.
From the aspect of the deterministic mPNP models, we attempt, using the unified formulation, to add the stochastic term to the deterministic mPNP equations and to derive the semi-phenomenological modifications \cite{bazant dynamics,eisenberg,mpnp channel, yochelis,mpnp rtil,mpnp self} from the first principle based on the liquid state theory.
In terms of the stochastic mPNP models, on the other hand, the unified formulation justifies the inclusion of modified terms proposed by the deterministic mPNP models \cite{bazant dynamics,eisenberg,mpnp channel, yochelis,mpnp rtil,mpnp self} into the stochastic equations \cite{demery,gole,andelman}.

The second purpose is to determine when density and charge oscillations emerge in non-equilibrium steady states.
To this end, we investigate stationary correlation functions which are averaged over the plane transverse to the applied electric field.
The unified mPNP model yields the stationary correlation functions at equal times, which enables us to explore crossovers from monotonic to oscillatory decay of density-density and charge-charge correlations.

%%%%%%%%%%%%%%%%%%%%%%%
{\itshape The organization of this paper}.---
In what follows, we first present the summarized results on both the unified form of mPNP models (Section II) and the correlation function analysis to investigate steady states (Section III) before going into the details.

On the one hand, Table 1 in Section II summarizes the obtained forms compared to previous formulations.
The essential achievement in terms of the theoretical formalism is clarified in Sec. IID.
As detailed in Appendix A, the hybrid framework of the field-theoretic approach, the equilibrium \cite{dft,ry} and dynamic \cite{witt} DFT, and the liquid state theory justifies the modified Poisson equations \cite{bazant dynamics,eisenberg,mpnp channel, yochelis,mpnp rtil,mpnp self,high pb,frusawa review,frydel review,finite gaussian,finite pb} and the generalized Debye-H\"uckel equation for the self-energy \cite{mpnp self,static self}.

On the other hand, Fig. 2 in Section III provides a schematic summary of electric-field-dependent decay length prior to the Kirkwood crossover in the external field direction.
Figure 2 illustrates not only the emergence of stripe states formed by segregation bands transverse to the applied field direction but also the intimate connection between the electric-field-induced shift of the decay length at the Kirkwood crossover and the underscreening behaviors \cite{under exp,under simu,under th,under andelman,under evans} observed in equilibrium electrolytes whose concentrations are higher than the conventional Kirkwood crossover point \cite{kirkwood original,kirkwood various,various smearing,gauss smearing,kirkwood fw}.
Section IIIE also presents 2D behaviors of oscillatory correlations above the Kirkwood crossover using heat maps, which corroborates the appearance of stripe states in the presence of relatively weak electric fields.

Section IV clarifies the detailed process to analytical and numerical results of the electric-field-induced Kirkwood crossover point and decay length using a couple of models typical for the liquid state theory.
In Section VI, we have discussions for clarifying what the obtained results imply.

\section{Formulation results on modifications of PNP model}
In the first place, this section summarizes the resulting formulation, according to Table 1 (Section IIA).
As seen from the equation type given in the leftmost column of Table 1, we have verified two modifications of PNP equations using the SDFT of the symmetric primitive model specified in Section IIB.
In Section IIC, we present fully modified PNP equations that incorporate density currents from Gaussian noise fields as well as a self-energy contribution into the PNP model.
Section IID describes the theoretical achievements in terms of the Dean-Kawasaki model.
In Section IIE, we investigate the linearized mPNP equations while neglecting the self-energy in order to obtain stationary correlation functions at equal times.

%%%%%%%%%%%%%%%%%%%%%%%%
%%%%%%%%%%%%%%%%%%%%%%%%
\subsection{Comparison with previous theories}
In Sections IIC and IID, we will present the self-energy-modified PNP equations \cite{mpnp self} and the linear mPNP equations \cite{demery,gole,andelman}. Table 1 compares these equation sets with previous approaches.

%%%%%%%%%%%%%
{\itshape Self-energy-modified PNP equations}.---
Previous theories \cite{bazant dynamics,eisenberg,mpnp channel, yochelis,mpnp rtil,mpnp self,witt,ddft} have made two modifications. One is to improve the Poisson equation for the Coulomb interaction potential $\psi(\bm{r},t)$ experienced by an ion (the higher-order Poisson equation (\ref{high poisson})) \cite{bazant dynamics,eisenberg,mpnp channel, yochelis,mpnp rtil,mpnp self,bsk,bazant bsk,high pb,frusawa review}.
The other is to make a self-energy correction to the Coulomb interaction term in the PNP equations (the self-energy term determined by the generalized Debye-H\"uckel equation (\ref{previous gdh})) \cite{mpnp self,static self}, thereby providing theoretical descriptions of ionic transport in agreement with simulation results;
however, these modifications are empirical, and correlation functions have been beyond the scope due to the absence of stochastic current.
Meanwhile, our self-energy-modified PNP equations, derived from the basic formulation of the SDFT (see Appendix A for details), verify the stochastic dynamics and encompass the above modifications.

{\itshape Linear mPNP equations}.---
Recently, the PNP model covers the stochastic dynamics of density fluctuations around a uniform state while neglecting the self-energy contribution \cite{demery,gole}.
Furthermore, a finite-spread Poisson equation has been used in an ad hoc manner depending on a charge smearing model adopted \cite{andelman}.
The stochastic mPNP equations allow us to evaluate correlation functions, yielding either the ion concentration-dependent electrical conductivity and the long-range fluctuation-induced force as mentioned in Section I.
In this study, we confirm the linear stochastic mPNP equations previously used as an approximation of the self-energy-modified PNP equations \cite{mpnp self}.
Accordingly, the use of the finite-spread Poisson equation \cite{demery,andelman,frusawa review,frydel review,finite gaussian,finite pb} is validated from the decomposition of the direct correlation function (DCF) to extract the weight function $\omega(\bm{k})$, implying that we can improve the finite-spread Poisson equation systematically by adopting a more appropriate function form of $\omega(\bm{k})$ other than eqn (\ref{omega two}).

%%%%%%%%%%%%%%%%%
%%%%%%%%% FIG0 %%%%%%%%%%%%%%
\begin{figure}[hbtp]
\begin{center}
\includegraphics[
width=7cm
]{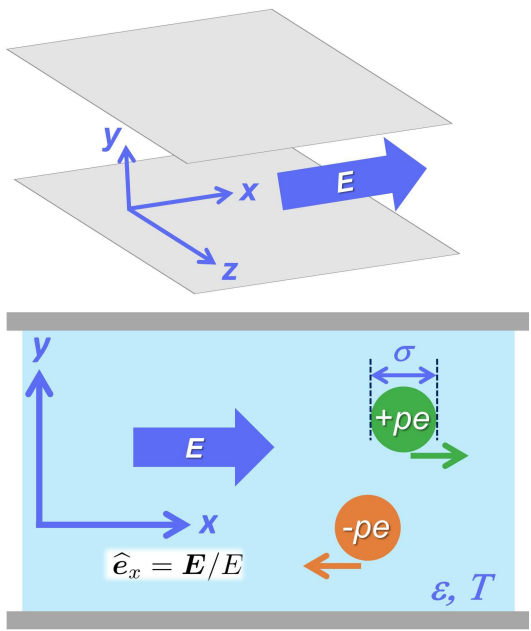}
\end{center}
\caption{A schematic of concentrated electrolytes under a static electric field $\bm{E}$ in Cartesian coordinates (top figure). The 3D primitive model is illustrated in the $xy$ plane (lower figure). The definition of symbols is provided in the main text.
}
\end{figure}
\subsection{Model}
The 3D primitive model introduces three parameters, $p$, $\sigma$ and $\epsilon$, for concentrated electrolytes with a static electric field $\bm{E}$ applied:
$p$--valent cations and anions are modeled by equisized charged hard spheres of diameter $\sigma$ immersed in a structureless and uniform dielectric medium with dielectric constant $\epsilon$ at a temperature $T$.
For later convenience in defining potential energies, we have defined the external field $\bm{E}$ as the conventional field multiplied by $e/k_BT$ (see also the statement after eqn (\ref{j0 ry})).
Figure 1 presents schematics of the electric-field-driven primitive model in Cartesian coordinates where the external electric field $\bm{E}$ with its strength of $E=|\bm{E}|$ is parallel to the unit vector $\hat{\bm{e}}_{x}=(1,\,0,0)^{\mathrm{T}}=\bm{E}/E$ in the direction of $x$-axis.
We can see two parallel plates in Fig. 1 as a reference for the later simplification; however, the interplate distance is much larger than the sphere diameter, and we suppose that the finite size effect due to the presence of the plates is negligible.

The primitive model is characterized by a pairwise interaction potential $v_{lm}(\bm{r})$ between charged hard spheres with a separation distance of $r=|\bm{r}|$:
$v_{11}(\bm{r})$, $v_{12}(\bm{r})$ and $v_{22}(\bm{r})$ represent cation-cation, cation-anion and anion-anion interactions, respectively.
We have
\begin{eqnarray}
\label{interaction potential}%%%%%%%%%%%%
v_{lm}(\bm{r})=
\left\{
\begin{array}{l}
\infty\quad(r< \sigma)\\
\\
\left.
(-1)^{l+m}p^2l_B\middle/r\quad(r\geq\sigma)
\right.,\\
\end{array}
\right.
\end{eqnarray}
where $l_B=e^2/(4\pi\epsilon k_BT)$ denotes the Bjerrum length, the length at which the bare Coulomb interaction between two monovalent ions is exactly $k_BT$.
It is noted that this paper defines all of the energetic quantities, including the pairwise interaction potentials, in units of $k_BT$.

%%%%%%%%%%%%%%%%%%%%%%%%
\subsection{Self-energy-modified PNP equations: a full set of the resulting formulation}
%%%%%%%%%%%%%%%%%%%%%%%%%%%%%
{\itshape The conservation equation with stochastic current}.---
In the conservation equation (\ref{conservation}) of the SDFT, the ionic current $\bm{J}_l(\bm{r},t)$ consists of three parts:
\begin{flalign}
&\bm{J}_l(\bm{r},t)=(-1)^{l-1}\mathcal{D}n_l(\bm{r},t)p\bm{E}-\mathcal{D}n_l(\bm{r},t)\nabla\mu_l[\bm{n}]
\nonumber\\
\label{pnp current}%%%%%%%%%%%
&\hphantom{\bm{J}_l(\bm{r},t)=\mathcal{D}n_l(\bm{r},t)z_l\bm{E}}
-\sqrt{2\mathcal{D}n_l(\bm{r},t)}\bm{\zeta}(\bm{r},t),
\end{flalign}
where $\mathcal{D}$ and $\mu_l[\bm{n}]$ denote, respectively, the diffusion constant and the chemical potential as a functional of $\bm{n}(\bm{r},t)=(n_1(\bm{r},t),n_2(\bm{r},t))^{\mathrm{T}}$, and $\bm{\zeta}(\bm{r},t)$ represents uncorrelated Gaussian noise fields defined below.
Incidentally, we have neglected an advection term \cite{donev, andelman}, $n_l(\bm{r},t)\bm{u}(\bm{r},t)$, on the right had side (rhs) of eqn (\ref{pnp current}) with $\bm{u}(\bm{r},t)$ denoting the solvent velocity field to satisfy the incompressibility condition $\nabla\cdot\bm{u}=0$, according to the treatment of the Dean-Kawasaki equation (see also Appendix A); this approximation is equivalent to supposing that $|\bm u|\ll |\mathcal{D}p\bm{E}|$.

%%%%%%%%%%%%%%%%%%%%%%%%%%%%%
{\itshape Concrete forms of density current $\bm{J}_l(\bm{r},t)$}.---
The first term on the right hand side (rhs) of eqn (\ref{pnp current}) represents the reference current directly determined by $\bm{E}$.
The chemical potential $\mu_l[\bm{n}]$ of the second term on the rhs of eqn (\ref{pnp current}) is given by
\begin{flalign}
\label{mu}%%%%%%%%%%%
&\mu_l[\bm{n}]=\ln n_l(\bm{r},t)+U_l[\bm{n}],
\end{flalign}
using an instantaneous interaction energy $U_l[\bm{n}]$ per cation ($l=1$) or anion ($l=2$).
Eqn (\ref{mu}) indicates that this part of the total current considers additional contributions from ideal entropy and Coulomb interactions considering steric effects.
The last term on the rhs of eqn (\ref{pnp current}) corresponds to the stochastic current arising from $\bm{\zeta}(\bm{r},t)$ characterized by
\begin{flalign}
\left\langle\bm{\zeta}(\bm{r},t)\bm{\zeta}(\bm{r}',t')^{\mathrm{T}}\right\rangle_{\zeta}=\delta(\bm{r}-\bm{r}')\delta(t-t'),
\label{noise}%%%
\end{flalign}
with the subscript "$\zeta$" representing the Gaussian noise averaging in space and time.

Following the mPNP equations proposed so far, $U_l[\bm{n}]$ is further divided into two parts:
\begin{flalign}
\label{pnp U}%%%%%%%%%%%
U_l[\bm{n}]&=(-1)^{l-1}\psi(\bm{r},t)+\frac{u(\bm{r},t)}{2},
\end{flalign}
where $(-1)^{l-1}\psi(\bm{r},t)$ and $u(\bm{r},t)/2$ denote, respectively, the instantaneous interaction potential and the instantaneous self-energy per ion.
We define $\psi(\bm{r},t)$ by
\begin{flalign}
\label{app potential}%%%%%%%%%%%
\psi(\bm{r},t)=-\int d^3\bm{r}'c(\bm{r}-\bm{r}')\,q(\bm{r}',t),
\end{flalign}
using the DCF $c(\bm{r}-\bm{r}')$ and an instantaneous charge density,
\begin{flalign}
\label{charge def}%%%%%%%% 
peq(\bm{r},t)=pe\{n_1(\bm{r},t)-n_2(\bm{r},t)\}.
\end{flalign}
By definition of the DCF, the potential function $\psi(\bm{r},t)$ defined in eqn (\ref{app potential}) corresponds to the conventional Coulomb potential multiplied by $pe/k_BT$ (see also eqn (\ref{fourier dcf})).
It follows that $q(\bm{r},t)$ in eqn (\ref{app potential}) is not an instantaneous charge density but is merely the concentration difference ($q=n_1-n_2$) as seen from eqn (\ref{charge def}).

Meanwhile, our resulting formulation provides the self-energy as follows:
\begin{flalign}
\label{self energy}%%%%%%%%%%%%%
\frac{u(\bm{r},t)}{2}=\frac{p^2}{2}\lim_{\bm{r}'\rightarrow\bm{r}}\left\{
G(\bm{r}-\bm{r}')-G_0(\bm{r}-\bm{r}')\right\},
\end{flalign}
where bare and dressed propagators, $G_0(\bm{r}-\bm{r}')$ and $G(\bm{r}-\bm{r}')$, are given by
\begin{flalign}
\label{g0 def}%%%%%%%
p^2G_0(\bm{r}-\bm{r}')&=-c(\bm{r}-\bm{r}'),\\
\label{g def}%%%%%%%%%%%%
p^2G(\bm{r}-\bm{r}')&=-h(\bm{r}-\bm{r}'),
\end{flalign}
with $h(\bm{r}-\bm{r}')$ denoting the total correlation function between ions of the same kind.

We investigate the following model forms as the DCFs:
\begin{flalign}
\label{omega two}%%%%%%%%%%%%%%%%%
\omega(\bm{k})&=
\left\{
\begin{array}{l}
e^{-(k\sigma)^2/2}
\\\\
\cos(k\sigma),
\end{array}
\right.
\end{flalign}
in the Fourier transform of the DCF given by
\begin{flalign}
\label{fourier dcf}%%%%%%%%%%%%%%
-c(\bm{k})=\frac{4\pi p^2l_B}{\bm{k}^2}\omega(-\bm{k}),
\end{flalign}
where $|\bm{k}|=k$.
While the former expression of $\omega(\bm{k})$ in eqn (\ref{omega two}) represents the Gaussian charge smearing model \cite{gauss smearing,finite gaussian} and has been used in the hypernetted chain approximation of one-component ionic fluids \cite{ng}, the latter form in eqn (\ref{omega two}) indicates the restriction of Coulomb interactions to the separation of $|\bm{r}-\bm{r}'|>\sigma$ with a cutoff at $|\bm{r}-\bm{r}'|=\sigma$ and is an approximate form of the modified MSA model \cite{mmsa smearing} as shown in Appendix A2.

%%%%%%%%%%%%%%%%%%%%%%%%%%%%%
{\itshape Finite-spread or higher-order Poisson equation}.---
Combining eqn (\ref{app potential}), (\ref{omega two}) and (\ref{fourier dcf}), we have
\begin{flalign}
\label{finite spread}%%%%%%%%%%
-\nabla^2\psi(\bm{r}-\bm{r}')
=4\pi p^2l_B\int d^3\bm{r}' 
\omega(\bm{r}-\bm{r}')\,q(\bm{r}',t),
\end{flalign}
which will be referred to as the finite-spread Poisson equation after the finite-spread Poisson-Boltzmann equation;
both equations consider the charge distribution inside a charged sphere using a weight function $\omega(\bm{r}-\bm{r}')$.
As shown in Appendix A2, eqn (\ref{finite spread}) transforms to
\begin{flalign}
\label{high poisson}%%%%%%%%%%%%%%%%%%%%
k_BT\epsilon\left(
\frac{\sigma^2}{2}\nabla^2-1
\right)\nabla^2\psi(\bm{r}-\bm{r}')=
(pe)^2 q(\bm{r},t),
\end{flalign}
when performing the low wavenumber expansion of $\omega(\bm{k})$, similarly to the transformation from the finite-spread Poisson-Boltzmann equation to the higher-order one for one-component fluids.
Eqn (\ref{high poisson}) will be referred to as the higher-order Poisson equation \cite{bazant dynamics,eisenberg,mpnp channel, yochelis,mpnp rtil,mpnp self,bsk,bazant bsk,high pb,frusawa review} for comparison with the finite-spread Poisson equation (\ref{finite spread}), though often called either the Poisson-Fermi equation or the Bazant-Storey-Kornyshev equation.

%%%%%%%%%%%%%%%%%%%%%%%%%%%%%
{\itshape A generalized Debye-H\"uckel equation}.---
It follows from eqn (\ref{g0 def}) to (\ref{fourier dcf}) that the DCF and the total correlation function, $c(\bm{r}-\bm{r}')$ and $h(\bm{r}-\bm{r}')$, obey a modified Poisson equation and a generalized Debye-H\"uckel equation, respectively:
we have
\begin{flalign}
\label{g0 poisson}%%%%%%%%%%%%
-\nabla^2 G_0(\bm{r}-\bm{r}')= 4\pi l_B\omega(\bm{r}-\bm{r}'),
\end{flalign}
whereas the Orstein-Zernike equation reads
\begin{flalign}
&-\nabla^2 G(\bm{r}-\bm{r}')+\int d^3\bm{r}" \omega(\bm{r}-\bm{r}")\kappa^2(\bm{r}")G(\bm{r}"-\bm{r}')
\nonumber\\
\label{gdh}%%%%%%%%%%%%%%%%
&\hphantom{-\nabla^2 G(\bm{r}-\bm{r}')+\int d^3\bm{r}" \omega}\quad\quad\quad
=4\pi l_B\omega(\bm{r}-\bm{r}'),
\end{flalign}
where a generalized Debye-H\"uckel length $\kappa^{-1}(\bm{r})$ has been defined as
\begin{flalign}
\label{gdh length}%%%%%%%%%
\kappa^{-1}(\bm{r})&=\left\{
4\pi l_Bp^2\rho(\bm{r},t)
\right\}^{-1/2},\\
\label{density def}%%%%%%%%%
\rho(\bm{r},t)&=n_1(\bm{r},t)+n_2(\bm{r},t).
\end{flalign}
Eqn (\ref{g0 poisson}) and (\ref{gdh}) for point charges ($\sigma=0$) reduce, respectively, to
\begin{flalign}
\label{poisson}%%%%%%%%%
-\nabla^2 G_0(\bm{r}-\bm{r}')
=4\pi l_B\delta(\bm{r}-\bm{r}')
\end{flalign}
and
\begin{flalign}
\label{previous gdh}%%%%%%%%%
-\nabla^2 G(\bm{r}-\bm{r}')+\kappa^2(\bm{r})G(\bm{r}-\bm{r}')
=4\pi l_B\delta(\bm{r}-\bm{r}')
\end{flalign}
because of
\begin{flalign}
\label{omega delta}%%%%%%%
\lim_{\sigma\rightarrow 0}\omega(\bm{r}-\bm{r}')=\delta(\bm{r}-\bm{r}'),
\end{flalign}
as confirmed from eqn (\ref{omega two}).
Eqn (\ref{previous gdh}) corresponds to the generalized Debye-H\"uckel equation previously used \cite{mpnp self,static self}.
%%%%%%%%%%%%%%%%%%%%%%%%%%%%%%

%%%%%%%%%%%%%%%%%%%%%
\subsection{Theoretical achievement in terms of the Dean-Kawasaki model}
In Appendices A and B, we prove that the Dean-Kawasaki model can be approximated by the formula given by eqn (\ref{pnp current}) to (\ref{g def}) for concentrated electrolytes:
the hybrid framework of the equilibrium DFT and field-theoretic treatment transforms the original Dean-Kawasaki equation to a tractable expression for binary ionic fluids without ad hoc modifications.
Here we clarify the theoretical achievement instead of going into the detailed formulations presented in Appendix A. 
It is found from eqn (\ref{mu def}), (\ref{appendix contour}), (\ref{appendix tr}) and (\ref{appendix u}) that the straightforward use of the original DK model provides exactly
\begin{flalign}
\label{original U}%%%%%%%%%%%
U_l[\bm{n}]&=(-1)^{l-1}\psi_0(\bm{r},t)-\frac{v_{ll}(\bm{0})}{2},\\
\psi_0(\bm{r},t)&=\int d^3\bm{r}'v_{ll}(\bm{r}-\bm{r}')\,q(\bm{r}',t).
\end{flalign}
It follows from eqn (\ref{pnp current}), (\ref{pnp U}), and (\ref{original U}) that the electrostatic contribution to ionic current reads
\begin{flalign}
&-\mathcal{D}\bm{n}_l(\bm{r},t)\nabla U_l[\bm{n}]\nonumber\\
\label{el current}%%%%%%%%%%%
&\quad=\mathcal{D}\bm{n}_l(\bm{r},t)\left\{(-1)^{l-1}\nabla\psi(\bm{r},t)+\frac{\nabla u(\bm{r},t)}{2}\right\}\\
\label{el current original}
&\quad=\mathcal{D}\bm{n}_l(\bm{r},t)\left\{(-1)^{l-1}\nabla\psi_0(\bm{r},t)\right\}
\end{flalign}
because of $\nabla v_{ll}(\bm{0})=\bm{0}$.
Comparison between eqn (\ref{el current}) and (\ref{el current original}) reveals that the SDFT based on the above hybrid framework justifies
\begin{flalign}
\label{gaussian}
\left|\nabla\left\{\psi_0(\bm{r},t)-\psi(\bm{r},t)\right\}\right|\approx
\left|\frac{\nabla c(\bm{0},t)}{2}\right|
\end{flalign}
in the Gaussian approximation of the auxiliary potential field (see Appendix A for details), where use has been made of the relation, $\nabla u(\bm{r})=\nabla \lim_{\bm{r}'-\bm{r}}c(\bm{r}-\bm{r}',t)\equiv \nabla c(\bm{0},t)$. 
Our achievement in terms of the theoretical formalism is essentially to validate eqn (\ref{gaussian}), which is the intrinsic reason why the SDFT successfully unifies various mPNP models.  

%%%%%%%%%%%%%%%%%%%%%%%%
\subsection{Linear \lowercase{m}PNP equations and the associated correlation functions}
%%%%%%%%%%%%%%%%%%%%%%%%%%%%%
{\itshape The matrix representation}.---
Let us introduce two vectors, $\bm{\theta}(\bm{r},t)$ and $\bm{\eta}(\bm{r},t)$, for having a compact form of the mPNP equation set:
\begin{flalign}
\label{theta def}%%%%%%%%%%
\bm{\theta}(\bm{r},t)&=
\begin{pmatrix}
\rho(\bm{r},t)\\
q(\bm{r},t)\\
\end{pmatrix},
\\
\label{eta def}%%%%%%%%%%
\bm{\eta}(\bm{r},t)&=
\begin{pmatrix}
\nabla\cdot\bm{\zeta}(\bm{r},t)\\
\nabla\cdot\bm{\zeta}'(\bm{r},t)\\
\end{pmatrix},
\end{flalign}
where $\rho(\bm{r},t)$ and $q(\bm{r},t)$ have been defined in eqn (\ref{density def}) and (\ref{charge def}), respectively, and $\bm{\zeta}'(\bm{r},t)$ is characterized by the same relation as eqn (\ref{noise}) of $\bm{\zeta}(\bm{r},t)$.
We perform the change of variables from $\bm{n}(\bm{r},t)$ to $\bm{\theta}(\bm{r},t)$ in the linearization of the mPNP equation set given by eqn (\ref{conservation}) and eqn (\ref{pnp current}) to (\ref{pnp U}) with the self-energy term (\ref{self energy}) being dropped.
Thus, we obtain the stochastic currents, $\bm{J}_{\rho}$ and $\bm{J}_q$, from linearizing the current given by eqn (\ref{pnp current}) (see Appendix A for details):
\begin{flalign}
\label{mat current rhoq}%%%%%%%%%%
\begin{pmatrix}
\bm{J}_{\rho}(\bm{r},t)\\
\bm{J}_q(\bm{r},t)\\
\end{pmatrix}
&=
\begin{pmatrix}
\bm{J}_1(\bm{r},t)+\bm{J}_2(\bm{r},t)\\
\bm{J}_1(\bm{r},t)-\bm{J}_2(\bm{r},t)\\
\end{pmatrix}
\nonumber\\
&=
-\mathcal{D}
\begin{pmatrix}
\nabla\rho(\bm{r},t)-q(\bm{r},t)p\bm{E}\\
\nabla q(\bm{r},t)+2\overline{n}\nabla\psi(\bm{r},t)-\rho(\bm{r},t)p\bm{E}\\
\end{pmatrix}\nonumber\\
&\qquad\qquad\qquad\qquad\qquad
-\sqrt{4\mathcal{D}\overline{n}}
\begin{pmatrix}
\bm{\zeta}(\bm{r},t)\\
\bm{\zeta}(\bm{r},t)\\
\end{pmatrix},
\end{flalign}
using the smeared density $\overline{n}$ of cations or anions.
We insert the expression (\ref{mat current rhoq}) into the conservation equation for $\rho(\bm{r},t)$ and $q(\bm{r},t)$:
\begin{flalign}
\label{mat dk rhoq}%%%%%%%%%%%%%%
\partial_t
\bm{\theta}(\bm{r},t)
&=-\nabla\cdot
\begin{pmatrix}
\bm{J}_{\rho}(\bm{r},t) \\
\bm{J}_q(\bm{r},t) \\
\end{pmatrix},
\end{flalign}
which is Fourier transformed to
\begin{flalign}
\label{mat dk m fourier}%%%%%%%%%%%%%%%
\partial_t
\bm{\theta}(\bm{k})
&=-\mathcal{D}
\bm{\mathcal{K}}(\bm{k})
\bm{\theta}(-\bm{k},t)+
\sqrt{4\mathcal{D}\overline{n}}\,\bm{\eta}(\bm{k}),
\end{flalign}
noting that the finite-spread Poisson equation (\ref{finite spread}) yields the Fourier transform of $2\overline{n}\nabla^2\psi(\bm{r},t)$ as follows:
\begin{flalign}
-2\overline{n}\bm{k}^2\psi(\bm{k})&=-8\pi p^2l_B\overline{n}\,\omega(\bm{k})q(-\bm{k})\nonumber\\
\label{fourier finite spread}%%%%%%%%%%%%
&=-\overline{\kappa}^2\omega(\bm{k})q(-\bm{k}),
\end{flalign}
where we have defined the smeared Debye-H\"uckel length,
\begin{flalign}
\label{smear dh}%%%%%%%%%%%%%%
\xi_{\mathrm{DH}}=\overline{\kappa}^{-1}\equiv
\left(8\pi p^2l_B\overline{n}\right)^{-1/2},
\end{flalign}
other than $\kappa^{-1}(\bm{r},t)$ defined by eqn (\ref{gdh length}).
In eqn (\ref{mat dk m fourier}), the matrix to determine restoring forces is expressed as
\begin{flalign}
\label{k matrix}%%%%%%%%%%%
\bm{\mathcal{K}}(\bm{k})&=
\begin{pmatrix}
\bm{k}^2& ik_{x}pE\\
ik_{x}pE&\mathcal{G}_1(\bm{k})\\
\end{pmatrix},
\\
\label{def g1}%%%%%%%%%%%
\mathcal{G}_1(\bm{k})&=\bm{k}^2+\overline{\kappa}^2\omega(\bm{k}).
\end{flalign}
In eqn (\ref{k matrix}), the anisotropy of the $\bm{k}$--space is associated with the direction of applied electric field (i.e., $\hat{\bm{e}}_{x}=\bm{E}/E$) and the relation,
\begin{flalign}
\label{k parallel}%%%%%%%%%%%
k_{x}=\bm{k}\cdot\hat{\bm{e}}_{x},
\end{flalign}
implies that
\begin{flalign}
\label{k decomp}%%%%%%%%%%%
\bm{k}&=k_{x}\hat{\bm{e}}_{x}+\bm{k}_{\perp},\\
\label{k component}%%%%%%%%%%%
\bm{k}_{\perp}&=(0,\,k_y,\,k_z)^{\mathrm{T}}.
\end{flalign}

%%%%%%%%%%%%%%%%%%%%%%%%%%
{\itshape Density-density and charge-charge correlation functions}.---
One of the benefits of stochastic equations is that correlation functions are calculated straightforwardly.
Here we consider density-density and charge-charge correlation functions at equal times, which are defined using the equal-time correlation matrix as follows:
\begin{flalign}
\bm{\mathcal{C}}(\bm{k},t)
&=\left<\bm{\theta}(\bm{k},t)\bm{\theta}(-\bm{k},t)^{\mathrm{T}}\right>_{\zeta}\nonumber\\
&=
\begin{pmatrix}
\left<\rho(\bm{k})\rho(-\bm{k},t)\right>_{\zeta}&\left<q(\bm{k})\rho(-\bm{k},t)\right>_{\zeta}\\
\left<\rho(\bm{k})q(-\bm{k},t)\right>_{\zeta}&\left<q(\bm{k})q(-\bm{k},t)\right>_{\zeta}\\
\end{pmatrix}\nonumber\\
\label{C def}%%%%%%%%%%%%%%%%
&=
\begin{pmatrix}
\mathcal{C}_{\rho\rho}(\bm{k},t)&\mathcal{C}_{q\rho}(\bm{k},t)\\
\mathcal{C}_{\rho q}(\bm{k},t)&\mathcal{C}_{qq}(\bm{k},t)\\
\end{pmatrix}.
\end{flalign}
In the matrix elements given by eqn (\ref{C def}), $\mathcal{C}_{\rho\rho}$ and $\mathcal{C}_{qq}$ are the target correlation functions.
To be precise, $p^2e^2\mathcal{C}_{qq}(\bm{k},t)$ is the charge-charge correlation function, according to the definition of eqn (\ref{charge def}).
Nevertheless, we will refer to $\mathcal{C}_{qq}(\bm{k},t)$ as the charge-charge correlation function for brevity in the following.

We focus on the steady-state solutions of the correlation functions:
\begin{flalign}
\label{steady1}%%%%%%%%%%%%
\mathcal{C}^{\mathrm{st}}_{\rho\rho}(\bm{k})=\lim_{t\rightarrow\infty}\mathcal{C}_{\rho\rho}(\bm{k},t),\\
\label{steady2}%%%%%%%%%%%
\mathcal{C}^{\mathrm{st}}_{qq}(\bm{k})=\lim_{t\rightarrow\infty}\mathcal{C}_{qq}(\bm{k},t).
\end{flalign}
As detailed below, these are written as
\begin{flalign}
\label{C solution2}%%%%%%%%%%%%%
\frac{1}{(2\pi)^3}
\begin{pmatrix}
\mathcal{C}^{\mathrm{st}}_{\rho\rho}(\bm{k})\\
\mathcal{C}^{\mathrm{st}}_{qq}(\bm{k})
\end{pmatrix}
&=\frac{2\overline{n}\,\mathcal{G}_2(\bm{k})}{\mathrm{det}\,\bm{\mathcal{P}}(\bm{k})}\,
\widetilde{\bm{\mathcal{P}}}(\bm{k})
\begin{pmatrix}
\bm{k}^2\\
\bm{k}^2\\
\end{pmatrix},
\end{flalign}
where we have
\begin{flalign}
\label{pole}%%%%%%%%%%%%
\frac{\mathcal{G}_2(\bm{k})}{\mathrm{det}\,\bm{\mathcal{P}}}
&=\frac{1}{\mathcal{G}_2(\bm{k})
\left\{
\bm{k}^2\mathcal{G}_1(\bm{k})
+k^2_{x}(pE)^2
\right\}},\\
\label{g2 def}%%%%%%%%%%%
\mathcal{G}_2(\bm{k})&=2\bm{k}^2+\overline{\kappa}^2\omega(\bm{k}),
\end{flalign}
and the adjugate matrix of $\bm{\mathcal{P}}(\bm{k})$, signified by $\widetilde{\bm{\mathcal{P}}}(\bm{k})$, reads
\begin{flalign}
\label{def cofactor B}%%%%%%%%%%
\widetilde{\bm{\mathcal{P}}}(\bm{k})=
\begin{pmatrix}
\mathcal{G}_1(\bm{k})\mathcal{G}_2(\bm{k})
+k^2_{x}(pE)^2&k^2_{x}(pE)^2\\
k^2_{x}(pE)^2&\bm{k}^2\mathcal{G}_2(\bm{k})+k^2_{x}(pE)^2\\
\end{pmatrix}.
\end{flalign}
Eqn (\ref{pole}) indicates that both $\mathcal{C}^{\mathrm{st}}_{\rho\rho}$ and $\mathcal{C}^{\mathrm{st}}_{qq}$ have identical poles under the external electric field $\bm{E}$, which will be further investigated in Sections III and V.

In the limit of $k\rightarrow 0$, we have
\begin{flalign}
\label{C solution limit}%%%%%%%%%%%%%
\frac{1}{(2\pi)^3}
\begin{pmatrix}
\mathcal{C}^{\mathrm{st}}_{\rho\rho}(\bm{0})\\
\mathcal{C}^{\mathrm{st}}_{qq}(\bm{0})
\end{pmatrix}
&=
\begin{pmatrix}
2\overline{n}\\
0\\
\end{pmatrix},
\end{flalign}
noting that $\omega(\bm{0})=1$ for $\omega(\bm{k})$ given by eqn (\ref{omega two}).
Then, we divide $\mathcal{C}^{\mathrm{st}}_{\rho\rho}(\bm{k})$ into two parts:
$\mathcal{C}^{\mathrm{st}}_{\rho\rho}(\bm{k})/(2\pi)^3=2\overline{n}+\Delta \mathcal{C}^{\mathrm{st}}_{\rho\rho}(\bm{k})/(2\pi)^3$ where
\begin{flalign}
\label{delta crr}%%%%%%%%%%
\frac{1}{(2\pi)^3}\Delta\mathcal{C}^{\mathrm{st}}_{\rho\rho}(\bm{k})
=\frac{-\overline{\kappa}^2\omega(\bm{k})k^2_{x}(pE)^2}
{\mathcal{G}_2(\bm{k})
\left\{
\bm{k}^2\mathcal{G}_1(\bm{k})
+k^2_{x}(pE)^2
\right\}}
\end{flalign}
is directly related to total correlation functions, or essential parts of density-density correlations.

Setting that $\omega(\bm{k})=1$ for simplicity, eqn (\ref{C solution2}) is approximated by
\begin{flalign}
&\frac{1}{(2\pi)^3}
\begin{pmatrix}
\mathcal{C}^{\mathrm{st}}_{\rho\rho}(\bm{k})\\
\mathcal{C}^{\mathrm{st}}_{qq}(\bm{k})
\end{pmatrix}
\nonumber\\
&=
2\overline{n}\left\{
\begin{pmatrix}
1\\
1\\
\end{pmatrix}
+\frac{k^2_{x}(pE\overline{\kappa}^{-1})^2}{\bm{k}^2+k^2_{x}(pE\overline{\kappa}^{-1})^2}
\begin{pmatrix}
-1\\
1\\
\end{pmatrix}
\right\}
\begin{pmatrix}
1\\
\left.\bm{k}^2\middle/\overline{\kappa}^2\right.\\
\end{pmatrix}
\label{ans large}%%%%%%%%%%%
\end{flalign}
in the low wavenumber region of $k\overline{\kappa}^{-1}\ll 1$; see Appendix D for the derivation.
It should be noted that eqn (\ref{ans large}) agrees with the expression previously obtained in a different manner and that the Fourier transform of eqn (\ref{ans large}) has been demonstrated to provide anisotropic long-range correlation functions exhibiting a power-law behavior with a dipolar character \cite{gole}.
At low field strength of $pE\overline{\kappa}^{-1}\ll 1$, eqn (\ref{ans large}) converges to
\begin{flalign}
\label{ans E0}%%%%%%%%%%%%%%
\frac{1}{(2\pi)^3}
\begin{pmatrix}
\mathcal{C}^{\mathrm{st}}_{\rho\rho}(\bm{k})\\
\mathcal{C}^{\mathrm{st}}_{qq}(\bm{k})
\end{pmatrix}
&\rightarrow2\overline{n}
\begin{pmatrix}
1\\
\left.\bm{k}^2\middle/\overline{\kappa}^2\right.\\
\end{pmatrix},
\end{flalign}
clarifying that low electric-field-driven electrolytes in steady states mimic weakly interacting ionic fluids without applied electric field on a large scale and that $\mathcal{C}^{\mathrm{st}}_{qq}(\bm{k})$ given in eqn (\ref{ans E0}) satisfies not only the electroneutrality but also the Stillinger-Lovett second-moment condition.

%%%%%%%%%%%%%%%%%%%%%%%
%%%%%%%%%%%%%%%%%%%%%%%%
\section{Correlation function analysis: electric-field-induced crossover to a damped oscillatory state}
The first two subsections will be devoted to what is implied by the complicated forms (\ref{C solution2}) to (\ref{delta crr}) of correlation functions, $\mathcal{C}^{\mathrm{st}}_{qq}(\bm{k})$ and $\Delta\mathcal{C}^{\mathrm{st}}_{\rho\rho}(\bm{k})$, especially focusing on the high wavenumber in the external field direction.
While Section IIIA provides the pole equations of the correlation functions, Section IIIB clarifies the electric-field-induced oscillations on target when considering the solutions to the pole equation (\ref{general pole2}), or an anisotropic crossover from monotonic to oscillatory decay of correlations along the direction of applied electric field.
Before going into the numerical details of the results obtained from the pole equation (\ref{general pole2}), Section IIIC aims to understand the relationship between the Kirkwood crossover at $E=0$ and $E\neq 0$ using Fig. 2, a schematic plot of the decay length $\xi^{(1)}_{\mathrm{Decay}}$.
After presenting the schematic summary, Section IIID explains how the Kirkwood crossover point under electric field is determined in the anisotropic approximation of the pole equation (\ref{general pole2}).
In the anisotropic approximation (\ref{anisotropic}), we can analytically investigate the electric-field-induced Kirkwood crossover (see Section IV for details). Fig. 3 in Section IIID gives the numerical results on the $E$--dependencies of both the decay length $\xi^{(*1)}_{\mathrm{Decay}}$ and the smeared Debye-H\"uckel length $\xi_{\mathrm{DH}}^{(*1)}$ at the Kirkwood crossover points.
Last, Section IIIE presents various results on the 2D inverse Fourier transforms of $\Delta\mathcal{C}^{\mathrm{st}}_{\rho\rho}(\bm{k})$.
We will see anisotropic density-density correlations reflecting the emergence of stripe states in a high-density region above the Kirkwood crossover, which corroborates the anisotropic approximation (\ref{anisotropic}).

%%%%%%%%%%%%%%%%%%%%%%%%
\subsection{Electric-field-induced synchronization between the emergence of density and charge oscillations}
It is found from the denominator on the rhs of eqn (\ref{pole}) that the obtained correlation functions given by eqn (\ref{C solution2}) to (\ref{delta crr}) provide the following pole equations:
\begin{flalign}
\label{pole original2}%%%%%%%%%%%
&\left(\bm{k}^{(1)}\right)^2+\overline{\kappa}^2\omega(\bm{k}^{(1)})
+\frac{\left(k^{(1)}_{x}\right)^2}{\left(\bm{k}^{(1)}\right)^2}(pE)^2=0,\\
\label{pole original1}%%%%%%%%%%%
&2\left(\bm{k}^{(2)}\right)^2+\overline{\kappa}^2\omega(\bm{k}^{(2)})=0,
\end{flalign}
which remarkably apply to both density-density and charge-charge correlation functions.
Namely, both density-density and charge-charge correlation functions exhibit the same behavior.

Let us discuss the concrete behaviors particularly in the anisotropic approximation of eqn (\ref{k component}) such that
\begin{flalign}
\label{anisotropic}%%%%%%%%%%%
\bm{k}^{(j)}\approx
k^{(j)}_{x}\hat{\bm{e}}_{x}
\end{flalign}
(see Section IIIB for details).
Focusing on the onset of oscillatory decay of correlations (or the Kirkwood crossover) at a fixed electric field, a summary provided in advance is threefold:
\begin{enumerate}
\item {\itshape Simultaneous emergence of density and charge oscillations}.---
The weight function $\omega(\bm{k})$ multiplied by $\overline{\kappa}^2$ allows us to have complex solutions to the pole equations (\ref{pole original2}) and (\ref{pole original1}), other than purely imaginary solutions.
The appearance of real solutions corresponds to the onset of oscillatory correlations.
Hence, we find that eqn (\ref{pole original2}) and (\ref{pole original1}), which are equally valid for density-density and charge-charge correlations, lead to simultaneous emergences of density and charge oscillations.
It is striking that the correlation function analysis directly predict the electric-field-induced synchronization between the emergence of density and charge oscillations.
The simultaneous occurrence of crossovers is in contrast to equilibrium crossover phenomena which emerge separately:
the equilibrium density-density and charge-charge correlation functions exhibit the Fisher-Widom \cite{under evans,kirkwood fw,fw} and Kirkwood \cite{under evans,kirkwood original,kirkwood various,various smearing,gauss smearing,kirkwood fw} crossovers, respectively.
%%%%%%%%%%%%%%%%%%
\item {\itshape Shifted crossover from monotonic to oscillatory decay of correlations}.---
We consider the case where a smallest value of the purely imaginary solution to either eqn (\ref{pole original2}) or (\ref{pole original1}) exists for
\begin{flalign}
\label{max kappa}%%%%%%%%%%
\overline{\kappa}\sigma\leq\overline{\kappa}^{(*j)}\sigma,
\end{flalign}
with the superscripts, (*1) and (*2), of the maxima denoting the upper bounds for eqn (\ref{pole original2}) and (\ref{pole original1}), respectively.
Namely, the solutions to eqn (\ref{pole original2}) and (\ref{pole original1}) become complex beyond $\overline{\kappa}^{(*1)}\sigma$ and $\overline{\kappa}^{(*2)}\sigma$, respectively.
It can be readily seen from eqn (\ref{pole original1}) that $\overline{\kappa}^{(*2)}\sigma$ is independent of $E$ but is larger than the conventional Kirkwood crossover value \cite{kirkwood original,kirkwood various,various smearing,gauss smearing,kirkwood fw} in the range of $1.0<\overline{\kappa}^{*}\sigma<1.2$ for symmetric electrolytes in equilibrium where the pole equation is $\bm{k}^2+\left(\overline{\kappa}^*\right)^2\omega(\bm{k})=0$:
it follows from eqn (\ref{pole original1}) that
\begin{flalign}
\label{conventional k}%%%%%%
\overline{\kappa}^{*}\sigma=\frac{\overline{\kappa}^{(*2)}\sigma}{\sqrt{2}}=\frac{\sigma}{\sqrt{2}\xi^{(*2)}_{\mathrm{DH}}}.
\end{flalign}
In contrast, eqn (\ref{pole original2}) implies that $\overline{\kappa}^{(*1)}\sigma$ depends on $E$ and is smaller than the above Kirkwood crossover value at $E=0$ due to additional screening effect measured by $pE$.
%%%%%%%%%%%%%%%%%%
\item {\itshape Finite decay length in the dilute limit}.---
Eqn (\ref{pole original2}) and (\ref{pole original1}) are reduced to
\begin{flalign}
\label{ze screening2}%%%%%%%%%%%
&f_1\left(k^{(1)}_{x}\right)\equiv\left(k^{(1)}_{x}\right)^2+\overline{\kappa}^2+(pE)^2=0,\\
\label{ze screening1}%%%%%%%%%%%
&f_2\left(k^{(2)}_{x}\right)\equiv\left(k^{(2)}_{x}\right)^2+0.5\overline{\kappa}^2=0,
\end{flalign}
respectively, when considering the anisotropic approximation (\ref{anisotropic}) and $\omega(\bm{k})=1$ for simplicity.
In the dilute limit of $\overline{\kappa}\rightarrow 0$, eqn (\ref{ze screening2}) yields a finite decay length $\xi^{(1)}_{\mathrm{Decay}}\approx(pE)^{-1}$ (see Section IIIB for detailed derivation), whereas eqn (\ref{ze screening1}) ensures the divergent behavior of the decay length $\xi^{(2)}_{\mathrm{Decay}}$ given by $\xi^{(2)}_{\mathrm{Decay}}=\sqrt{2}\overline{\kappa}^{-1}$.
It would be difficult to detect the former decay length $\xi^{(1)}_{\mathrm{Decay}}\approx(pE)^{-1}$ without the oscillatory behavior in a dilute solution;
however, the existence of non-vanishing decay length helps us understand the physics of the decay mode on target.
\end{enumerate}

%%%%%%%%%%%%%%%%%
Before proceeding to the electric-field-induced Kirkwood crossover in the anisotropic approximation (\ref{anisotropic}), we examine what is indicated by the hidden decay length $\xi^{(1)}_{\mathrm{Decay}}\approx(pE)^{-1}$ in terms of competing electrokinetics between electrophoresis and free diffusion.
We have an electrophoresis time, $L/(\mathcal{D}pE)$, for a variable length $L$ because the electrophoretic velocity is given by $(\mathcal{D}/k_BT)pEk_BT=\mathcal{D}pE$, remembering that the force $pE$ exerted on a single ion by the applied electric field is defined in units of $k_BT$ and that the mobility is given by $\mathcal{D}/k_BT$ according to the Einstein relation.
It follows that the equality between a required time for electrophoresis and free diffusion reads
\begin{flalign}
\label{equal}%%%%%%%
\frac{L}{\mathcal{D}pE}=\frac{L^2}{\mathcal{D}},
\end{flalign}
which is equivalent to the above relation $\xi^{(1)}_{\mathrm{Decay}}\approx(pE)^{-1}$ when $L=\xi^{(1)}_{\mathrm{Decay}}$.

Hence, eqn (\ref{equal}) indicates an electrokinetic crossover occurring at $L=\xi^{(1)}_{\mathrm{Decay}}$.
In the smaller scale of $L<\xi^{(1)}_{\mathrm{Decay}}$, free diffusion is dominant, and the electrophoretic migration path is blurred by diffusion.
Meanwhile, for an electrophoresis dominant length scale $L>\xi^{(1)}_{\mathrm{Decay}}$, fluctuations in diffusion processes become negligible in comparison with electrophoretic migration: spatial distribution of charged spheres in a steady state is mainly determined by particles migrating uniformly.
This electrokinetic aspect of a steady state provides an explanation of the hidden decay length $\xi^{(1)}_{\mathrm{Decay}}\approx(pE)^{-1}$ that remains finite even in the dilute limit of $\overline{\kappa}\rightarrow 0$.

%%%%%%%%%%%%%%%%%%%%%%%%%
\subsection{Electric-field-induced Kirkwood crossover on target}
Let $\bm{r}_{\perp}$ be a transverse vector $\bm{r}_{\perp}=(0,y,z)^{\mathrm{T}}$ similar to $\bm{k}_{\perp}$ defined by eqn (\ref{k component}) (see also Fig. 1).
The Fourier transform then reads
\begin{flalign}
\label{fourier1}%%%%%%%%%
&\Delta\mathcal{C}^{\mathrm{st}}_{\rho\rho}(\bm{k})\nonumber\\
&=\int d^2\bm{r}_{\perp}\int dx\,\Delta\mathcal{C}^{\mathrm{st}}_{\rho\rho}(\bm{r})\,e^{-ik_{x}x\,-i\bm{k}_{\perp}\cdot\bm{r}_{\perp}},
\\
&\mathcal{C}^{\mathrm{st}}_{qq}(\bm{k})\nonumber\\
\label{fourier2}%%%%%%%%
&=\int d^2\bm{r}_{\perp}\int dx\,\mathcal{C}^{\mathrm{st}}_{qq}(\bm{r})\,e^{-ik_{x}x\,-i\bm{k}_{\perp}\cdot\bm{r}_{\perp}}.
\end{flalign}
To clarify the "hidden" decay length given in the preceding subsection, we need to see the real-space representations of $\Delta\mathcal{C}^{\mathrm{st}}_{\rho\rho}(\bm{k})$ and $\mathcal{C}^{\mathrm{st}}_{qq}(\bm{k})$ defined by
\begin{flalign}
\label{ani c}%%%%%%%%%%%
\frac{1}{2\pi}
\begin{pmatrix}
\Delta\mathcal{C}^{\mathrm{st}}_{\rho\rho}(k_{x})\\
\mathcal{C}^{\mathrm{st}}_{qq}(k_{x})
\end{pmatrix}
=
\frac{1}{(2\pi)^3}
\int d^2\bm{k}_{\perp}
\begin{pmatrix}
\Delta\mathcal{C}^{\mathrm{st}}_{\rho\rho}(\bm{k})\\
\mathcal{C}^{\mathrm{st}}_{qq}(\bm{k})
\end{pmatrix}
(2\pi)^2\delta(\bm{k}_{\perp}),
\end{flalign}
following the anisotropic approximation (\ref{anisotropic}).
Accordingly, eqn (\ref{fourier1}) and (\ref{fourier2}) are reduced, respectively, to
\begin{flalign}
\label{aniso fourier1}%%%%%%%%%
&\Delta\mathcal{C}^{\mathrm{st}}_{\rho\rho}(k_{x})
=\int dx\,\,\overline{\Delta\mathcal{C}^{\mathrm{st}}_{\rho\rho}}(x)\,e^{-ik_{x}x\,},
\\
\label{aniso fourier2}%%%%%%%%
&\mathcal{C}^{\mathrm{st}}_{qq}(k_{x})
=\int dx\,\,\overline{\mathcal{C}^{\mathrm{st}}_{qq}}(x)\,e^{-ik_{x}x\,},
\end{flalign}
using smeared correlation functions which are integrated over a cross section transverse to the applied electric field:
\begin{flalign}
\label{smear1}%%%%%%%%%
&\overline{\Delta\mathcal{C}^{\mathrm{st}}_{\rho\rho}}(x)
=\int d^2\bm{r}_{\perp}\Delta\mathcal{C}^{\mathrm{st}}_{\rho\rho}(\bm{r}),
\\
\label{smear2}%%%%%%%%
&\overline{\mathcal{C}^{\mathrm{st}}_{qq}}(x)
=\int d^2\bm{r}_{\perp}\mathcal{C}^{\mathrm{st}}_{qq}(\bm{r}).
\end{flalign}
Correspondingly, the pole equations (\ref{pole original2}) and (\ref{pole original1}) are simplified, respectively, as
\begin{flalign}
\label{general pole2}%%%%%%%
&(k^{(1)}_{x}\sigma)^2+(\overline{\kappa}\sigma)^2\omega(k^{(1)}_{x})+(pE\sigma)^2=0,\\
\label{general pole1}%%%%%%%
&(k^{(2)}_{x}\sigma)^2+0.5(\overline{\kappa}\sigma)^2\omega(k^{(2)}_{x}) =0,
\end{flalign}
both of which are different not only from the Debye-H\"uckel-type equation, $\mathcal{G}_1(\bm{k})=0$, used in equilibrium electrolytes but also from the approximate forms (\ref{ze screening2}) and (\ref{ze screening1}) where $\omega(\bm{k})=1$.

The complex solutions $k^{(j)}_{x}\sigma$ ($j=1,\,2$) to eqn (\ref{general pole2}) and (\ref{general pole1}) are related, respectively, to the wavelengths $\mu^{(j)}$ and decaying lengths $\xi^{(j)}_{\mathrm{Decay}}$ of stationary correlation functions at equal times as follows:
\begin{flalign}
\label{complex solution}%%%%%%%%
&k^{(j)}_{x}\sigma=x^{(j)}+iy^{(j)},\\
\label{complex length}%%%%%%%%%%
&(x^{(j)},\,y^{(j)})=\left(\frac{2\pi\sigma}{\mu^{(j)}},\>\frac{\sigma}{\xi^{(j)}_{\mathrm{Decay}}}
\right).
\end{flalign}
Thus, we have clarified that the above expressions (\ref{aniso fourier1}) and (\ref{aniso fourier2}) of the anisotropic Fourier transforms satisfy the pole equations (\ref{general pole2}) and (\ref{general pole1}) with eqn (\ref{complex solution}) and (\ref{complex length}).
This leads to the averaged correlation functions expressed as
\begin{flalign}
\label{long smear1}%%%%%%%%%%%%
&\overline{\Delta\mathcal{C}^{\mathrm{st}}_{\rho\rho}}(x)
=\sum_{j=1}^2A_je^{-x\,/\xi^{(j)}_{\mathrm{Decay}}}\cos\left(\frac{2\pi x\,}{\mu^{(j)}}+\delta_a^{(j)}\right),\\
\label{long smear2}%%%%%%%%%
&\overline{\mathcal{C}^{\mathrm{st}}_{qq}}(x)
=\sum_{j=1}^2B_je^{-x\,/\xi^{(j)}_{\mathrm{Decay}}}\cos\left(\frac{2\pi x\,}{\mu^{(j)}}+\delta_b^{(j)}\right),
\end{flalign}
where it is noted that both of these density-density and charge-charge correlation functions have the same wavelengths of oscillations in addition to the identical decay lengths, reflecting the above electric-field-induced synchronization.

The above definition of averaged correlation functions given by eqn (\ref{smear1}) and (\ref{smear2}) enables us to investigate correlations between coarse-grained planes perpendicular to the electric field without consideration of lane formation.
In particular, we focus on the pole equation (\ref{general pole2}) that predicts an electric-field-induced shift of the Kirkwood crossover from a monotonic decay state to a damped oscillatory state.

%%%%%%%%% FIG2 %%%%%%%%%%%%%%
\begin{figure}[hbtp]
\begin{center}
\includegraphics[
width=6.5cm
]{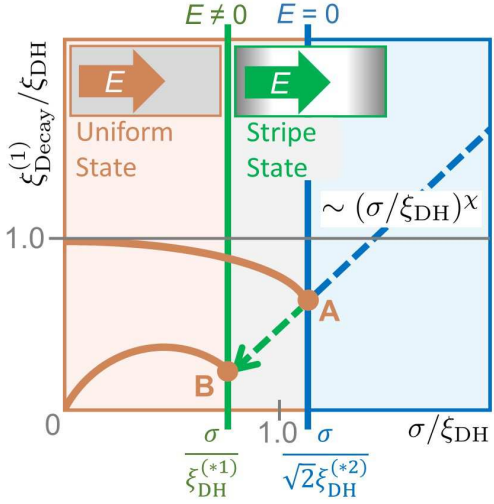}
\end{center}
\caption{
A schematic summary of results obtained in this study is depicted in terms of the decay length $\xi^{(1)}_{\mathrm{Decay}}$ of either monotonic decay or damped oscillatory on a log-log plot of $\xi^{(1)}_{\mathrm{Decay}}/\xi_{\mathrm{DH}}=\overline{\kappa}\xi_{\mathrm{Decay}}$ vs. $\sigma/\xi_{\mathrm{DH}}=\overline{\kappa}\sigma$ where $\xi_{\mathrm{DH}}=\overline{\kappa}^{-1}$ denotes the smeared Debye-H\"uckel screening length defined by eqn (\ref{smear dh}).
Our numerical results of $\xi^{(1)}_{\mathrm{Decay}}$ will be given in Figs. 3 and 8. In Fig. 2, these are shown using the solid brown lines terminated at nodes A and B, and the dashed green arrow from node A to node B, or from the Kirkwood crossover at $E=0$ to that under electric field ($E\neq 0$).
While the solid brown line terminated at node A ($E=0$) converges to $\xi^{(1)}_{\mathrm{Decay}}/\xi_{\mathrm{DH}}=1$ in the dilute limit, the solid brown line terminated at node B ($E\neq0$) approaches zero in the dilute limit because of the finite decay length $\xi^{(1)}_{\mathrm{Decay}}=(pE)^{-1}$ (see also a discussion given at the end of Section IIIA).
As illustrated by the upper inset, the green vertical line through node B marks $\sigma/\xi_{\mathrm{DH}}=\sigma/\xi_{\mathrm{DH}}^{(*1)}$, or the onset of shifted Kirkwood crossover from a uniform state to a stripe state {\itshape without consideration of lane formation} \cite{lowen,band}.
For comparison, we add the dashed blue line to show underscreening behavior in concentrated electrolytes without applied electric field beyond the conventional Kirkwood crossover indicated by the blue vertical line through node A; this blue line represents $\sigma/\xi_{\mathrm{DH}}=\sigma/(\sqrt{2}\xi_{\mathrm{DH}}^{(*2)})$, which is the conventional Kirkwood crossover value \cite{kirkwood original,kirkwood various,various smearing,gauss smearing,kirkwood fw} in the range of 1.0 to 1.2 for symmetric electrolytes as described prior to eqn (\ref{conventional k}).
Recent studies \cite{under simu,under th,under andelman,under evans} have demonstrated that $\xi^{(1)}_{\mathrm{Decay}}/\xi_{\mathrm{DH}}\sim(\sigma/\xi_{\mathrm{DH}})^{\chi}$ with the exponent of $\chi>1$ in a damped oscillatory state.
It will be seen from Figs. 3(b) and 8(b) that a similar scaling relation holds for the dashed green arrow.
}
\end{figure}.

More precisely, our focus is on the electric-field-induced Kirkwood crossover between the two regions specified below.
For $\overline{\kappa}\sigma\leq\overline{\kappa}^{(*1)}\sigma$, both solutions to eqn (\ref{general pole2}) and (\ref{general pole1}) are purely imaginary: eqn (\ref{long smear1}) and (\ref{long smear2}) read
\begin{flalign}
\label{region1 smear1}%%%%%%%%%%%%
&\overline{\Delta\mathcal{C}^{\mathrm{st}}_{\rho\rho}}(x)
=\sum_{j=1}^2A'_je^{-x\,/\xi^{(j)}_{\mathrm{Decay}}},\\
\label{region1 smear2}%%%%%%%%%
&\overline{\mathcal{C}^{\mathrm{st}}_{qq}}(x)
=\sum_{j=1}^2B'_je^{-x\,/\xi^{(j)}_{\mathrm{Decay}}},
\end{flalign}
respectively, where $A'_j=A_j\cos\left(\delta_a^{(j)}\right)$ and $B'_j=B_j\cos\left(\delta_b^{(j)}\right)$.
It is difficult to detect $\xi^{(1)}_{\mathrm{Decay}}$ because of $\xi^{(1)}_{\mathrm{Decay}}<\xi^{(2)}_{\mathrm{Decay}}\approx \sqrt{2}\overline{\kappa}^{-1}$ in a dilute ionic fluid, which is what we have meant by the "hidden" decay length.
In the range of $\overline{\kappa}\sigma>\overline{\kappa}^{(*1)}\sigma$, on the other hand, the solution to eqn (\ref{general pole2}) becomes complex while the solution to (\ref{general pole1}) is purely imaginary: eqn (\ref{region1 smear1}) and (\ref{region1 smear2}) transform to
\begin{flalign}
\overline{\Delta\mathcal{C}^{\mathrm{st}}_{\rho\rho}}(x)
=
&A_1e^{-x\,/\xi^{(1)}_{\mathrm{Decay}}}\cos\left(\frac{2\pi x\,}{\mu^{(1)}}+\delta_a^{(1)}\right)\nonumber\\
\label{region2 smear1}%%%%%%%%%
&+A'_2e^{-x\,/\xi^{(2)}_{\mathrm{Decay}}},\\
\overline{\mathcal{C}^{\mathrm{st}}_{qq}}(x)
=
&B_1e^{-x\,/\xi^{(1)}_{\mathrm{Decay}}}\cos\left(\frac{2\pi x\,}{\mu^{(1)}}+\delta_b^{(1)}\right)\nonumber\\
\label{region2 smear2}%%%%%%%%%
&+B'_2e^{-x\,/\xi^{(2)}_{\mathrm{Decay}}},
\end{flalign}
respectively.

The electric-field-induced Kirkwood crossover on target is thus represented by the changes of the correlation functions from eqn (\ref{region1 smear1}) and (\ref{region1 smear2}) to eqn (\ref{region2 smear1}) and (\ref{region2 smear2}), which occurs at $\overline{\kappa}\sigma=\kappa^{(*1)}\sigma$.
We further predict the Fisher-Widom crossover \cite{under evans,kirkwood fw,fw} that the density and charge oscillations become obvious in the range of $\kappa^{(*1)}\sigma<\overline{\kappa}\sigma<\kappa^{(*2)}\sigma$ where the two decay lengths, $\xi^{(1)}_{\mathrm{Decay}}$ and $\xi^{(2)}_{\mathrm{Decay}}$, approach each other;
however, it is beyond the scope of this paper to determine the full phase diagram using the steady-state extensions of the Kirkwood and Fisher-Widom crossovers \cite{under evans} related to eqn (\ref{general pole2}) and (\ref{general pole1}).

%%%%%%%%%%%%%%%%%%%%%%%%%%%%%
%%%%%%%%%%%%%%%%%%%%%%%%%%%%%
\subsection{Relationship between $E$--dependent solutions to eqn (\ref{general pole2}) and the equilibrium decay length}
Figure 2 shows a schematic representation of numerical results presented in Figs. 3 and 8.
In Fig. 2, the ratio of $\xi_{\mathrm{Decay}}^{(1)}$ to the smeared Debye-H\"uckel screening length $\xi_{\mathrm{DH}}$ (i.e., $\xi_{\mathrm{Decay}}^{(1)}/\xi_{\mathrm{DH}}$) is shown on a log-log plot as a function of $\sigma/\xi_{\mathrm{DH}}$.

First, it is seen from Fig. 2 that the equilibrium Kirkwood crossover point \cite{kirkwood original,kirkwood various,various smearing,gauss smearing,kirkwood fw} located at node A shifts gradually along the green arrow with the increase of electric field strength:
the dashed green arrow from node A to node B represents the numerical results shown in Figs. 3(b) and 8(b).
Incidentally, node B is merely an electric-field-induced Kirkwood crossover point at an arbitrary field strength.

Next, we explain the solid brown curves in Fig. 2 terminated at nodes A and B.
These curves represent the $\overline{\kappa}\sigma$--dependencies of $\xi_{\mathrm{Decay}}^{(1)}$ in a uniform state without and with applied electric field, respectively.
On the one hand, $\xi_{\mathrm{Decay}}^{(1)}$ at $E=0$ is identified with $\xi_{\mathrm{DH}}$ in the dilute limit of $\overline{\kappa}\sim n^{1/2}\rightarrow 0$ and decreases more rapidly than $\xi_{\mathrm{DH}}$ with increase of $\overline{n}$ in a uniform state prior to the Kirkwood crossover in equilibrium.
On the other hand, there are two features as seen from the brown curve under the applied electric field ($E\neq 0$):
the dilute limit of $\xi_{\mathrm{Decay}}^{(1)}/\xi_{\mathrm{DH}}$ approaches zero because of the finiteness of the decay length $\xi_{\mathrm{Decay}}^{(1)}$ in the limit of $\xi_{\mathrm{DH}}\rightarrow\infty$ as mentioned before, whereas the downward trend of $\xi_{\mathrm{Decay}}^{(1)}/\xi_{\mathrm{DH}}$, similar to the above behavior at $E=0$, is observed near the electric-field-induced Kirkwood crossover.

Third, let us turn our attention to the dashed blue line in Fig. 2 representing a typical underscreening behavior beyond the Kirkwood line (the vertical blue line through node A) with no electric field applied.
The dashed blue line in Fig. 2 depicts the following relation for a decay length $\xi_{\mathrm{Decay}}$:
\begin{flalign}
\label{decay scaling}%%%%%
&\frac{\xi_{\mathrm{Decay}}}{\xi_{\mathrm{DH}}}\sim
\left(\frac{\sigma}{\xi_{\mathrm{DH}}}\right)^{\chi},\\
\label{decay exponent}%%%%%
&1<\chi\leq 2,
\end{flalign}
according to previous simulation and theoretical studies \cite{under simu,under th,under andelman,under evans}; the experimental results of $\chi\approx 3$ in RTILs are beyond the scope of this study.
It follows from eqn (\ref{decay exponent}) that eqn (\ref{decay scaling}) reads
\begin{flalign}
&\xi_{\mathrm{Decay}}\sim \overline{n}^{\frac{1-\chi}{2}},\nonumber\\
&1-\chi<0.
\label{decay n}%%%%%
\end{flalign}
Eqn (\ref{decay n}) implies that the decay length $\xi_{\mathrm{Decay}}$ of damped oscillations for charge-charge correlations becomes longer despite increasing $\overline{n}$, which has been referred to as underscreening behavior without an electric field.

Remarkably, the scaling relation given by eqn (\ref{decay scaling}) and (\ref{decay n}) applies to the $\overline{\kappa}\sigma$--dependence of $\xi_{\mathrm{Decay}}^{(*1)}/\xi_{\mathrm{DH}}^{(*1)}$ at the Kirkwood crossover;
the exponent $\chi$ appears close to 1.4 as will be shown in Fig. 3(b) \cite{under simu,under th,under andelman,under evans}.
Reflecting this similarity between the exponents $\chi$ of the underscreening behavior and the electric-field-induced shift for $\xi_{\mathrm{Decay}}^{(*1)}$, the dashed blue line in Fig. 2 is drawn as an extension of the green arrow from node A ($E=0$) to node B ($E\neq 0$).

Last, we focus on the vertical green line through node B in Fig. 2, indicating the condition of the electric-field-induced Kirkwood crossover from a uniform state to a stripe state.
In the stripe state, we can observe a damped oscillatory decay of both density-density and charge-charge correlation functions along the direction of applied electric field in the anisotropic approximation (\ref{anisotropic}).
It is to be noted here that the stripe state is specified using the averaged correlation functions, $\overline{\Delta\mathcal{C}^{\mathrm{st}}_{\rho\rho}}(x)$ and $\overline{\mathcal{C}^{\mathrm{st}}_{qq}}(x)$, which by definition smear out density and charge distributions on cross sections perpendicular to the applied electric field (see eqn (\ref{smear1}) and (\ref{smear2})).

The emergence of anisotropic density and charge modulations is consistent with the previous results as follows:
Theoretically, the SDFT using the Gaussian charge smearing model \cite{gauss smearing,finite gaussian} has provided numerical results of two-dimensional correlation functions showing a tendency to form alternating chains of cations and anions along the field direction \cite{demery}.
Also, according to simulation and experimental studies on oppositely charged colloids, the electric-field-driven mixtures have been found to form bands non-parallel to the field direction, other than lanes in the electric field direction, under some conditions on various dynamic phase diagrams of steady states with AC or DC fields applied \cite{lowen,band}.
Our present findings of the emergence of stripe state thus shed light on these anisotropic inhomogeneities as related to crossover phenomena of steady-state correlations.

%%%%%%%%% FIG3 %%%%%%%%%%%%%%
\begin{figure}[hbtp]
\begin{center}
\includegraphics[
width=7.6cm
]{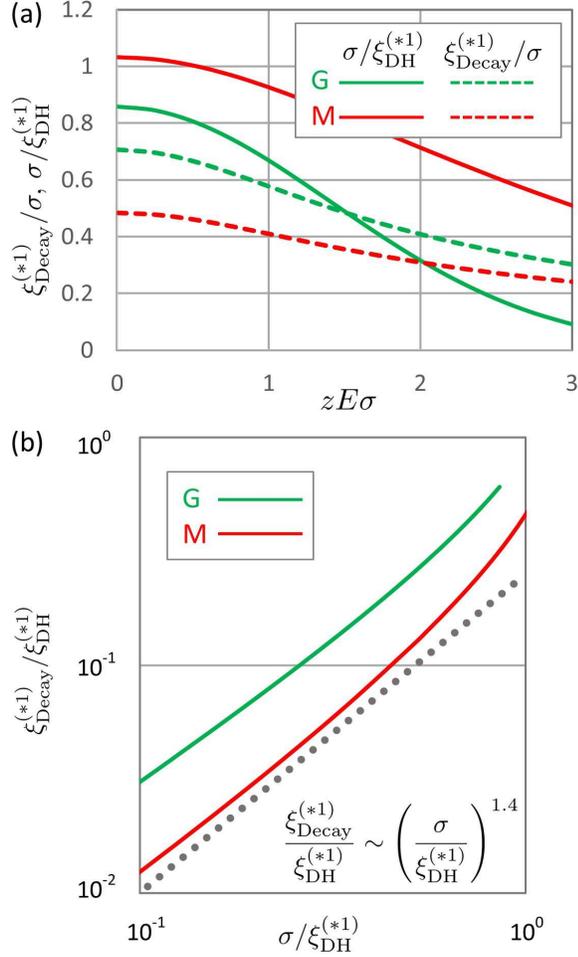}
\end{center}
\caption{
Comparison between the electric-field-dependent length results which are obtained from the Gaussian charge smearing model (abbreviated as "G" and represented by green lines) and the modified MSA model (abbreviated as "M" and represented by red lines).
Section IVB presents the detailed formulation to obtain the results given in this figure.
(a) The electric-field dependencies of the Debye-H\"uckel length $\xi_{\mathrm{DH}}^{(*1)}=1/\overline{\kappa}^{(*1)}$ and the decay length $\xi^{(*1)}_{\mathrm{Decay}}$ at the Kirkwood crossover are shown by the plot of $\sigma/\xi_{\mathrm{DH}}^{(*1)}$ and $\xi^{(*1)}_{\mathrm{Decay}}/\sigma$ against $pE\sigma$, an energetic measure of electric field strength in units of $k_BT$.
While the solid lines represent $\sigma/\xi_{\mathrm{DH}}^{(*1)}$, the dashed lines $\xi^{(*1)}_{\mathrm{Decay}}/\sigma$. The related equations are as follows: green solid line (eqn (\ref{lambert cross1})); green dashed line (eqn (\ref{xi el})); red solid and dashed lines (eqn (\ref{u func}) and (\ref{v func})).
(b) A log-log plot of $\sigma/\xi_{\mathrm{DH}}^{(*1)}$--dependencies of $\xi^{(*1)}_{\mathrm{Decay}}/\xi_{\mathrm{DH}}^{(*1)}$.
The dotted line, as a guide to the eye, indicates a scaling relation $\xi^{(*1)}_{\mathrm{Decay}}/\xi_{\mathrm{DH}}^{(*1)}\sim (\sigma/\xi_{\mathrm{DH}}^{(*1)})^{1.4}$.
}
\end{figure}
%%%%%%%%%%%%%%%%%%%%%%%%%
%%%%%%%%%%%%%%%%%%%%%%%%
\subsection{Numerical solutions to eqn (\ref{general pole2})}
Figure 3(a) shows the electric-field effects on the Kirkwood crossover in terms of the smeared Debye-H\"uckel length $\xi_{\mathrm{DH}}^{(*1)}$ and a decay length $\xi_{\mathrm{Decay}}^{(*1)}$ at the Kirkwood crossover.
As detailed in Section IV, we have obtained these results using both the Gaussian charge smearing model \cite{gauss smearing,finite gaussian} (or the HNC approximation for one-component ionic fluids \cite{ng}) and the modified MSA model for the DCF \cite{mmsa smearing}, or its essential function $\omega(\bm{k})$ given by eqn (\ref{omega two}).
The former model is depicted by green lines, whereas the latter by red lines.
All of the results in Fig. 3(a) exhibit downward trends in accordance with analytical observations made in Section IV.

Furthermore, Fig. 3(a) allows us to make quantitative comparisons between the present two models for the DCF.
First, it is confirmed from the values of $\xi_{\mathrm{DH}}^{(*1)}$ at $E=0$ in Fig. 3(a) that the numerical results correctly reproduce the Kirkwood crossover points previously obtained for the Gaussian charge smearing model \cite{various smearing,gauss smearing} and the modified MSA model \cite{under th, under andelman,under evans,various smearing}.
Second, Fig. 3(a) shows that the electric-field-induced shifts of $\sigma/\xi_{\mathrm{DH}}^{(*1)}$ are similar to each other.
Remembering that $\overline{n}^*=1/\{8\pi l_B(\xi^{(*1)}_{\mathrm{DH}})^2\}$ by definition (\ref{smear dh}), it is seen from the variations of $\sigma/\xi_{\mathrm{DH}}^{(*1)}$ in Fig. 3(a) that, irrespective of the models adopted, the crossover densities at $pE\sigma=3.0$ are evaluated to be less than half of those at $E=0$.
Last, we turn our attention to the relationship between $\xi_{\mathrm{Decay}}^{(*1)}$ and $1/\xi_{\mathrm{DH}}^{(*1)}$ as a function of either $\overline{n}^*$ or $E$.
For a fixed strength of applied electric field, the decay length $\xi_{\mathrm{Decay}}^{(*1)}$ becomes shorter as the ionic solution density $\overline{n}^*$, or $\sigma/\xi_{\mathrm{DH}}^{(*1)}$, becomes larger, which is consistent with the previous results conventionally found for concentrated electrolytes prior to the Kirkwood crossover without an applied electric field \cite{under andelman,under evans,kirkwood original,kirkwood various}.
The electric-field dependencies, on the other hand, exhibit an opposite relationship between $\xi_{\mathrm{Decay}}^{(*1)}$ and $1/\xi_{\mathrm{DH}}^{(*1)}$:
the downward trends in Fig. 3(a) indicate that both $\xi_{\mathrm{Decay}}^{(*1)}$ and $1/\xi_{\mathrm{DH}}^{(*1)}$ are smaller as $E$ is larger.

Figure 3(b) demonstrates this opposite tendency using a log-log plot of $\xi_{\mathrm{Decay}}^{(*1)}/\xi_{\mathrm{DH}}^{(*1)}$ vs. $\sigma/\xi_{\mathrm{DH}}^{(*1)}$:
it is seen from Fig. 3(b) that
\begin{equation}
\frac{\xi_{\mathrm{Decay}}^{(*1)}}{\xi_{\mathrm{DH}}^{(*1)}}
\sim
\left(\frac{\sigma}{\xi_{\mathrm{DH}}^{(*1)}}\right)^{\chi}
\label{scaling}%%%%%%%%
\end{equation}
for an exponent $\chi$ larger than unity.
The dotted line is a guide to the eye, indicating that $\chi$ is close to 1.4 and is consistent with the relation $1<\chi\leq 1.5$ previously obtained from simulation results on underscreening behaviors in RTILs beyond the Kirkwood line with no electric field applied \cite{under simu,under th,under andelman,under evans}.

%%%%%%%%%%%%%%%%%%%%%%%%
\subsection{The 2D inverse Fourier transforms for assessing the anisotropic approximation (\ref{anisotropic})}
The last subsection of Sec. III presents the results of the 2D inverse Fourier transforms using heat maps, which would help us not only to understand the above analytical results concretely but also to assess the anisotropic approximation (\ref{anisotropic}).
As a consequence of the inverse Fourier transforms, Figs. 4 to 6 provide real-space behaviors of the density-density correlation function in a high-density region such that $\overline{\kappa}\sigma$ is beyond not only the Kirkwood crossover \cite{kirkwood original,kirkwood various,various smearing,gauss smearing,kirkwood fw} but also the FIsher-Widom-like crossover \cite{under evans,kirkwood fw,fw}: $\overline{\kappa}\sigma>\overline{\kappa}^{(*2)}\sigma(>\overline{\kappa}^{(*1)}\sigma)$ is investigated (see also the discussion at the end of Sec. IIB) .
Figure 4 demonstrates that stripe states illustrated in Fig. 2 are observed more clearly as $\overline{\kappa}\sigma$ is larger at a fixed strength of electric field.
Figure 5(a) shows the correlation functions in the $x$--direction at a fixed $y$--coordinate.
We validate the anisotropic approximation (\ref{anisotropic}) from comparing Fig. 5(a) with Fig. 5(b) obtained from the 1D inverse Fourier transforms (see also eqn (\ref{aniso fourier1}) and (\ref{smear1}) for the definition of $\overline{\Delta\mathcal{C}^{\mathrm{st}}_{\rho\rho}}(x)$). 
Furthermore, Fig. 6 indicates the breaking down of stripe states: we can observe the emergence of a lane structure with the increase of electric field strength from $pE\sigma=0.1$ to $1.0$.

We perform the 2D inverse Fourier transform of $\Delta\mathcal{C}^{\mathrm{st}}_{\rho\rho}(k_x,\,k_y)$ by setting $k_z=0$ similar to the expressions (\ref{ani c}), (\ref{aniso fourier1}) and (\ref{smear1}):
\begin{flalign}
\label{smear fourier}%%%%%%%%%
\Delta\mathcal{C}^{\mathrm{st}}_{\rho\rho}(k_x,\,k_y)
&=\iint dxdy\,\overline{\Delta\mathcal{C}^{\mathrm{st}}_{\rho\rho}}(x,y)\,e^{-ik_{x}x-ik_yy},
\\
\label{smear z}%%%%%
\overline{\Delta\mathcal{C}^{\mathrm{st}}_{\rho\rho}}(x,y)
&\equiv\int dz\,\Delta\mathcal{C}^{\mathrm{st}}_{\rho\rho}(\bm{r}).
\end{flalign}
Correspondingly, the inverse Fourier transform provides the mean correlation function $\overline{\Delta\mathcal{C}^{\mathrm{st}}_{\rho\rho}}(x,\,y)$ as follows:
\begin{flalign}
\label{2d inverse}%%%%%%%
\overline{\Delta\mathcal{C}^{\mathrm{st}}_{\rho\rho}}(x,y)
=\frac{1}{(2\pi)^2}\iint dk_xdk_y\,\Delta\mathcal{C}^{\mathrm{st}}_{\rho\rho}(k_x,\,k_y)
e^{ik_{x}x+ik_yy},
\end{flalign}
which is relevant as long as the translational symmetry of $\Delta\mathcal{C}^{\mathrm{st}}_{\rho\rho}(\bm{r})$ is preserved with  respect to the $z$--direction and the correlation functions on the $xy$ cross-sections are indistinguishable at two different $z$ values.
The setup in Fig. 1 is one plausible example to satisfy such translational symmetry.
The upper figure of Fig. 1 indicates that the plate-plate distance is sufficiently smaller than the size in the $z$--direction, and yet we suppose that the finite-size effects are negligible because the plate-plate distance is much larger than the sphere diameter as mentioned in Sec. IIA.
These premises allow us to investigate the 2D inverse Fourier transforms of the 3D primitive model.

%%%%%%%%% FIG4 %%%%%%%%%%%%%%
\begin{figure}[hbtp]
\begin{center}
\includegraphics[
width=7.8cm
]{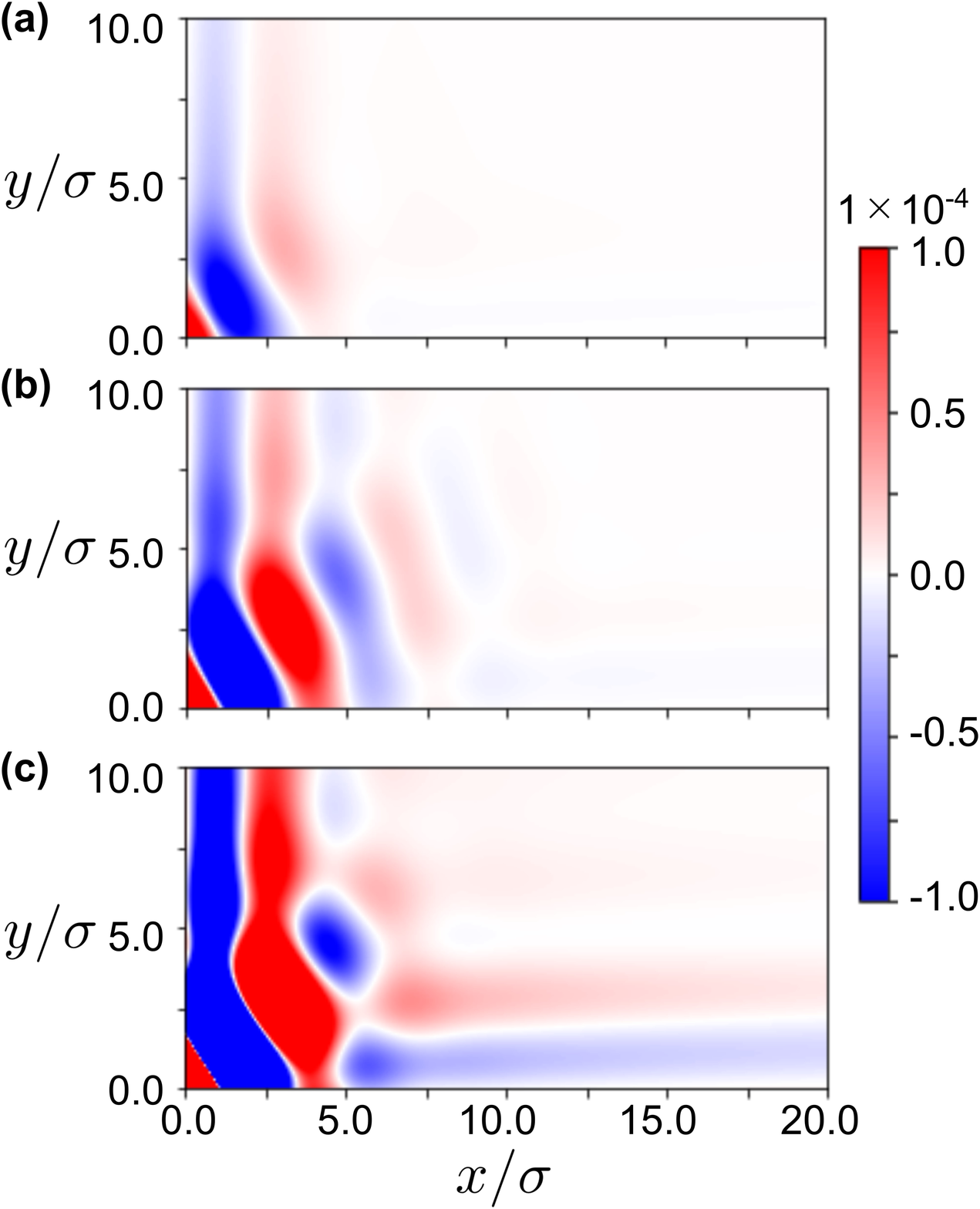}
\end{center}
\caption{Comparison between the 2D results of $\overline{\Delta\mathcal{C}^{\mathrm{st}}_{\rho\rho}}(x,y)$ for different conditions on ionic condition and electric field strength. The color bar on the right hand side, which is common to the three heat maps, represents the value of $\overline{\Delta\mathcal{C}^{\mathrm{st}}_{\rho\rho}}(x,y)$ at a location $(x/\sigma,\,y/\sigma)$ measured in units of sphere diameter $\sigma$. While the difference between Figs. 4(a) and 4(b) is ionic concentration, or $\overline{\kappa}\sigma$, at an identical electrical field, an electric field effect is seen from comparing Figs. 4(b) and 4(c) at a same ionic concentration: (a) $(\overline{\kappa}\sigma,\,pE\sigma)=(2.2,\,0.5)$; (b) $(\overline{\kappa}\sigma,\,pE\sigma)=(2.6,\,0.5)$; (c) $(\overline{\kappa}\sigma,\,pE\sigma)=(2.6,\,1.5)$.}
\end{figure}
%%%%%%%%%%%%%%%%%%%%%%%%%%%

Figure 4 shows how the density-density correlation behaviors vary depending on the ionic concentration and electric field strength. The difference between Figs. 4(a) and 4(b) is the ionic concentration at the same electric field of $pE\sigma=0.5$.
Meanwhile, the difference between Figs. 4(b) and 4(c) is the strength of electric field at the same ionic condition of $\overline{\kappa}\sigma=2.6$.
Figure 4(b) can be a reference result for investigating the effects of ionic concentration and electric field strength.
Figure 4(b) exhibits the oscillatory decay behaviors in the external field direction on an $xy$ plane, which is typical of density-density correlations in the stripe state.

When $\overline{\kappa}\sigma$ is reduced from 2.6 to 2.2 without changing the electric field strength, we obtain the result of Fig. 4(a).
Comparison between Figs. 4(a) and 4(b) indicates the following.
First, we can observe the oscillatory decays in the external field direction for both values of $\overline{\kappa}\sigma$ when setting the electric field strength to be $pE\sigma=0.5$. 
Furthermore, Fig. 4(a) shows that the correlation function becomes almost zero for $x\geq 5\sigma$: the density-density correlation function becomes equal to $2\overline{n}\delta(\bm{r})$ for $x\geq 5\sigma$  in contrast to the long-range correlations seen in Fig. 4(b). The different behaviors of density-density correlations suggest that the smaller $\overline{\kappa}\sigma$ is, the shorter the decay length becomes. In other words, comparison between Figs. 4(a) and 4(b) reveals an underscreening behavior \cite{under exp,under simu,under th,under andelman,under evans} beyond the Kirkwood condition as depicted in the schematic of Fig. 2.
Figure 4(c) further demonstrates that alignment of segregation band to the external field direction becomes clear by increasing the electric field strength to $pE\sigma=1.5$ at $\overline{\kappa}\sigma=2.6$.

%%%%%%%%% FIG5 %%%%%%%%%%%%%%
\begin{figure}[hbtp]
\begin{center}
\includegraphics[
width=7.9cm
]{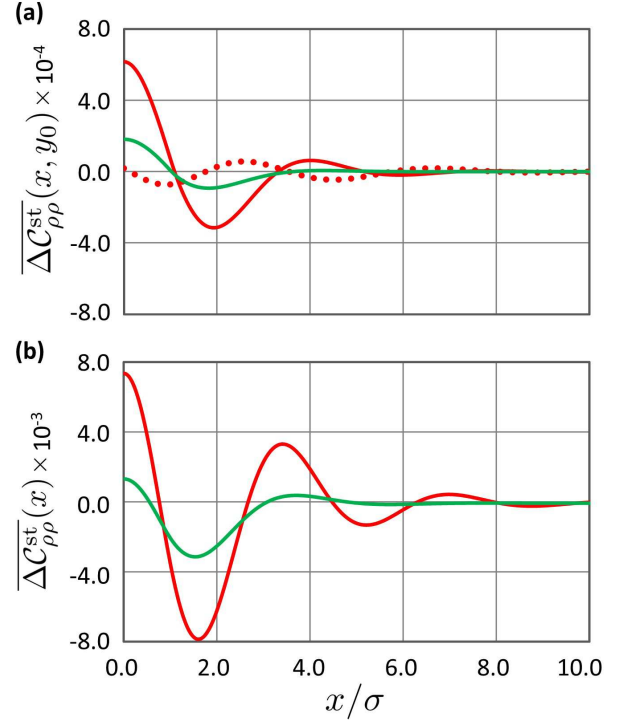}
\end{center}
\caption{Comparison between the 1D results of $\overline{\Delta\mathcal{C}^{\mathrm{st}}_{\rho\rho}}(x,y_0)$ at a fixed $y$--coordinate value of $y_0$ (Fig. 5(a)) and $\overline{\Delta\mathcal{C}^{\mathrm{st}}_{\rho\rho}}(x)$ defined by eqn (\ref{smear1}) (Fig. 5(b)). The density-density correlation functions are plotted as functions of $x/\sigma$, the separation distance in the applied field direction measured in units of sphere diameter $\sigma$. At an identical electric field strength $pE\sigma=0.5$, two ionic concentrations, $\overline{\kappa}\sigma=2.2$ and 2.6, are considered in both figures: (a) the green and red solid lines depict the behaviors of $\overline{\Delta\mathcal{C}^{\mathrm{st}}_{\rho\rho}}(x,y_0=0)$ at $\overline{\kappa}\sigma=2.2$ (green) and 2.6 (red), respectively, whereas the red dotted line represents $\overline{\Delta\mathcal{C}^{\mathrm{st}}_{\rho\rho}}(x,y_0=5\sigma)$ at $\overline{\kappa}\sigma=2.6$; (b) the green and red solid lines depict the behaviors of $\overline{\Delta\mathcal{C}^{\mathrm{st}}_{\rho\rho}}(x)$ at $\overline{\kappa}\sigma=2.2$ and 2.6, respectively.}
\end{figure}
%%%%%%%%%%%%%%%%%%%%%%%%%

Figure 4 has found an external field condition ($pE\sigma=0.5$) that creates an anisotropic density modulation reflecting the stripe state as depicted in Fig. 2.
This finding has justified the anisotropic approximation (\ref{anisotropic}) from a qualitative point of view.
We make below a quantitative assessment of the anisotropic approximation (\ref{anisotropic}).
To this end, we further investigate the extent to which the one-variable correlation function represents the results of Fig. 4 using Figs. 5 and 6.

Figure 5 compares the $x$-dependencies of the 2D correlation function $\overline{\Delta\mathcal{C}^{\mathrm{st}}_{\rho\rho}}(x,y_0)$ at $y_0$, a fixed $y$-coordinate value, with the behaviors of the one-variable correlation function $\overline{\Delta\mathcal{C}^{\mathrm{st}}_{\rho\rho}}(x)$ defined by eqn (\ref{smear1}).
Both solid lines in Fig. 5(a) show the $x$-dependencies at $y_0=0$.
The same external field condition $pE\sigma=0.5$ is used in both results of Figs. 5(a) and 5(b), and the ionic conditions for the green and red lines are identical in Figs. 5(a) and 5(b): the green and red lines represent the results at $\overline{\kappa}\sigma=2.2$ and 2.6, respectively.

It is noted that the value of the vertical axis in Fig. 5(a) is one-tenth of that in Fig. 5(b) due to the different definitions of the two correlation functions.
Nevertheless, the behaviors bear resemblances.
First, these two functions, $\overline{\Delta\mathcal{C}^{\mathrm{st}}_{\rho\rho}}(x,y_0)$ and $\overline{\Delta\mathcal{C}^{\mathrm{st}}_{\rho\rho}}(x)$, exhibit oscillatory decay behaviors, and we will make a quantitative comparison using Fig. 6.
They also share the feature of correlation value that becomes smaller with the decrease of ionic concentration from $\overline{\kappa}\sigma=2.6$ to 2.2.
Furthermore, we observe that the oscillations disappear faster for the green line than for the red line in both Figs. 5(a) and 5(b), which corresponds to the underscreening behavior suggested by Fig. 4.

The solid lines in Fig. 5(a) are the results at a fixed $y$--coordinate: $y_0=0$.
The specific value of $y_0$ raises the question as to whether or not the above similarity of solid lines in Figs. 5(a) and 5(b) is a coincidence.
To address this question, the red dashed line shows the $x$-dependency of the two-variable function at $y_0/\sigma=5$ when $pE\sigma=0.5$ and $\overline{\kappa}\sigma=2.2$.
We can see that the period of the dashed red line is close to that of the solid red line.
However, the initial phase is different from that at $y_0/\sigma=0$, and the correlation value is reduced considerably even at $x/\sigma=0$ as $y_0/\sigma$ varies from 0 to 5.
The latter difference implies that the $x$-dependency of $\overline{\Delta\mathcal{C}^{\mathrm{st}}_{\rho\rho}}(x,y_0)$ near $y_0/\sigma=0$ greatly contributes to the one-variable correlation function $\overline{\Delta\mathcal{C}^{\mathrm{st}}_{\rho\rho}}(x)$ defined by eqn (\ref{smear1}), which is why the two solid lines in Fig. 4(a) reproduce the one-variable function behaviors in Fig. 4(b).
%%%%%%%%% FIG5 %%%%%%%%%%%%%%
\begin{figure}[hbtp]
\begin{center}
\includegraphics[
width=8cm
]{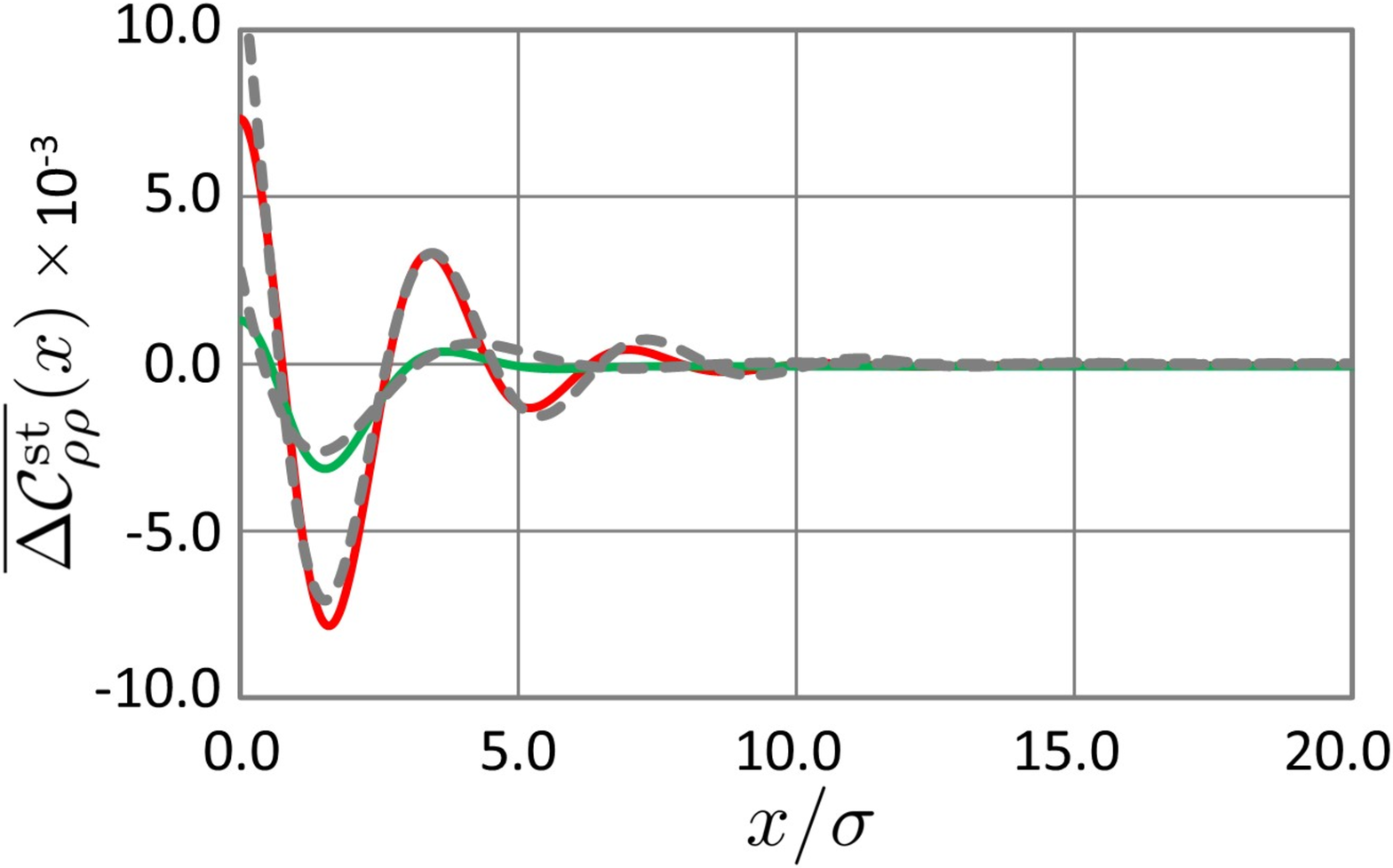}
\end{center}
\caption{The green and red lines are the same results as those in Fig. 5(b): we show $\overline{\Delta\mathcal{C}^{\mathrm{st}}_{\rho\rho}}(x)$ over the range, $0\leq x/\sigma\leq20$, at $\overline{\kappa}\sigma=2.2$ (green) and 2.6 (red) in the presence of applied electric field ($pE\sigma=0.5$). The dashed lines correspond to the best fit of eqn (\ref{single smear1}).}
\end{figure}
%%%%%%%%%%%%%%%%%%%%%%%%%%%
%%%%%%%%%%%%%%%%%%%%%%%%

Let us consider a simple asymptotic form determined by a single decay length $\xi_{\mathrm{Decay}}$ and oscillation period $\mu$:
\begin{flalign}
\label{single smear1}
\overline{\Delta\mathcal{C}^{\mathrm{st}}_{\rho\rho}}(x)
=A\,e^{-x\,/\xi_{\mathrm{Decay}}}\cos\left(\frac{2\pi x}{\mu}+\delta_a\right),
\end{flalign}
which is fitted to the results of Fig. 6 instead of eqn (\ref{long smear1}).
While the solid lines in Fig. 6, which are the same as those of Fig. 5(b), are shown over the range $0\leq x/\sigma\leq 20$, the dashed lines in Fig. 6 correspond to the best fit of eqn (\ref{single smear1}).
The best-fit parameter sets are as follows: $(A,\,\xi_{\mathrm{Decay}},\,\mu,\,\delta_a)=(0.6\times 10^{-2},\,2.0,\,5.6,\,1.1)$ at $\overline{\kappa}\sigma=2.2$, whereas $(A,\,\xi_{\mathrm{Decay}},\,\mu,\,\delta_a)=(1.3\times 10^{-2},\,2.5,\,3.8,\,0.4)$ at $\overline{\kappa}\sigma=2.6$.
The best-fit periods, $\mu=5.6$ and 3.8, reflect the oscillatory behaviors seen from Fig. 6.
Meanwhile, the best-fit decay length $\xi_{\mathrm{decay}}$ extends from $2\sigma$ to $2.5\sigma$ with the increase of $\overline{\kappa}\sigma$ from 2.2 to 2.6, which is a quantitative result of underscreening behavior.
Evaluating the exponent $\chi$ defined by eqn (\ref{scaling}) from this increase in $\xi_{\mathrm{Decay}}$, we have $2<\chi<3$;
it is interesting to note that the present exponent is larger than the equilibrium exponent ($1<\chi\leq 1.5$) previously obtained from the MSA of the 3D primitive model but is close to the exponent experimentally obtained \cite{under exp}.

%%%%%%%%% FIG6 %%%%%%%%%%%%%%
\begin{figure}[hbtp]
\begin{center}
\includegraphics[
width=6.9cm
]{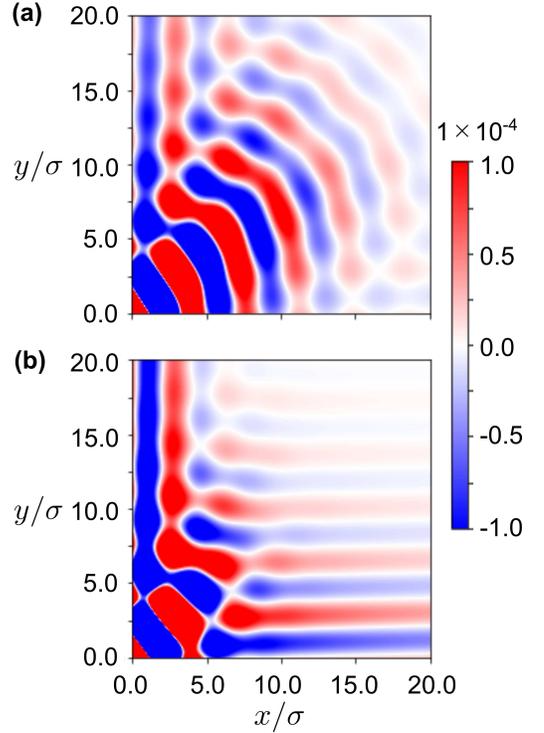}
\end{center}
\caption{Comparison between the 2D results of $\overline{\Delta\mathcal{C}^{\mathrm{st}}_{\rho\rho}}(x,y)$ at different electric field strengths: (a) $pE\sigma=0.1$ and (b) $pE\sigma=1.0$. The same ionic condition $\overline{\kappa}\sigma=2.78$ is adopted in both results.}
\end{figure}
%%%%%%%%%%%%%%%%%%%%%%%%%%%
%%%%%%%%%%%%%%%%%%%%%%%%

At the end of Section II, we consider the 2D inverse Fourier transforms of  $\overline{\Delta\mathcal{C}^{\mathrm{st}}_{\rho\rho}}(x,y)$ when increasing $\kappa\sigma$ to 2.78.
The strengths of the applied electric field are $pE\sigma=0.1$ (Fig. 7(a)) and $pE\sigma=1.0$ (Fig. 7(b)).
The heat maps in Fig. 7 reveal the oscillatory 2D patterns due to the suppression of decaying behaviors. 
On the one hand, even at the weak electric field strength of $pE\sigma=0.1$, Fig. 7(a) shows that segregation bands of ions with the same sign are deformed along the $x$--axis, the external field direction, though the stripe state remains a good approximation in the region of $y/\sigma\leq 5$.
On the other hand, at $pE\sigma=1.0$, Fig. 7(b) demonstrates the emergence of lane structure formed by aligned bands.

\section{Details on the correlation function analysis presented in Section III}
We perform the correlation function analysis, especially focusing on the pole equation (\ref{general pole2}).
In Section IVA, we see that discriminant analysis of a quadratic equation becomes available to investigate the solution to eqn (\ref{general pole2}), irrespective of the function forms of $\omega(\bm{k})$, as a result of the small $k_{x}\sigma$--expansion of the key function $\omega(\bm{k})$.
Section IVB provides concrete results of both the Gaussian charge smearing model and the modified MSA model for clarifying how the results in Fig. 3 are obtained.
In Section IVC, we relax the condition $\bm{k}_{\perp}=\bm{0}$ which has been referred to as the anisotropic approximation (see eqn (\ref{anisotropic})).
Then, the density-density correlation function analysis indicates that a long-range correlation in the perpendicular direction to $\bm{E}$ is enhanced on approaching the target mode in the external field direction (i.e., $k_{x}\rightarrow k_{x}^{(1)}$).
%%%%%%%%%%%%%%%%%%
\subsection{A general approximation of eqn (\ref{general pole2}) for evaluating the Kirkwood crossover point}
Expanding $\omega(\bm{k})$ with respect to $k_{x}^{(1)}\sigma$, we have a general form,
\begin{flalign}
\omega(k_{x}^{(1)})\approx
1-\alpha_1 (k_{x}^{(1)}\sigma)^2+\alpha_2(k_{x}^{(1)}\sigma)^4,
\label{f approx}%%%%%%%%
\end{flalign}
as seen from eqn (\ref{omega two});
for instance, $\alpha_1=1/2$ and $\alpha_2=1/8$ for the Gaussian charge smearing model \cite{gauss smearing,finite gaussian}, and $\alpha_1=1/2$ and $\alpha_2=1/24$ for the modified MSA model \cite{mmsa smearing}.
Eqn (\ref{general pole2}) then reduces to the quadratic equation for $\mathcal{S}\equiv (k_{x}^{(1)}\sigma)^2$:
\begin{flalign}
\label{approx kirkwood eq1}%%%%%%%%%%
&\alpha_2\overline{\kappa}^2\sigma^2 \mathcal{S}^2
+(1-\alpha_1\overline{\kappa}^2\sigma^2) \mathcal{S}
+\overline{\kappa}^2\sigma^2+(pE)^2\sigma^2=0.
\end{flalign}
It follows from eqn (\ref{complex solution}) that
\begin{flalign}
\label{ks xy}%%%%%
(k_{x}^{(1)}\sigma)^2=x^2-y^2+2ixy
\end{flalign}
where $x$ and $y$ are related to the decay length $\xi_{\mathrm{Decay}}^{(1)}$ and wavelength $\mu^{(1)}$ as defined in eqn (\ref{complex length}).
Eqn (\ref{complex length}) and (\ref{ks xy}) imply that discriminant analysis of eqn (\ref{approx kirkwood eq1}) is found useful to determine the Kirkwood crossover point where the decay of the correlation functions changes from monotonic ($\mu^{(1)}=0$) to oscillatory ($\mu^{(1)}\neq 0$).
As mentioned in eqn (\ref{max kappa}), the imaginary solution $2ixy$ disappears at $\overline{\kappa}^{(*1)}\sigma$ because of $x=0$, or $\mu^{(1)}\rightarrow\infty$:
the Kirkwood crossover occurs when exceeding $\overline{\kappa}^{(*1)}\sigma$.

%%%%%%%%%%%%%%%%%%%%%%%%%
We find approximate forms of the solution to the discriminant equation of eqn (\ref{approx kirkwood eq1}) as follows:
\begin{flalign}
\left(\overline{\kappa}^{(*1)}\sigma\right)^2
&=\frac{1-\sqrt{\alpha_2}(pE)^2\sigma^2}{\alpha_1+2\sqrt{\alpha_2}}
\qquad\,(pE\sigma\ll 1)\label{approx k sol1}%%%%%%%%%
\\
&\rightarrow 0
\hphantom{\frac{1}{\alpha_1+2\sqrt{\alpha_2}}}
\qquad\quad\quad(pE\sigma\gg 1);
\label{approx k sol2}%%%%%%%%
\end{flalign}
see Appendix E for these derivations.
Plugging the modified MSA coefficients, $\alpha_1=1/2$ and $\alpha_2=1/24$, into the relation (\ref{approx k sol1}) for $E=0$, we have
\begin{flalign}
\overline{\kappa}^{(*1)}\sigma=\left(\frac{1}{\alpha_1+2\sqrt{\alpha_2}}\right)^{1/2}\approx 1.05,
\label{comp kirkwood}%%%%%%%%
\end{flalign}
which is in good agreement with the Kirkwood crossover values previously obtained for the primitive model in the absence of applied electric field \cite{under th,under andelman,under evans,kirkwood original,kirkwood various}.

Eqn (\ref{approx k sol1}) and (\ref{approx k sol2}) imply that the Debye-H\"uckel length $\xi^*_{\mathrm{DH}}=1/\overline{\kappa}^{(*1)}$ at the Kirkwood crossover becomes longer as $E$ is larger.
Namely, the crossover density $\overline{n}^*=1/\{8\pi l_B(\xi^*_{\mathrm{DH}})^2\}$ becomes lower with the increase of $E$;
eqn (\ref{approx k sol2}) predicts that both charge-charge and density-density oscillations are observed even in a dilute electrolyte upon applying a high electric field.

%%%%%%%%%%%%%%%%
\subsection{Analytical and numerical results}
{\itshape Gaussian charge smearing model} \cite{gauss smearing,finite gaussian}.---
First, we consider the Gaussian charge smearing model.
This model is represented by $\omega(k_{x}^{(1)}\sigma)=e^{-(k_{x}^{(1)}\sigma)^2/2}$ in eqn (\ref{omega two}).
Then, eqn (\ref{general pole2}) is rewritten as
\begin{flalign}
\label{pre lambert}%%%%%%%%%%%%%%
2\tau e^{\frac{(k_{x}^{(1)}\sigma)^2}{2}}+\overline{\kappa}^2\sigma^2=0,
\\
\label{tau def}%%%%%%%%%%%%%
2\tau=(k_{x}^{(1)}\sigma)^2+(pE)^2\sigma^2.
\end{flalign}
It is convenient to transform eqn (\ref{pre lambert}) and (\ref{tau def}) to
\begin{flalign}
e^{\tau}\tau&=-\frac{\overline{\kappa}^2\sigma^2}{2}e^{\frac{(pE)^2\sigma^2}{2}},
\label{pre lambert2}%%%%%%%%%%%%%
\\
2\tau&=x^2-y^2+(pE)^2\sigma^2+2ixy,
\label{pre lambert complex}%%%%%%%%%%%%
\end{flalign}
which can be rewritten as
\begin{flalign}
\tau=\mathcal{W}\left(-\frac{\overline{\kappa}^2\sigma^2}{2}e^{\frac{(pE)^2\sigma^2}{2}}\right),
\label{lambert sol}%%%%%%%%%%
\end{flalign}
using the Lambert function $\mathcal{W}$ \cite{gauss smearing} defined by $\tau=\mathcal{W}(\tau e^{\tau})$.

Focusing on the principal branch of the Lambert function \cite{gauss smearing}, it is found that the Kirkwood crossover point satisfies the relations,
\begin{flalign}
\label{1/e}%%%%%%%%%%%%
&\frac{(\overline{\kappa}^{(*1)}\sigma)^2}{2}e^{\frac{(pE)^2\sigma^2}{2}}=1/e,
\\
\label{tauc}%%%%%%%%%
&\tau^*=-1,
\end{flalign}
similar to those at $E=0$.
Eqn (\ref{1/e}) transforms to
\begin{flalign}
\label{lambert cross1}%%%%%%%%%%%
\overline{\kappa}^{(*1)}\sigma=
e^{-\frac{(pE)^2\sigma^2}{4}}\sqrt{\frac{2}{e}}
\end{flalign}
or
\begin{flalign}
\label{lambert cross2}%%%%%%%%%%%%
\overline{n}^*=
\left(\frac{1}{4\pi p^2l_B\sigma^2}\right)
e^{-\frac{(pE)^2\sigma^2}{2}-1}
\end{flalign}
for the crossover density $\overline{n}^*$.
Eqn (\ref{lambert cross1}) verifies the above approximate result (\ref{approx k sol2}), whereas eqn (\ref{lambert cross2}) enables us to make an analytical prediction that the increase of $E$ results in the decrease of $\overline{n}^*$.
Inserting eqn (\ref{tauc}) into eqn (\ref{pre lambert complex}), we have
\begin{flalign}
\label{tauc-1}%%%%%%%%%%%%
-2=-\left(\frac{\sigma}{\xi^{(*1)}_{\mathrm{Decay}}}\right)^2+(pE)^2\sigma^2,
\end{flalign}
or
\begin{flalign}
\label{xi el}%%%%%%%%
\frac{\xi^{(*1)}_{\mathrm{Decay}}}{\sigma}=\frac{1}{\sqrt{2+(pE)^2\sigma^2}},
\end{flalign}
because of $x=0$ at the Kirkwood crossover.
Eqn (\ref{lambert cross2}) and (\ref{xi el}) state that, as $E$ is larger, Coulomb interactions are more short-ranged despite the decrease in the Kirkwood crossover density $\overline{n}^*$ given by eqn (\ref{lambert cross2}).
In other words, our target mode $k^{(*1)}_{x}$ describes an aspect of electric-field-induced screening which is enhanced by the applied electric field (see also the last paragraph of Sec. IIIA for the underlying physics).

%%%%%%%%% FIG3 %%%%%%%%%%%%%%
\begin{figure}[hbtp]
\begin{center}
\includegraphics[
width=8.5cm
]{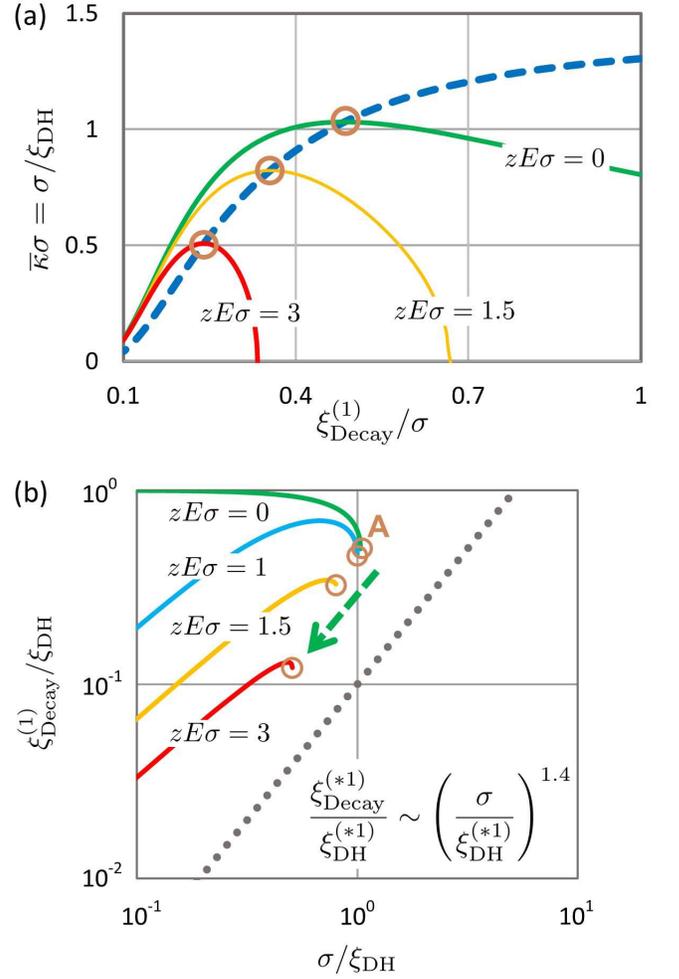}
\end{center}
\caption{Comparison between Fig. 2 (a schematic summary) and numerical results of the modified MSA model.
(a) The graphical representation of the solution to $u(y^*)=v(y^*)$ given by eqn (\ref{u func}) and (\ref{v func}), respectively.
Eqn (\ref{u func}) provides $\overline{\kappa}\sigma=\sqrt{u(y=\sigma/\xi^{(1)}_{\mathrm{Decay}})}$, the $\xi^{(1)}_{\mathrm{Decay}}$--dependence of $\overline{\kappa}\sigma$, which varies depending on the electric field strength measured by $pE\sigma$.
The colored solid lines represent these dependencies for $pE\sigma=0$ (green), $pE\sigma=1.5$ (orange), and $pE\sigma=3$ (red).
Meanwhile, the blue dashed line shows another dependence of $\overline{\kappa}\sigma$ on $\xi^{(1)}_{\mathrm{Decay}}/\sigma$ which is given by $\overline{\kappa}\sigma=\sqrt{v(y)}$ (see eqn (\ref{v func})).
The three intersection points are indicated by brown circles, giving both the Debye-H\"uckel lengths $\xi^{(*1)}_{\mathrm{DH}}$ at the electric-field-induced crossovers and the Kirkwood decay lengths $\xi^{(*1)}_{\mathrm{Decay}}$ at different field strengths.
(b) Numerical results summarized in Fig. 2.
In this figure, the $\overline{\kappa}\sigma$--dependencies of $\xi^{(1)}_{\mathrm{Decay}}$ are depicted by the solid lines colored green ($pE\sigma=0$), blue ($pE\sigma=1$), orange ($pE\sigma=1.5$), and red ($pE\sigma=3$), from top to bottom, on a log-log plot of $\xi^{(1)}_{\mathrm{Decay}}/\xi_{\mathrm{DH}}=\overline{\kappa}\xi^{(1)}_{\mathrm{Decay}}$ vs. $\sigma/\xi_{\mathrm{DH}}=\overline{\kappa}\sigma$.
The brown circles mark the termination points of these lines representing the Kirkwood crossover points at each electric-field strength, and the rightmost circle A corresponds to node A in Fig. 2, the Kirkwood crossover point at $E=0$.
The location shift of brown circles with the increase of $E$ is indicated by the dashed green arrow, showing an electric-field-induced shift of the Kirkwood crossover.
For comparison, the dotted line delineates the scaling relation, $\xi^{(*1)}_{\mathrm{Decay}}/\xi^{(*1)}_{\mathrm{DH}}\sim(\sigma/\xi^{(*1)}_{\mathrm{DH}})^{1.4}$, as well as that in Fig. 3(b).
}
\end{figure}

%%%%%%%%%%%%%%%%

%%%%%%%%%%%%%%%%%%%%%%
%%%%%%%%%%%%%%%%%%%%%%%%%
{\itshape The modified MSA model} \cite{mmsa smearing}.---
Next, we adopt $\omega(\bm{k})=\cos(k_{x}^{(1)}\sigma)$, according to the strong-coupling approximation of the modified MSA model (see Sec. IVB).
Bearing in mind that $\cos(k_{x}^{(1)}\sigma)=\cos x\cosh y$, eqn (\ref{general pole2}) reads
\begin{flalign}
\overline{\kappa}^2\sigma^2\cos x\cosh y&=y^2-(pE)^2\sigma^2-x^2,
\label{real eq cos}\\
\overline{\kappa}^2\sigma^2\sin x\sinh y&=2xy,
\label{ima eq sin}
\end{flalign}
for the real and imaginary parts, respectively.
The Kirkwood crossover occurs in the limit of $x\rightarrow 0$ (or $\mu^{(1)}\rightarrow\infty$).
In this limit , eqn (\ref{real eq cos}) and (\ref{ima eq sin}) reduce to 
\begin{flalign}
(\overline{\kappa}^{(*1)}\sigma)^2&=u(y^*)=\frac{(y^*)^2-(pE)^2\sigma^2}{\cosh y^*},
\label{u func}\\
&=v(y^*)=\frac{2y^*}{\sinh y^*},
\label{v func}
\end{flalign}
for $y^*=\sigma/\xi^{(*1)}_{\mathrm{Decay}}$.

Figure 4(a) shows the curves of $\overline{\kappa}\sigma=\sqrt{u(y)}$ and $\overline{\kappa}\sigma=\sqrt{v(y)}$ as a function of $y=\sigma/\xi^{(1)}_{\mathrm{Decay}}$ including the Kirkwood crossover value $y^*$.
While there is a single line of $\overline{\kappa}\sigma=\sqrt{v(y)}$ in Fig. 4(a), the curves of $\overline{\kappa}\sigma=\sqrt{u(y)}$ are depicted using different values of $pE\sigma=0,$ 1.5 and 3.
We can see from Fig. 4(a) that the intersection points of these curves (three brown circles located at the intersections in Fig. 4(a)) is determined by $u(y^*)=v(y^*)$ and is located at the maximum of $\overline{\kappa}\sigma=\sqrt{u(y)}$ as a function of $y$;
actually, it is easily confirmed that $u(y^*)=v(y^*)$ is nothing but the maximum condition for $\overline{\kappa}\sigma=\sqrt{u(y)}$.
We find from a series of intersection points $(y^*,\sqrt{u(y^*)})$ for different field strengths in Fig. 4(a) that the modified MSA model \cite{mmsa smearing} exhibits a similar trend observed in the above Gaussian charge smearing model \cite{gauss smearing}:
the maxima of $\overline{\kappa}\sigma=\sqrt{u(y)}$ decrease with increase of $E$.
That is, the Debye-H\"uckel length $\xi^{(*1)}_{\mathrm{DH}}$ at the electric-field-induced crossover is larger as the decay length $\xi^{(*1)}_{\mathrm{Decay}}$ at the Kirkwood crossover is smaller due to the increase of $E$.
These dependencies are in qualitative agreement with eqn (\ref{lambert cross1}) and (\ref{xi el}) of the Gaussian charge smearing model.

{\itshape Relationship between the results in Fig. 4 and the present results given by eqn (\ref{lambert cross1}), (\ref{xi el}), (\ref{u func}) and (\ref{v func})}.---
Thus, we have obtained the formulation to find the results in Fig. 4(a).
On the one hand, eqn (\ref{lambert cross1}) and (\ref{xi el}) yield $\sigma/\xi^{(*1)}_{\mathrm{DH}}$ (or $\overline{\kappa}^{(*1)}\sigma$) and $\xi_{\mathrm{Decay}}^{(*1)}/\sigma$ of the Gaussian charge smearing model, respectively.
On the other hand, eqn (\ref{u func}) and (\ref{v func}) are solved numerically to find $k_{x}^{(*1)}\sigma$, or the inverse of $\xi_{\mathrm{Decay}}^{(*1)}/\sigma$ in the modified MSA \cite{mmsa smearing}, and we can easily calculate $\overline{\kappa}^{(*1)}\sigma$ from $\xi_{\mathrm{Decay}}^{(*1)}/\sigma$ using eqn (\ref{u func}).
The same results as those of Fig. 4(a) are presented on a log-log plot in Fig. 4(b), further indicating that, in the range of $10^{-1}<\sigma/\xi^{(*1)}_{\mathrm{DH}}<10^0$, the $\sigma/\xi^{(*1)}_{\mathrm{DH}}$--dependencies of $\xi_{\mathrm{Decay}}^{(*1)}/\xi^{(*1)}_{\mathrm{DH}}$ exhibit a scaling relation $\xi_{\mathrm{Decay}}^{(*1)}/\xi^{(*1)}_{\mathrm{DH}}\sim(\sigma/\xi^{(*1)}_{\mathrm{DH}})^{\chi}$ with $\chi$ being close to 1.4, which is similar to eqn (\ref{decay scaling}) previously found for concentrated equilibrium electrolytes \cite{under simu,under th,under andelman,under evans}.

%%%%%%%%%%%%%%%%%%%%
%%%%%%%%%%%%%%%%%%%%%%%
\section{Discussion and conclusions}
So far, we have demonstrated the usefulness of the SDFT on concentrated electrolytes under steady electric fields in two respects.
First, a hybrid framework of the equilibrium DFT and field-theoretic approach justifies the previously modified terms of the PNP equations \cite{bazant dynamics,eisenberg,mpnp channel, yochelis,mpnp rtil,mpnp self,witt,ddft} in addition to a stochastic extension of the density dynamics \cite{demery,gole,andelman}.
Second, the stochastic set of density dynamics equations allows us to calculate the correlation functions. As a result, we can predict electric-field-induced oscillations which appear prior to the conventional onset of oscillatory decay of correlations, or the Kirkwood crossover without an applied electric field \cite{under th,under andelman,under evans,kirkwood original,kirkwood various,various smearing,gauss smearing,kirkwood fw}.

While Table 1 has provided a more detailed summary of the former modifications, we would like to make additional three remarks related to the latter results schematically illustrated in Fig. 2.

{\itshape (i) Correlation function analysis}.---
In this paper, it has been proved that we can extend the mPNP model to consider the stochastic process, thereby allowing us to obtain stationary equal-time correlation functions which include the key function $\omega(\bm{k})$ as seen from eqn (\ref{C solution2}) to (\ref{delta crr}).
It should be noted that the Kirkwood crossover does not occur without $\omega(\bm{k})$ given by eqn (\ref{omega two});
therefore, it is indispensable to incorporate either the finite-spread Poisson equation (\ref{finite spread}) or the generalized Debye-H\"uckel equation (\ref{gdh}) into the stochastic mPNP models for predicting the onset of oscillatory decay of correlations.

{\itshape (ii) Stripe states}.---
As illustrated in Fig. 2, the shifting behavior of the electric-field-induced Kirkwood crossover behavior bears a similarity to that of underscreening previously found by simulation and theoretical studies on concentrated electrolytes \cite{under simu,under th,under andelman,under evans}.
To be noted, any lane formation \cite{lowen}, or any modulation perpendicular to the applied field direction, is beyond the scope of this study.
Namely, an oscillatory state along the field direction (a stripe state) as given in Fig. 2 is stationary as long as lane formation is not favored.
Nevertheless, the stripe state is consistent with some previous findings of inhomogeneous steady states such as alternating chains of cations and anions along the applied field direction in electrolytes \cite{demery} and non-parallels bands in oppositely charged colloidal mixtures \cite{lowen,band} (see the last paragraph of Section IIIC).

{\itshape (iii) Fisher-Widom crossover between two Kirkwood crossovers}.---
Above the Kirkwood crossover condition of $\overline{\kappa}\sigma>\overline{\kappa}^{(*1)}\sigma$, we have smeared correlation functions, $\overline{\Delta\mathcal{C}^{\mathrm{st}}_{\rho\rho}}(x)$ and $\overline{\mathcal{C}^{\mathrm{st}}_{qq}}(x)$, which are given by the sum of oscillatory and monotonic decay functions (i.e., eqn (\ref{region2 smear1}) and (\ref{region2 smear2})).
Furthermore, the monotonic decay parts of correlation functions subsequently become oscillatory when $\overline{\kappa}\sigma$ goes beyond $\overline{\kappa}^{(*2)}\sigma$ which is related to the equilibrium Kirkwood crossover value $\overline{\kappa}^*\sigma$ as eqn (\ref{conventional k}).
This crossover phenomenon suggests the possibility of simultaneous occurrence of the Fisher-Widom crossover \cite{under evans,kirkwood fw,fw} for density-density and charge-charge correlations in the range of $\overline{\kappa}^{(*1)}\sigma\leq\overline{\kappa}\sigma\leq\overline{\kappa}^{(*2)}\sigma$ though the full phase diagram of steady states for electric-field-driven electrolytes remains to be determined (see Section IIIB).
For the specific understanding of anisotropic density modulations in stripe states, Sec. IIIE presents various results on the 2D density-density correlations beyond the Fisher-Widom crossover.
As seen from Figs. 4 and 7, there are some electric field conditions that create stripe states formed by segregation bands transverse to the external field direction.

It is still necessary to investigate whether experimental and simulation studies can find an electric-field-induced shift of the Kirkwood crossover from monotonic to oscillatory decay of density-density and charge-charge correlations in the applied electric field direction.
Therefore, let us make three comparisons in terms of realizability.

Although the primitive model has been used for investigating concentrated electrolytes, we would like to see the interaction parameters of RTILs and colloidal nano-particle dispersions as well as concentrated electrolytes.
For instance, let us consider $(p,\sigma,\epsilon,l_B)=(1,\,0.7,\,10,\,5.6)$ for RTILs \cite{rtil dielectric} and $(10,\,10,\,80,\,0.7)$ for colloidal nano-particle dispersions as adequate parameters of valence $p$, diameter $\sigma$ [nm], dielectric constant $\epsilon$, and the Bjerrum length $l_B$ [nm] at room temperature $T=300$ K.
Accordingly, we have $p^2l_B/\sigma=8$ (RTILs) and $p^2l_B/\sigma=7$ (nano-particle dispersions), and the use of eqn (\ref{app potential}) and (\ref{fourier dcf}) can be justified because the relation (\ref{f condition}) barely holds.

Next, we would like to evaluate a realistic range of electric field strength.
At $pE\sigma=1.5$, we have $E\approx 5.5\times 10^7$ V/m for the RTILs (i.e., $(p,\sigma)=(1,\,0.7)$) and $E\approx 3.8\times 10^5$ V/m for the nano-particles (i.e., $(p,\sigma)=(10,\,10)$).
These are plausible values according to previous simulation and experimental studies as follows:
molecular dynamics simulations of RTILs have revealed that $E\sim 10^7$ V/m corresponds to a boundary value beyond which RTILs are reorganized into nematic-like order and exhibit anisotropic dynamics \cite{rtil field}, whereas, for colloidal dispersions, a magnitude of $E\sim 10^5$ V/m is within the possible range when referring to segregation of oppositely charged colloidal particles into bands perpendicular to the field direction of an applied alternating current electric field \cite{lowen,band}.

Last, let us evaluate the Kirkwood crossover densities at $pE\sigma=1.5$.
We have obtained that $\overline{\kappa}^*\sigma$ is equal to 1.04 ($pE\sigma=0$) and that $\overline{\kappa}^{(*1)}\sigma\approx0.82$ ($pE\sigma=1.5$) when adopting the modified MSA model \cite{mmsa smearing}.
It follows that the Kirkwood crossover density varies from 0.30 M ($pE\sigma=0$) to 0.19 M ($pE\sigma=1.5$) for an RTIL (1-butyl-3-methylimidazolium bis(trifluoromethylsulfonyl)imide) diluted with propylene carbonate where we set $(p,\sigma,\epsilon,l_B)=(1,\,0.4,\,65,\,0.88)$.
The former density (0.30 M) agrees well with experimental and simulation results \cite{under exp,under simu} with no electric field applied, whereas the validity of density difference ($0.30-0.19=0.11$ M) due to the external electric field needs to be assessed in future.

\section*{Conflicts of interest}
There are no conflicts to declare.
%%%%%%%%%%%%%%%%%%%%%%%%
\appendix
%%%%%%%%%%%%%%
%%%%%%%%%%%%%%
\section{Details on modifications of the PNP model presented in Section II}
We provide the detailed formulations of the results in Section II by dividing this section into four parts:
general formulation of two-component fluids (Appendix A1), two modifications of the Poisson equation (Appendix A2), formulations of stochastic currents for electric-field-driven electrolytes (Appendix A3), and outline of deriving stationary correlation functions at equal times (Appendix A4).
In Appendix A1, the functional-integral representation of the Dean-Kawasaki model reveals that the Gaussian approximation of the free energy difference between non-equilibrium and equilibrium free energies yields the self-energy modified current of each component in mixtures.
In Appendix A2, we validate the approximate form (\ref{app potential}) of interaction potential $\psi(\bm{r},t)$ for the primitive model using the modified MSA \cite{mmsa smearing} model, and also demonstrate for the modified MSA model that the finite-spread Poisson equation obtained from this expression (\ref{app potential}) leads to the higher-order Poisson equation due to the small $k\sigma$--expansion.
In Appendix A3, we show that the self-energy-modified current given by eqn (\ref{pnp current}) to (\ref{pnp U}) is obtained from combining the results in Appendix A2 and that linearization of this current corresponds to the first-order expansion of non-equilibrium chemical potential around a uniform density $\overline{n}$.
Appendix A4 explains that the stationary condition (\ref{stationary C}) imposed on a general matrix form of equal-time correlation functions yields density-density and charge-charge correlation functions given by eqn (\ref{C solution2}) to (\ref{delta crr}).
%%%%%%%%%%%%%%

\subsection{General formulation}
{\itshape Stochastic current in the Dean-Kawasaki model}.---
The stochastic equations for the density fields $n_l(\bm{r},t)$ $(l=1,\,2)$ have been formulated based on the Dean-Kawasaki model.
We have, according to the Dean-Kawasaki model, the conservation equation (\ref{conservation}) for $n_l(\bm{r},t)$ by introducing the stochastic current $\bm{J}_l(\bm{r},t)$:
the Dean-Kawasaki model provides a general form of the stochastic current $\bm{J}_l(\bm{r},t)$ expressed as
\begin{flalign}
\label{j def}%%%%%%%%%
&\bm{J}_l(\bm{r},t)=-\mathcal{D}n_l(\bm{r},t)\nabla\mu_l(\bm{r},t)-\sqrt{2\mathcal{D}n_l(\bm{r},t)}\bm{\zeta}_l(\bm{r},t),\\
\label{mu def}%%%%%%%%%
&\mu_l(\bm{r},t)=\frac{\delta\mathcal{A}[\bm{n}]}{\delta n_l(\bm{r},t)}.
\end{flalign}
We can see from eqn (\ref{j def}) and (\ref{mu def}) that there are two features of the Dean-Kawasaki model, compared with the dynamic DFT based on the deterministic density-functional equation:
(i) the deterministic current, the first term on the rhs of eqn (\ref{j def}), is nonlinear with respect to $n_l(\bm{r},t)$ in general and is determined by a constrained free energy $\mathcal{A}[\bm{n}]$, instead of the equilibrium free energy functional \cite{dft,ry}; 
(ii) addition of the stochastic current, the second term on the rhs of eqn (\ref{j def}), allows us to describe non-equilibrium systems with multiplicative noise.

%%%%%%%%%%%%%%%%%%
{\itshape Functional-integral representation of constrained free energy $\mathcal{A}[\bm{n}]$}.---
It has been shown that the constrained free energy $\mathcal{A}[\bm{n}]$ as a functional of given density fields $\bm{n}(\bm{r},t)=(n_1(\bm{r},t),n_2(\bm{r},t))^{\mathrm{T}}$ can be expressed by considering fluctuating potential fields $\bm{\phi}(\bm{r},t)=(\phi_1(\bm{r},t),\phi_2(\bm{r},t))^{\mathrm{T}}$, which are conjugate to $\bm{n}(\bm{r},t)$, in addition to an adjusted potential field $\varphi^{\mathrm{dft}}_l(\bm{r},t)$ similar to that of the equilibrium DFT \cite{dft,ry}.
Extending the previous result \cite{frusawa review,frusawa sdft} to the expression for two-component systems (see Appendix B for details), we have
\begin{flalign}
\label{A def}%%%%%%%%%%%%
e^{-\mathcal{A}[\bm{n}]}&=\prod_{l=1}^2\int
D\phi_l\Delta[n_l]\,e^{-F[\bm{n},\bm{\phi}]},
\end{flalign}
with the following constraint imposed by the canonical ensemble:
\begin{eqnarray}
\label{delta number}%%%%%%%%%%%%
\Delta[n_l]=
\left\{
\begin{array}{l}
1\quad(\int d^3\bm{r}n_l(\bm{r})=N)\\
\\
0\quad(\int d^3\bm{r}n_l(\bm{r})\neq N),\\
\end{array}
\right.
\end{eqnarray}
where the total number of either anions or cations is equally $N$.
The free-energy functional $F[\bm{n},\bm{\phi}]$ in the exponent of eqn (\ref{A def}) is defined using the grand potential of the primitive model with an imaginary external field $i\phi(\bm{r})$ applied, and can be divided into two parts (see Appendix B for details):
\begin{flalign}
\label{f sep}%%%%%%%%%
F[\bm{n},\bm{\phi}]&=F[\bm{n},\bm{0}]+\Delta F[\bm{n},\bm{\phi}].
\end{flalign}
The free-energy functional $F[\bm{n},0]$ in the absence of fluctuating potential reduces to the intrinsic Helmholtz free energy, a key thermodynamic quantity in the equilibrium DFT \cite{dft,ry}.
It follows that $F[\bm{n},0]$ is related to the chemical potential $\mu_{\mathrm{eq}}$ in equilibrium through the following stationary equation:
\begin{flalign}
\frac{\delta F[\bm{n},\bm{0}]}{\delta n_l(\bm{r},t)}
=\mu_{\mathrm{eq}}-\varphi^{\mathrm{dft}}_l(\bm{r},t)\equiv\mu_l^0[\bm{n}],
\label{f stationary}%%%%%%%%%%%%%%%%%
\end{flalign}
where a non-equilibrium chemical potential $\mu_l^0[\bm{n}]$ is a functional of $\bm{n}(\bm{r},t)$ because the external potential distribution $\varphi^{\mathrm{dft}}_l(\bm{r},t)$ is adjusted to identify $n_l(\bm{r},t)$ with the equilibrium density as is the case with the equilibrium DFT (see Appendix B for details) \cite{dft,ry}.

%%%%%%%%%%%%%
{\itshape Decomposition of the stochastic current given by eqn (\ref{j def})}.---
It follows from eqn (\ref{mu def}) to (\ref{f stationary}) that
\begin{flalign}
\label{da dn}%%%%%%%%%%%%%%%%%
\mu_l[\bm{n}]
&=\mu_l^0[\bm{n}]
+\mu_l^{\delta}[\bm{n}]-\mu_N,\\
\mu_l^{\delta}[\bm{n}]&=
\left\langle\frac{\delta
\Delta F[\bm{n},\bm{\phi}]}{\delta n_l(\bm{r},t)}\right\rangle_{\phi}\nonumber\\
\label{delta f av}%%%%%%%%%%%%
&\equiv\frac{\prod_{l=1}^2\int
D\phi_l\,\left(\frac{\delta
\Delta F}{\delta n_l}\right)
e^{-\Delta F[\bm{n},\bm{\phi}]}}
{\prod_{l=1}^2\int
D\phi_l\,e^{-\Delta F[\bm{n},\bm{\phi}]}},
\end{flalign}
where $\mu_N$ corresponds to the Lagrange multiplier to enforce the constraint $\Delta[n_l]$ given by eqn (\ref{delta number}).
Correspondingly, the stochastic current $\bm{J}_l(\bm{r},t)$ can be decomposed into three parts:
\begin{flalign}
\label{j sep}%%%%%%%
\bm{J}_l(\bm{r},t)&=\bm{J}^0_l(\bm{r},t)+\bm{J}_l^{\delta}(\bm{r},t)-\sqrt{2\mathcal{D}n_l(\bm{r},t)}\bm{\zeta}_l(\bm{r},t),
\end{flalign}
where eqn (\ref{j def}), (\ref{da dn}) and (\ref{delta f av}) provide
\begin{flalign}
\bm{J}^0_l(\bm{r},t)&=-\mathcal{D}n_l(\bm{r},t)\nabla\mu_l^0[\bm{n}],
\label{j0}\\%%%%
\bm{J}_l^{\delta}(\bm{r},t)&=-\mathcal{D}n_l(\bm{r},t)\nabla\mu_l^{\delta}[\bm{n}].
\label{delta j}%%%
\end{flalign}
While the expression (\ref{j0}) indicates that $\bm{J}^0_l(\bm{r},t)$ is the conventional current used in the deterministic density-functional equation, the additional current $\bm{J}_l^{\delta}(\bm{r},t)$ is obtained from eqn (\ref{j def}), (\ref{da dn}) and (\ref{delta f av}).

Here we adopt the Ramakrishnan-Yussouf functional \cite{ry} as the equilibrium free energy $F[\bm{n},\bm{0}]$, yielding
\begin{flalign}
\mu_l^0[\bm{n}]&=\ln n_l(\bm{r},t)+(-1)^{l-1}z\Psi(\bm{r})\nonumber\\
\label{ry diff}%%%%%%%%
&\qquad\qquad-\int d^3\bm{r}' \sum_{m=1}^2c_{lm}(\bm{r}-\bm{r}')\,n_m(\bm{r}',t),
\end{flalign}
with $c_{lm}(\bm{r}-\bm{r}')$ denoting the DCF between the $l$--th and $m$--th ions.
Eqn (\ref{ry diff}) yields $\bm{J}^0_l(\bm{r},t)$ for electric-field-driven electrolytes in the presence of an applied steady potential $\Psi(\bm{r})$ as well as the interaction potential,
\begin{flalign}
\label{in potential}%%%%%%%%%%%%%%
\psi_l(\bm{r},t)=&-\int d^3\bm{r}' \sum_{m=1}^2c_{lm}(\bm{r}-\bm{r}')\,n_m(\bm{r}',t),
\end{flalign}
which is a time-varying potential due to the time dependence of $n_m(\bm{r}',t)$.
Combining eqn (\ref{j0}), (\ref{ry diff}) and (\ref{in potential}), we have
\begin{flalign}
\bm{J}^0_l(\bm{r},t)
&=\mathcal{D}n_l(\bm{r},t)(-1)^{l-1}p\bm{E}\nonumber\\
\label{j0 ry}%%%%%%%%%%%%
&\qquad-\mathcal{D}n_l(\bm{r},t)\nabla\left\{\ln n_l(\bm{r},t)+\psi_l(\bm{r},t)\right\},
\end{flalign}
where the applied electric field $\bm{E}\equiv-\nabla\Psi(\bm{r})$, which is multiplied by the elementary charge $e$, generates an external force $(-1)^{l-1}p\bm{E}$ exerted on a cation ($l=1$) or an anion ($l=2$).

{\itshape Self-energy contribution} \cite{mpnp self,static self}.---
We evaluate the free-energy difference $\Delta F[\bm{n}.\bm{\phi}]$ in the Gaussian approximation, or the Gaussian expansion around the equilibrium free-energy functional $F[\bm{n},\bm{0}]$ with the density distributions being fixed at $\bm{n}(\bm{r},t)$.
Namely, $\Delta F[\bm{n}.\bm{\phi}]$ is expressed by the quadratic term of fluctuating $\phi$--fields:
\begin{flalign}
\label{f diff}%%%%%%%%%%%
\Delta F[\bm{n},\bm{\phi}]&=\frac{1}{2}\iint d^3\bm{r}d^3\bm{r}'
\bm{\phi}(\bm{r})^{\mathrm{T}}\bm{\mathcal{N}}(\bm{r}-\bm{r}')\bm{\phi}(\bm{r}'),
\end{flalign}
where the $\bm{\mathcal{N}}$--matrix is given by
\begin{flalign}
\label{mat n def}%%%%%%%%%%%%
\bm{\mathcal{N}}(\bm{r}-\bm{r}')&=
\begin{pmatrix}
\mathcal{N}_{11}(\bm{r}-\bm{r}') & \mathcal{N}_{12}(\bm{r}-\bm{r}') \\
\mathcal{N}_{21}(\bm{r}-\bm{r}') & \mathcal{N}_{22}(\bm{r}-\bm{r}') \\
\end{pmatrix},
\\
\label{n def}%%%%%%%%%%%
\mathcal{N}_{lm}(\bm{r}-\bm{r}')&=n_l(\bm{r})\left\{
\delta_{lm}\delta(\bm{r}-\bm{r}')+h_{lm}(\bm{r}-\bm{r}')n_m(\bm{r}')
\right\},
\end{flalign}
using the total correlation functions $h_{lm}(\bm{r}-\bm{r}')$ between the $l$--th and $m$--th ions.
As detailed in Appendix C, combination of eqn (\ref{delta f av}) and (\ref{f diff}) yields eqn (\ref{self energy}):
\begin{flalign}
\mu_l^{\delta}[\bm{n}]&=\frac{u(\bm{r},t)}{2}\nonumber\\
\label{self mu}%%%%%%%%
&=\frac{1}{2}\lim_{\bm{r}\rightarrow\bm{r}'}
\left\{
c_{ll}(\bm{r}-\bm{r}')-h_{ll}(\bm{r}-\bm{r}')
\right\}.
\end{flalign}
It follows from eqn (\ref{delta j}) and (\ref{self mu}) that
\begin{flalign}
\bm{J}_l^{\delta}(\bm{r},t)&=-\mathcal{D}n_l(\bm{r},t)\frac{\nabla u(\bm{r},t)}{2}\nonumber\\
\label{self current}%%%%%%%%%
&=-\mathcal{D}n_l(\bm{r},t)\frac{\nabla c_{ll}(\bm{0})}{2},
\end{flalign}
where use has been made of the identity, $h_{ll}(\bm{0})=-1$ independent of $n_l(\bm{r},t)$, in the last equality.

%%%%%%%%%%%%%%%%%%%%%%%%
%%%%%%%%%%%%%%%%%%%%%%%%
\subsection{Modified Poisson equations given by eqn (\ref{finite spread}) and (\ref{high poisson})}
{\itshape The DCF form given by eqn (\ref{omega two}) and (\ref{fourier dcf})}.---
In the modified MSA \cite{mmsa smearing} , the DCF is of the following form:
\begin{flalign}
\label{dcf msa}%%%%%%%%%%
-c_{lm}(\bm{k})&=\frac{4\pi}{\bm{k}^2}f_{lm}(\bm{k}),
\\
\label{two fij}%%%%%%%%
f_{lm}(\bm{k})&=f_{lm}^{\mathrm{c}}(\bm{k})+f_{lm}^{\mathrm{h}}(\bm{k}),
\\
\label{omegac}%%%%%%%%%%%
f_{lm}^{\mathrm{c}}(\bm{k})&=(-1)^{l+m}p^2l_B\cos(k\sigma),
\\
\label{omegas}%%%%%%%%%%
f_{lm}^{\mathrm{h}}(\bm{k})&=-\sigma\left\{
\cos(k\sigma)-\frac{\sin(k\sigma)}{k\sigma}
\right\},
\end{flalign}
where $f_{lm}(\bm{k})$ is separated into two parts, $f_{lm}^{\mathrm{c}}$ and $f_{lm}^{\mathrm{h}}$.
Eqn (\ref{two fij}) reduces to
\begin{flalign}
\label{f approx}
f_{lm}(\bm{k})\approx f_{lm}^{\mathrm{c}}(\bm{k})=(-1)^{l+m}p^2l_B\cos(k\sigma)
\end{flalign}
when
\begin{flalign}
\label{f condition}%%%%%%%%%%%%
\frac{p^2l_B}{\sigma}\gg 1.
\end{flalign}
Namely, the above expression of the DCF given by eqn (\ref{omega two}) and (\ref{fourier dcf}) is verified in the modified MSA \cite{mmsa smearing} under the condition (\ref{f condition}).
It is noted that the approximation (\ref{f approx}) applies to the primitive model because $f_{lm}^{\mathrm{c}}(\bm{k})$ represents Coulomb interactions including steric effects \cite{andelman}.
The relation (\ref{f condition}) corresponds to the strong-coupling condition for one-component plasma, implying that the strong Coulomb interactions justify the negligibility of $f_{lm}^{\mathrm{h}}$ given by eqn (\ref{omegas}).
In this paper, we have supposed that, in general, the simplified form (\ref{fourier dcf}) applies to aqueous electrolytes if only because of $p^2l_B/\sigma>1$ (see Section V for a more detailed comparison between the relation (\ref{f condition}) and experimental conditions).

%%%%%%%%%%%%%%%%%%%%%%%%
{\itshape Finite-spread Poisson equation: derivation of eqn (\ref{finite spread})} \cite{demery,andelman,frusawa review,frydel review,finite gaussian,finite pb}.---
The two interaction potentials, $\psi_1(\bm{r},t)$ and $\psi_2(\bm{r},t)$, have been defined in eqn (\ref{in potential});
however, the approximate form of the DCF given by eqn (\ref{fourier dcf}) justifies that
\begin{flalign}
\label{in potential1}%%%%%%%%%%
\psi(\bm{r},t)&=\psi_1(\bm{r},t)=-\psi_2(\bm{r},t).
\end{flalign}
Thus, the expression (\ref{app potential}) of $\psi(\bm{r},t)$ has been verified by eqn (\ref{in potential1}), and the approximate form (\ref{fourier dcf}) follows the notations of
\begin{flalign}
\label{cll=c}%%%%%%%%%%
c_{11}(\bm{r},t)&=c_{22}(\bm{r},t)=c(\bm{r},t),\\
\label{clm=c}%%%%%%%%%%
c_{12}(\bm{r},t)&=c_{21}(\bm{r},t)=-c(\bm{r},t),
\end{flalign}
thereby leading to the finite-spread Poisson equation (\ref{finite spread}).

%%%%%%%%%%%%%%%%%%%%%%
{\itshape Higher-order Poisson equation: derivation of eqn (\ref{high poisson})} \cite{bazant dynamics,eisenberg,mpnp channel, yochelis,mpnp rtil,mpnp self,bsk,bazant bsk,high pb,frusawa review}.---
We perform the small $k\sigma$--expansion of $\omega(\bm{k})$ in eqn (\ref{fourier dcf}), yielding
\begin{flalign}
\label{approx f}%%%%%%%%%%%
\omega(\bm{k})\approx1-\frac{(k\sigma)^2}{2},
\end{flalign}
irrespective of the model forms given by eqn (\ref{omega two}).
It follows from eqn (\ref{app potential}), (\ref{charge def}), (\ref{omega two}) and (\ref{fourier dcf}) that the Fourier transform of the Poisson equation reads
\begin{flalign}
\label{fourier poisson}%%%%%%%%%%%%%%
\bm{k}^2\psi(\bm{k})
&=4\pi p^2l_B\omega(-\bm{k})\,q(\bm{k},t)\nonumber\\
&=\frac{(pe)^2}{k_BT\epsilon}\left\{
1-\frac{(k\sigma)^2}{2}
\right\}\,q(\bm{k},t),
\end{flalign}
which is further reduced to
\begin{flalign}
\label{fourier poisson2}%%%%%%%%%%%%%%%
k_BT\epsilon\left(
1+\frac{\bm{k}^2\sigma^2}{2}
\right)\bm{k}^2\psi(\bm{k})
=(pe)^2\,q(\bm{k},t),
\end{flalign}
in the small $k\sigma$--expansion: $\left(1-\bm{k}^2\sigma^2/2\right)^{-1}\approx1+\bm{k}^2\sigma^2/2$.
The real-space representation of eqn (\ref{fourier poisson2}) is the higher-order Poisson equation (\ref{high poisson}).

%%%%%%%%%%%%%%%%%%%%%%%%
%%%%%%%%%%%%%%%%%%%%%%%%
\subsection{Stochastic \lowercase{m}PNP currents given by eqn (\ref{pnp current}) and (\ref{mat current rhoq})}
{\itshape Confirming the self-energy-modified PNP current given by eqn (\ref{pnp current})}.---
Eqn (\ref{in potential1}) reads
\begin{flalign}
\label{psi l}%%%%%%%%%%%
\psi_l(\bm{r})=(-1)^{l-1}\psi(\bm{r}).
\end{flalign}
Plugging this expression (\ref{psi l}) into eqn (\ref{j0 ry}), combination of eqn (\ref{j sep}), (\ref{j0 ry}) and (\ref{self current}) provides
\begin{flalign}
&\bm{J}_l(\bm{r},t)
=\mathcal{D}n_l(\bm{r},t)(-1)^{l-1}p\bm{E}\nonumber\\
&-\mathcal{D}n_l(\bm{r},t)\nabla\left\{\ln n_l(\bm{r},t)+(-1)^{l-1}\psi(\bm{r},t)+\frac{u(\bm{r},t)}{2}\right\}\nonumber\\
\label{pnp current confirm}%%%%%%%%%%%%
&\qquad\qquad\qquad\qquad\qquad
-\sqrt{2\mathcal{D}n_l(\bm{r},t)}\bm{\zeta}_l(\bm{r},t).
\end{flalign}
While eqn (\ref{self mu}) and (\ref{cll=c}) with the notation of $h_{ll}(\bm{r})=h(\bm{r})$ verify the self-energy $u(\bm{r},t)$ given by eqn (\ref{self energy}) to (\ref{g def}), it has been confirmed in the preceding subsection that $\psi(\bm{r},t)$ satisfies eqn (\ref{high poisson}).
Thus, we have proved that eqn (\ref{pnp current confirm}) is of the same form as eqn (\ref{pnp current}) with eqn (\ref{mu}) and (\ref{pnp U}).

%%%%%%%%%%%%%%%%%%%%%%
{\itshape Derivation of linear mPNP current given by eqn (\ref{mat current rhoq})} \cite{demery,gole,andelman}.---
Let $\nu_l(\bm{r},t)$ be the density difference:
\begin{flalign}
\label{difference}%%%%%%%%%%%%
\nu_l(\bm{r},t)=n_l(\bm{r},t)-\overline{n}.
\end{flalign}
To linearize the self-energy-modified current given by eqn (\ref{pnp current}), we expand the chemical potential $\mu_l$ around $n_1(\bm{r},t)=n_2(\bm{r},t)=\overline{n}$ (or $q\equiv 0$) to the first order in $\nu_l(\bm{r},t)$:
\begin{flalign}
\mu_l[\bm{n}]
&=\ln n_l(\bm{r},t)+(-1)^{l-1}\psi(\bm{r},t)+\frac{u(\bm{r},t)}{2}
\nonumber\\
&=\mu_l[\overline{n}]
+\sum_{m=1}^2\int d^3\bm{r}'\left.
\frac{\delta\mu_l[\bm{n}]}{\delta n_m(\bm{r}',t)}\right|_{n_m=\overline{n}}\nu_m(\bm{r}',t)
\nonumber\\
&=\mu_l[\overline{n}]+\frac{\nu_l(\bm{r},t)}{\overline{n}}\nonumber\\
\label{mu expansion}%%%%%%%%%%%%%%%
&\qquad+\sum_{m=1}^2\int d^3\bm{r}'\left.
\frac{\delta U_l[\bm{n}]}{\delta n_m(\bm{r}',t)}\right|_{n_m=\overline{n}}\nu_m(\bm{r}',t),
\end{flalign}
with $\mu_l[\overline{n}]$ denoting $\mu_l[\overline{n}]=\mu_l[(\overline{n},\overline{n})^{\mathrm{T}}]$.
Since we have
\begin{flalign}
&\sum_{m=1}^2\int d^3\bm{r}'\left.
\frac{\delta U_l[\bm{n}]}{\delta n_m(\bm{r}',t)}\right|_{n_m=\overline{n}}\nu_m(\bm{r}',t)
\nonumber\\
&\qquad=-(-1)^{l-1}\int d{\bf r}'c(\bm{r}-\bm{r}')q(\bm{r}',t)\nonumber\\
\label{u diff}%%%%%%%%%%
&\qquad=(-1)^{l-1}\psi(\bm{r},t),
\end{flalign}
neglecting the contribution from the triplet DCF, eqn (\ref{mu expansion}) simply reads
\begin{flalign}
\label{mu expansion2}%%%%%%%%%%%%%%%
\mu_l[\bm{n}]=\mu_l[\overline{n}]+(-1)^{l-1}\psi(\bm{r},t).
\end{flalign}
Combining eqn (\ref{pnp current}) and (\ref{mu expansion2}), the mPNP current becomes, to the lowest order,
\begin{flalign}
\begin{pmatrix}
\bm{J}_1 \\
\bm{J}_2 \\
\end{pmatrix}
&=-\mathcal{D}
\begin{pmatrix}
\nabla\nu_1(\bm{r},t)+\overline{n}\nabla\psi(\bm{r},t)-n_1(\bm{r},t)p\bm{E}\\
\nabla\nu_2(\bm{r},t)-\overline{n}\nabla\psi(\bm{r},t)+n_2(\bm{r},t)p\bm{E}\\
\end{pmatrix}\nonumber\\
&\qquad\qquad\qquad\qquad\qquad
-\sqrt{2\mathcal{D}\overline{n}}
\begin{pmatrix}
\bm{\zeta}_1(\bm{r},t)\\
\bm{\zeta}_2(\bm{r},t)\\
\end{pmatrix}.
\label{mat current 12}%%%%%%%%%%%%%%%%%%%
\end{flalign}
We also note that
\begin{flalign}
\label{grad rho}%%%%%%%%
\nabla\rho(\bm{r},t)&=\nabla\left\{\nu_1(\bm{r},t)+\nu_2(\bm{r},t)\right\},\\
\label{grad q}%%%%%%%%%%%%%%%
\nabla q(\bm{r},t)&=\nabla\left\{\nu_1(\bm{r},t)-\nu_2(\bm{r},t)\right\}.
\end{flalign}
Thus, eqn (\ref{mat current 12}) to (\ref{grad q}) lead to the stochastic currents, $\bm{J}_{\rho}=\bm{J}_1+\bm{J}_2$ and $\bm{J}_{q}=\bm{J}_1-\bm{J}_2$, given by eqn (\ref{mat current rhoq}).

%%%%%%%%%%%%%%%%%%%%%
%%%%%%%%%%%%%%%%%%%%%
\subsection{Equal-time correlation functions given by eqn (\ref{C solution2}) to (\ref{def cofactor B})}
{\itshape Stationary condition of equal-time correlation matrix}.---
The compact form (\ref{mat dk m fourier}) of the stochastic equation is solved to obtain \cite{demery,andelman}
\begin{flalign}
\bm{\theta}(\bm{k},t)=\left\{
\int_{-\infty}^tds\,
e^{-\mathcal{D}\bm{\mathcal{K}}(\bm{k})(t-s)}
\right\}
\sqrt{4\mathcal{D}\overline{n}}\,\bm{\eta}(\bm{k}).
\label{theta solution}%%%%%%%%%
\end{flalign}
It follows from eqn (\ref{noise}) and (\ref{eta def}) that
\begin{flalign}
\label{eta matrix}%%%%%%%%%%%
\left<
\bm{\eta}(\bm{k},t)\bm{\eta}(-\bm{k},t)^{\mathrm{T}}\right>
=
(2\pi)^3
\begin{pmatrix}
\bm{k}^2\delta(t-t')&0\\
0&\bm{k}^2\delta(t-t')\\
\end{pmatrix}.
\end{flalign}
Plugging eqn (\ref{theta solution}) and (\ref{eta matrix}) into the definition (\ref{C def}) of the equal-time correlation matrix, we have
\begin{flalign}
\bm{\mathcal{C}}(\bm{k},t)=
\iint_{-\infty}^tds\,ds'\,
e^{-\mathcal{D}\bm{\mathcal{K}}(t-s)}
\mathcal{D}\bm{\mathcal{R}}\,
e^{-\mathcal{D}\bm{\mathcal{K}}^{\dagger}(t-s')},
\label{C solution}%%%%%%%%%%
\end{flalign}
where the relation (\ref{eta matrix}) generates the following matrix:
\begin{flalign}
\bm{\mathcal{R}}(\bm{k})
=(2\pi)^3
\begin{pmatrix}
4\overline{n}\bm{k}^2&0\\
0&4\overline{n}\bm{k}^2\\
\end{pmatrix}.
\label{R def}%%%%%%%%%%%%
\end{flalign}
It has been shown that the stationary condition $d\bm{\mathcal{C}}(\bm{k},t)/dt=0$ for the expression (\ref{C solution}) reads \cite{demery,andelman}
\begin{flalign}
\label{stationary C}%%%%%%%%%%
\bm{\mathcal{K}}\bm{\mathcal{C}}+\bm{\mathcal{C}}\bm{\mathcal{K}}^{\dagger}=\bm{\mathcal{R}}.
\end{flalign}
The four matrix elements of $\bm{\mathcal{C}}$, or the four kinds of correlation functions in eqn (\ref{C def}), can be determined by four simultaneous equations arising from the above stationary condition (\ref{stationary C}) (see Appendix D for details).

%%%%%%%%%%%%%%%%%%%
{\itshape Density-density and charge-charge correlation functions at equal times: derivation scheme of eqn (\ref{C solution2}) to (\ref{delta crr})}.---
We are concerned with the stationary density-density and charge-charge correlation functions at equal times, $\mathcal{C}^{\mathrm{st}}_{\rho\rho}$ and $\mathcal{C}^{\mathrm{st}}_{qq}$, among the matrix elements of $\bm{\mathcal{C}}$.
As detailed in Appendix D, the solution to eqn (\ref{stationary C}) reads
\begin{flalign}
\label{C pre solution}%%%%%%%%%%%%%%%%%%%%%%
\bm{\mathcal{P}}(\bm{k})
\begin{pmatrix}
\mathcal{C}^{\mathrm{st}}_{\rho\rho}(\bm{k})\\
\mathcal{C}^{\mathrm{st}}_{qq}(\bm{k})
\end{pmatrix}
=(2\pi)^3\,\mathcal{G}_2(\bm{k})
\begin{pmatrix}
2\overline{n}\bm{k}^2\\
2\overline{n}\bm{k}^2\\
\end{pmatrix},
\end{flalign}
where $\mathcal{G}_2(\bm{k})$ has been defined in eqn (\ref{g2 def}), and the matrix elements of $\bm{\mathcal{P}}(\bm{k})$ is given by
\begin{flalign}
\bm{\mathcal{P}}(\bm{k})&=
\begin{pmatrix}
\bm{k}^2\mathcal{G}_2(\bm{k})+k^2_{x}(pE)^2&-k^2_{x}(pE)^2\\
-k^2_{x}(pE)^2&\mathcal{G}_1(\bm{k})\mathcal{G}_2(\bm{k})
+k^2_{x}(pE)^2\\
\end{pmatrix}.
\label{P def}%%%%%%%
\end{flalign}
Obviously, eqn (\ref{C pre solution}) and (\ref{P def}) transform to eqn (\ref{C solution2}) to (\ref{delta crr}).

%%%%%%%%%%%%%%%%%%%
%%%%%%%%%%%%%%%%%%%
\section{Details on the constrained free energy $\mathcal{A}[\bm{n}]$ of a given density distribution $\bm{n}(\bm{r},t)$}
We consider the overdamped dynamics of ions with the total number $N$ of charged spheres being fixed. Hence,  the constrained free-energy functional $\mathcal{A}[\bm{n}]$ of a given density distribution $n_l(\bm{r},t)$ ($l=1,\,2$) is defined for the canonical ensemble using the contour integral over a complex variable $w=e^{\mu}$ \cite{frusawa sdft}:
\begin{flalign}
&e^{-\mathcal{A}[\bm{n}]}\nonumber\\
&=\left(
\frac{1}{2\pi i}\oint \frac{dw}{w^{N+1}}
\right)^2\nonumber\\
&\qquad\times\left(
\mathrm{Tr}\,e^{-U[\widehat{\bm{n}}]}\prod_{l=1}^2\prod_{{\bf r}}\delta\left[
\widehat{n}_l({\bf r},t)-n_l({\bf r},t)\right]
\right),
\label{appendix contour}%%%%%%%%%
\\
&\mathrm{Tr}=
\left(\sum_{N=0}^{\infty}\frac{1}{N!}\right)^2\prod_{l=1}^2
\int d{\bf r}^l_1\cdots\int d{\bf r}^l_N,
\label{appendix tr}%%%%%%%%%%%%%
\\
&U[\widehat{\bm{n}}]=
\sum_{l=1}^2\sum_{m=1}^2\biggl[
\frac{1}{2}\iint d^3\bm{r}d^3\bm{r}'\widehat{n}_l(\bm{r})v(\bm{r}-\bm{r}')\widehat{n}_m(\bm{r})
\nonumber\\
&\qquad\qquad\qquad\qquad\qquad\qquad\qquad
-\int d^3\bm{r}\widehat{n}_l(\bm{r})\mu
\biggr].
\label{appendix u}%%%%%%%%
\end{flalign}
In eqn (\ref{appendix contour}), the Dirac delta functional represents the constraint on the original density distribution $\widehat{n}_l(\bm{r})=\sum_{i=1}^N\delta[\bm{r}-\bm{r}_i^l(t)]$.
It has been shown for one-component fluid that the constrained free energy functional is expressed by the functional integral over a fluctuating potential field $\phi_l(\bm{r})$. Similarly, we have
\begin{flalign}
&e^{-\mathcal{A}[\bm{n}]}\nonumber\\
&=\left(
\frac{1}{2\pi i}\oint \frac{dw}{w^{N+1}}
\right)^2\prod_{l=1}^2
\int D\phi_l e^{-F[\bm{n},\bm{\phi}]+\int d^3\bm{r}n_l(\bm{r})\mu}\nonumber\\
&=\prod_{l=1}^2
\int D\phi_l\Delta[n_l]
\,e^{-F[\bm{n},\bm{\phi}]},
\label{appendix canonical}%%%%%%%%%%%%%
\end{flalign}
where
\begin{align}
&F[\bm{n},\bm{\phi}]-\sum_{l=1}^2\int d{\bf r}n_l({\bf r})\mu\nonumber\\
&=\Omega[\bm{\psi}^{\mathrm{dft}}-i\bm{\phi}]
-\sum_{l=1}^2\int d{\bf r}\,n_l({\bf r})\left\{\psi_l^{\mathrm{dft}}({\bf r})-i\phi_l({\bf r})\right\},
\label{appendix f nphi}%%%%%%%%%%%%
\end{align}
and the superscript "dft" denotes the equilibrium DFT \cite{dft} according to which the external field $\psi_l^{\mathrm{dft}}(\bm{r})$ is used for ensuring that the equilibrium density found from the grand potential $\Omega[\psi]$ is equated with $n_l({\bf r})$:
\begin{align}
\left.
\frac{\delta\Omega[\psi]}{\delta\psi({\bf r})}\right|_{\psi=\psi_l^{\mathrm{dft}}}
=n_l({\bf r}).
\label{appendix omega rho}%%%%%%%%%%%%%%
\end{align}
The free-energy functional $F[\bm{n},\bm{0}]$ in the absence of fluctuating $\phi$--field corresponds to the intrinsic Helmholtz free energy that is related to the grand potential $\Omega[\bm{\psi}^{\mathrm{dft}}]$ through the Legendre transform using the external field $\psi_l^{\mathrm{dft}}(\bm{r})$:
\begin{align}
&F[\bm{n},\bm{0}]-\sum_{l=1}^2\int d{\bf r}n_l({\bf r})\mu\nonumber\\
&\qquad\qquad=\Omega[\psi_{\mathrm{dft}}]
-\sum_{l=1}^2\int d{\bf r}\,n_l({\bf r})\psi_l^{\mathrm{dft}}({\bf r}),
\label{appendix f n0}%%%%%%%%%%%%
\end{align}
as well as the equilibrium DFT.

%%%%%%%%%%%%%%%%%%%%%%%%%
\section{Derivation of $\mu_l^{\delta}$ given by eqn (\ref{self mu})}
Let us introduce the potential-potential correlation matrix $\bm{\Phi}$ that represents the set of potential-potential correlation functions defined by
\begin{flalign}
\bm{\Phi}(\bm{r}-\bm{r}')&=\left\langle\bm{\phi}(\bm{r})\bm{\phi}^{\mathrm{T}}(\bm{r}')\right\rangle_{\phi}\nonumber\\
&=
\begin{pmatrix}
\left\langle\Phi_{11}(\bm{r}-\bm{r}')\right\rangle_{\phi} & \left\langle\Phi_{12}(\bm{r}-\bm{r}')\right\rangle_{\phi} \\
\left\langle\Phi_{21}(\bm{r}-\bm{r}')\right\rangle_{\phi} & \left\langle\Phi_{22}(\bm{r}-\bm{r}')\right\rangle_{\phi} \\
\end{pmatrix},
\label{appendix mat Phi def}%%%%%%%%%%%%%%%
\\
\end{flalign}
where $\left\langle\Phi_{lm}(\bm{r}-\bm{r}')\right\rangle_{\phi}$ is related to the DCF function $c_{lm}(\bm{r}-\bm{r}')$ as
\begin{flalign}
\left\langle\Phi_{lm}(\bm{r}-\bm{r}')\right\rangle_{\phi}
&=\left\langle\phi_l(\bm{r})\phi_m(\bm{r}')\right\rangle_{\phi}\nonumber\\
&=\frac{\delta_{lm}\delta(\bm{r}-\bm{r}')}{n_l(\bm{r})}-c_{lm}(\bm{r}-\bm{r}').
\label{appendix Phi def}%%%%%%%%%%%%%%
\end{flalign}
The average, $\left\langle\phi_l(\bm{r})\phi_m(\bm{r}')\right\rangle_{\phi}$, has been performed over the fluctuating potential field as
\begin{flalign}
\left\langle\phi_l(\bm{r})\phi_m(\bm{r}')\right\rangle_{\phi}
=\frac{\prod_{l=1}^2\int
D\phi_l\,\phi_l(\bm{r})\phi_m(\bm{r}')
e^{-\Delta F[\bm{n},\bm{\phi}]}}
{\prod_{l=1}^2\int
D\phi_l\,e^{-\Delta F[\bm{n},\bm{\phi}]}},
\label{appendix phi av}%%%%%%%%%%%%
\end{flalign}
following the expression (\ref{delta f av}). Hence, eqn (\ref{appendix phi av}) yields eqn (\ref{appendix Phi def}) as far as the Gaussian functional form (\ref{f diff}) for $\Delta F[\bm{n},\bm{\phi}]$ is concerned.
It is found from eqn (\ref{n def}) and (\ref{appendix Phi def}) that the Ornstein-Zernike equations for two-component liquids read
\begin{flalign}
\bm{\mathcal{N}}(\bm{k})\bm{\Phi}(-\bm{k})=\bm{I},
\label{appendix matrix oz}%%%%%%%
\end{flalign}
indicating that the Fourier-transformed matrix $\bm{\Phi}(-\bm{k})$ is the inverse of density-density correlation matrix $\bm{\mathcal{N}}(\bm{k})$.

Considering that $\mathcal{N}_{12}=\mathcal{N}_{21}$ for the density-density correlation function $\mathcal{N}_{lm}(\bm{r}-\bm{r}')$ defined by eqn (\ref{n def}), eqn (\ref{f diff}) leads to
\begin{flalign}
&\left\langle\frac{\delta \Delta F[\bm{n},\bm{\phi}]}{\delta n_1(\bm{r})}\right\rangle_{\phi}\nonumber\\
&=\left\langle\frac{\delta}{\delta n_1(\bm{r})}\iint d^3\bm{r}d^3\bm{r}'
\frac{1}{2}\mathcal{N}_{11}(\bm{r}-\bm{r}')\Phi_{11}(\bm{r}-\bm{r}')\right\rangle_{\phi}\nonumber\\
&+\left\langle\frac{\delta}{\delta n_1(\bm{r})}\iint d^3\bm{r}d^3\bm{r}'\mathcal{N}_{12}(\bm{r}-\bm{r}')\Phi_{12}(\bm{r}-\bm{r}')\right\rangle_{\phi}.
\label{appendix deriv0}%%%%%%%%%%
\end{flalign}
To evaluate the above functional derivatives, we introduce the following notations:
it follows from eqn (\ref{n def}) that
\begin{flalign}
&\frac{\delta}{\delta n_l(\bm{r})}\iint d^3\bm{r}d^3\bm{r}'\mathcal{N}_{lm}(\bm{r}-\bm{r}')
=\int d^3\bm{r}'\mathcal{N}^{(1)}_{lm}(\bm{r}-\bm{r}'),\nonumber\\
&\mathcal{N}^{(1)}_{lm}=
\delta_{lm}\left\{\delta(\bm{r}-\bm{r}')+h_{lm}(\bm{r}-\bm{r}')n_m(\bm{r}')\right\}\nonumber\\
&\hphantom{\mathcal{N}^{(1)}_{lm}=\delta_{lm}\left\{\delta(\bm{r}-\bm{r}')\right.}
+h_{lm}(\bm{r}-\bm{r}')n_m(\bm{r}'),
\label{appendix n deriv}%%%%%%%%%%%%%
\end{flalign}
with which we have
\begin{flalign}
&\left\langle\frac{\delta}{\delta n_l(\bm{r})}\iint d^3\bm{r}d^3\bm{r}'
\mathcal{N}_{lm}(\bm{r}-\bm{r}')\Phi_{lm}(\bm{r}-\bm{r}')\right\rangle_{\phi}
\nonumber\\
&=\frac{\prod_{l=1}^2\int
D\phi_l\,e^{-\Delta F[\bm{n},\bm{\phi}]}\left[
\frac{\delta}{\delta n_l(\bm{r})}\iint d^3\bm{r}d^3\bm{r}'
\mathcal{N}_{lm}\Phi_{lm}
\right]}
{\prod_{l=1}^2\int
D\phi_l\,e^{-\Delta F[\bm{n},\bm{\phi}]}}
\nonumber\\
&=\int d^3\bm{r}'\left(\mathcal{N}^{(1)}_{lm}\left\langle\Phi_{lm}\right\rangle_{\phi}
+\mathcal{N}_{lm}\left\langle\Phi^{(1)}_{lm}\right\rangle_{\phi}\right),
\label{appendix deriv av}%%%%%%%%%%%%%
\end{flalign}
using $\Phi^{(1)}_{lm}$, a derivative of $\Phi_{lm}$, which is defined by eqn (\ref{appendix n deriv}) and (\ref{appendix deriv av}) themselves.
At the same time, it is also useful to consider the following functional derivative:
\begin{flalign}
&\int d^3\bm{r}'\left\langle\Phi_{lm}\right\rangle_{\phi}^{(1)}\nonumber\\
&=\frac{\delta}{\delta n_l(\bm{r})}\left(\iint d^3\bm{r}d^3\bm{r}'\left\langle\Phi_{lm}\right\rangle_{\phi}\right)\nonumber\\
&=\frac{\delta}{\delta n_l(\bm{r})}\left\{
\frac{\delta_{lm}\delta(\bm{r}-\bm{r}')}{n_l(\bm{r})}-c_{lm}(\bm{r}-\bm{r}')
\right\}\nonumber\\
&=-\frac{\delta_{lm}\delta(\bm{r}-\bm{r}')}{n_l^2(\bm{r})},
\label{appendix dcf deriv}%%%%%%%%%%%%%%%%%
\end{flalign}
where use has been made of the expression (\ref{appendix Phi def}) in the last two equalities.

The unknown functional $\left\langle\Phi^{(1)}_{lm}\right\rangle_{\phi}$ is related to $\left\langle\Phi_{lm}\right\rangle_{\phi}^{(1)}$, given by eqn (\ref{appendix dcf deriv}), as
\begin{flalign}
&\int d^3\bm{r}'\left\langle\Phi_{lm}\right\rangle_{\phi}^{(1)}\nonumber\\
&=\frac{\delta}{\delta n_l(\bm{r})}\left(\iint d^3\bm{r}d^3\bm{r}'\left\langle\Phi_{lm}\right\rangle_{\phi}\right)\nonumber\\
&=\frac{\delta}{\delta n_l(\bm{r})}\left\{
\frac{\iint d^3\bm{r}d^3\bm{r}'\prod_{l=1}^2\int D\phi_l\,\Phi_{lm}
e^{-\Delta F[\bm{n},\bm{\phi}]}}
{\prod_{l=1}^2\int D\phi_l\,e^{-\Delta F[\bm{n},\bm{\phi}]}}
\right\}\nonumber\\
&=\int d^3\bm{r}'\left[\left\langle\Phi_{lm}^{(1)}\right\rangle_{\phi}
-\left\langle\Phi_{lm}\frac{\delta\Delta F[\bm{n},\bm{\phi}]}{\delta n_l(\bm{r})}\right\rangle_{\phi}\right.\nonumber\\
&\hphantom{\int d^3\bm{r}'\left[\left\langle\Phi_{lm}^{(1)}\right\rangle_{\phi}\right.}
\left.
+\left\langle\Phi_{lm}\right\rangle_{\phi}\left\langle\frac{\delta\Delta F[\bm{n},\bm{\phi}]}{\delta n_l(\bm{r})}\right\rangle_{\phi}
\right]
\nonumber\\
&\approx\int d^3\bm{r}'\left\langle\Phi_{lm}^{(1)}\right\rangle_{\phi},
\label{appendix phiphi deriv}%%%%%%%%%%%%%%%%
\end{flalign}
stating that $\left\langle\Phi_{lm}^{(1)}\right\rangle_{\phi}$ can be equated with $\left\langle\Phi_{lm}\right\rangle_{\phi}^{(1)}$ in the approximation of $\left\langle AB \right\rangle_{\phi}\approx \left\langle A\right\rangle_{\phi}\left\langle B\right\rangle_{\phi}$ for $A=\Phi_{lm}$ and $B=\delta\Delta F[\bm{n},\bm{\phi}]/\delta n_l(\bm{r})$.
Thus, we obtain
\begin{flalign}
&\int d^3\bm{r}'\mathcal{N}_{lm}\left\langle\Phi_{lm}^{(1)}\right\rangle\nonumber\\
&=\int d^3\bm{r}'n_l(\bm{r})\left\{
\delta_{lm}\delta(\bm{r}-\bm{r}')+h_{lm}(\bm{r}-\bm{r}')n_m(\bm{r}')
\right\}\nonumber\\
&\qquad\qquad\qquad\qquad\qquad\qquad\quad
\times\left\{-\frac{\delta_{lm}\delta(\bm{r}-\bm{r}')}{n_l^2(\bm{r})}\right\}\nonumber\\
&=-\delta_{lm}\left\{\frac{1}{n_l(\bm{r})}+h_{lm}(\bm{0})\right\},
\label{appendix deriv av2}%%%%%%%%%%%%%%%%
\end{flalign}
from plugging eqn (\ref{appendix dcf deriv}) and (\ref{appendix phiphi deriv}) into the second term in the last line of eqn (\ref{appendix deriv av}).

Meanwhile, we have
\begin{flalign}
&\int d^3\bm{r}'\mathcal{N}^{(1)}_{lm}\left\langle\Phi_{lm}\right\rangle_{\phi}\nonumber\\
&=\int d^3\bm{r}'\biggl[\delta_{lm}\left\{\delta(\bm{r}-\bm{r}')+h_{lm}(\bm{r}-\bm{r}')n_m(\bm{r}')
\right\}\nonumber\\
&+h_{lm}(\bm{r}-\bm{r}')n_m(\bm{r}')\biggr]\left\{
\frac{\delta_{lm}\delta(\bm{r}-\bm{r}')}{n_l(\bm{r})}-c_{lm}(\bm{r}-\bm{r}')
\right\}.
\label{appendix deriv av3}%%%%%%%%%%%%%%%%%%%%%%
\end{flalign}
Hence, the combination of eqn (\ref{appendix deriv av2}) and (\ref{appendix deriv av3}) gives
\begin{flalign}
&\frac{1}{2}\int d^3\bm{r}'\left(\mathcal{N}^{(1)}_{11}\left\langle\Phi_{11}\right\rangle_{\phi}
+\mathcal{N}_{11}\left\langle\Phi_{11}\right\rangle_{\phi}^{(1)}\right)\nonumber\\
&=\frac{1}{2}\biggl[\frac{1}{n_1(\bm{r})}+2h_{11}({\bf 0})-c_{11}({\bf 0})\nonumber\\
&\qquad\quad-2\int d^3\bm{r}'h_{11}(\bm{r}-\bm{r}')n_1(\bm{r}')c_{11}(\bm{r}-\bm{r}')\nonumber\\
&\qquad\quad-\left\{\frac{1}{n_1(\bm{r})}+h_{11}(\bm{0})\right\}\biggr]\nonumber\\
&=\frac{1}{2}\left\{h_{11}({\bf 0})-c_{11}({\bf 0})\right\}\nonumber\\
&\qquad\quad
-\int d^3\bm{r}'h_{11}(\bm{r}-\bm{r}')n_1(\bm{r}')c_{11}(\bm{r}-\bm{r}'),
\label{appendix 11}%%%%%%%%%%%%%%%%
\end{flalign}
and
\begin{flalign}
&\int d^3\bm{r}'\left(\mathcal{N}^{(1)}_{12}\left\langle\Phi_{12}\right\rangle_{\phi}+\mathcal{N}_{12}\left\langle\Phi_{12}\right\rangle_{\phi}^{(1)}\right)
\nonumber\\
&\quad\quad
=-\int d^3\bm{r}'h_{12}(\bm{r}-\bm{r}')n_2(\bm{r}')c_{12}(\bm{r}-\bm{r}').
\label{appendix 12}%%%%%%%%%%%%%%
\end{flalign}
Considering the Ornstein-Zernike equation,
\begin{flalign}
h_{11}({\bf 0})=c_{11}({\bf 0})+\sum_{m=1}^2\int d^3\bm{r}'h_{1m}(\bm{r}-\bm{r}')n_m(\bm{r}')c_{1m}(\bm{r}-\bm{r}'),
\end{flalign}
the sum of eqn (\ref{appendix 11}) and (\ref{appendix 12}) leads to
\begin{flalign}
\overline{\frac{\delta \Delta F[\bm{n}]}{\delta n_1(\bm{r})}}
&=\frac{1}{2}\left\{h_{11}({\bf 0})-c_{11}({\bf 0})\right\}\nonumber\\
&\qquad\quad-\int d^3\bm{r}'h_{11}(\bm{r}-\bm{r}')n_1(\bm{r}')c_{11}(\bm{r}-\bm{r}')\nonumber\\
&\qquad\quad-\int d^3\bm{r}'h_{12}(\bm{r}-\bm{r}')n_2(\bm{r}')c_{12}(\bm{r}-\bm{r}')\nonumber\\
&=\frac{1}{2}\left\{c_{11}({\bf 0})-h_{11}({\bf 0})\right\}.
\end{flalign}
Thus, the resulting form (\ref{self mu}) of $\mu_l^{\delta}$ has been verified.

%%%%%%%%%%%%%%%%%%%%%%%%%
%%%%%%%%%%%%%%%%%%%%%%%%%%%
\section{Solving the steady-state equation (\ref{stationary C})}
\subsection{Derivation of eqn (\ref{C pre solution}) and (\ref{P def})}
We calculate the matrix elements of $\bm{\mathcal{K}}\bm{\mathcal{C}}$ and $\bm{\mathcal{C}}\bm{\mathcal{K}}^{\dagger}$, using a simplified form of
\begin{flalign}
\bm{\mathcal{K}}(\bm{k})&=
\begin{pmatrix}
\alpha&i\gamma\\
i\gamma&\alpha+\beta\\
\end{pmatrix},
\label{k mat el}%%%%%%%%%%%%%%%%%%%%%
\end{flalign}
with $\alpha=\bm{k}^2$, $\beta=\overline{\kappa}^2\omega(\bm{k})$ and $\gamma=k_{x}pE$.
It follows that
\begin{flalign}
\bm{\mathcal{K}}\bm{\mathcal{C}}
&=
\begin{pmatrix}
\alpha\,\mathcal{C}^{\mathrm{st}}_{\rho\rho}+i\gamma\mathcal{C}^{\mathrm{st}}_{\rho q}
&\alpha\,\mathcal{C}^{\mathrm{st}}_{q\rho}+i\gamma\mathcal{C}^{\mathrm{st}}_{q q}\\
(\alpha+\beta)\mathcal{C}^{\mathrm{st}}_{\rho q}+i\gamma\mathcal{C}^{\mathrm{st}}_{\rho\rho}
&(\alpha+\beta)\mathcal{C}^{\mathrm{st}}_{qq}+i\gamma\mathcal{C}^{\mathrm{st}}_{q\rho}\\
\end{pmatrix},\label{stationary mat el1}%%%%%%%%%%%%
\\
\bm{\mathcal{C}}\bm{\mathcal{K}}^{\dagger}&=
\begin{pmatrix}
\alpha\,\mathcal{C}^{\mathrm{st}}_{\rho\rho}-i\gamma\mathcal{C}^{\mathrm{st}}_{q\rho}
&(\alpha+\beta)\mathcal{C}^{\mathrm{st}}_{q\rho}-i\gamma\mathcal{C}^{\mathrm{st}}_{\rho\rho}\\
\alpha\,\mathcal{C}^{\mathrm{st}}_{\rho q}-i\gamma\mathcal{C}^{\mathrm{st}}_{qq}
&(\alpha+\beta)\mathcal{C}^{\mathrm{st}}_{qq}-i\gamma\mathcal{C}^{\mathrm{st}}_{\rho q}\\
\end{pmatrix}.
\label{stationary mat el2}%%%%%%%%%%%%%%%%%
\end{flalign}
The sum of eqn (\ref{stationary mat el1}) and (\ref{stationary mat el2}) provides the steady-state equation (\ref{stationary C}) which consists of the four kinds of equations for correlation functions as follows:
\begin{flalign}
\left\{
\begin{array}{l}
2\alpha\mathcal{C}^{\mathrm{st}}_{\rho\rho}+i\gamma\left(\mathcal{C}^{\mathrm{st}}_{\rho q}-\mathcal{C}^{\mathrm{st}}_{q\rho}\right)=(2\pi)^34\overline{n}\bm{k}^2\\
2(\alpha+\beta)\mathcal{C}^{\mathrm{st}}_{qq}-i\gamma\left(\mathcal{C}^{\mathrm{st}}_{\rho q}-\mathcal{C}^{\mathrm{st}}_{q\rho}\right)=(2\pi)^34\overline{n}\bm{k}^2\\
(2\alpha+\beta)\mathcal{C}^{\mathrm{st}}_{q\rho}+i\gamma\left(\mathcal{C}^{\mathrm{st}}_{qq}-\mathcal{C}^{\mathrm{st}}_{\rho\rho}\right)=0\\
(2\alpha+\beta)\mathcal{C}^{\mathrm{st}}_{\rho q}-i\gamma\left(\mathcal{C}^{\mathrm{st}}_{qq}-\mathcal{C}^{\mathrm{st}}_{\rho\rho}\right)=0.\\
\end{array}
\right.
\label{appendix four eq}%%%%%%%
\end{flalign}
It is easy to find from the last two equations of the above set that $\mathcal{C}^{\mathrm{st}}_{\rho q}=-\mathcal{C}^{\mathrm{st}}_{q\rho}$ and
\begin{flalign}
\mathcal{C}^{\mathrm{st}}_{\rho q}-\mathcal{C}^{\mathrm{st}}_{q\rho}=\frac{2i\gamma}{2\alpha+\beta}\left(\mathcal{C}^{\mathrm{st}}_{qq}-\mathcal{C}^{\mathrm{st}}_{\rho\rho}\right).
\label{appendix c qrho}%%%%%%%%%%%%
\end{flalign}
Substituting eqn (\ref{appendix c qrho}) into the first two equations of eqn (\ref{appendix four eq}), we have
\begin{flalign}
\left\{
\begin{array}{l}
\alpha\mathcal{C}^{\mathrm{st}}_{\rho\rho}-\frac{\gamma^2}{2\alpha+\beta}\left(\mathcal{C}^{\mathrm{st}}_{qq}-\mathcal{C}^{\mathrm{st}}_{\rho\rho}\right)=(2\pi)^32\overline{n}\bm{k}^2\\
(\alpha+\beta)\mathcal{C}^{\mathrm{st}}_{qq}+\frac{\gamma^2}{2\alpha+\beta}\left(\mathcal{C}^{\mathrm{st}}_{qq}-\mathcal{C}^{\mathrm{st}}_{\rho\rho}\right)=(2\pi)^32\overline{n}\bm{k}^2,\\
\end{array}
\right.
\end{flalign}
which reads
\begin{flalign}
\frac{1}{2\alpha+\beta}
\bm{\mathcal{P}}(\bm{k})
\begin{pmatrix}
\mathcal{C}^{\mathrm{st}}_{\rho\rho}\\
\mathcal{C}^{\mathrm{st}}_{qq}
\end{pmatrix}
=(2\pi)^32\overline{n}
\begin{pmatrix}
\bm{k}^2\\
\bm{k}^2\\
\end{pmatrix},
\label{appendix steady}%%%%%%%%%%%%%%%%%%
\end{flalign}
and
\begin{flalign}
\bm{\mathcal{P}}(\bm{k})&=
\begin{pmatrix}
\alpha(2\alpha+\beta)+\gamma^2&-\gamma^2\\
-\gamma^2&(\alpha+\beta)(2\alpha+\beta)+\gamma^2\\
\end{pmatrix}.
\label{appendix P def}%%%%%%%%%%%%%%%%%%%%%
\end{flalign}
In the matrix elements, we note that $2\alpha+\beta=\mathcal{G}_2(\bm{k})$ and $\alpha+\beta=\mathcal{G}_1(\bm{k})$. Hence, the above expressions (\ref{appendix steady}) and (\ref{appendix P def}) are found to be equivalent to eqn (\ref{C pre solution}) and (\ref{P def}).

%%%%%%%%%%%%%%%%%%%%
\subsection{Derivation of eqn (\ref{ans large})}
Eqn (\ref{C solution2}) to (\ref{def cofactor B}) are combined into a single form,
\begin{flalign}
&
\frac{1}{(2\pi)^3}\begin{pmatrix}
\mathcal{C}^{\mathrm{st}}_{\rho\rho}(\bm{k})\\
\mathcal{C}^{\mathrm{st}}_{qq}(\bm{k})
\end{pmatrix}\nonumber\\
&=\frac{2\overline{n}\bm{k}^2}{\mathcal{G}_2(\bm{k})
\left\{
\bm{k}^2\mathcal{G}_1(\bm{k})
+k^2_{x}(pE)^2
\right\}}
\begin{pmatrix}
\mathcal{G}_1(\bm{k})\mathcal{G}_2(\bm{k})+2k^2_{x}(pE)^2\\
\bm{k}^2\mathcal{G}_2(\bm{k})+2k^2_{x}(pE)^2\\
\end{pmatrix}.
\label{appendix c answer}%%%%%%%%%%%%%%%%%
\end{flalign}
The three propagators, $\mathcal{G}_1(\bm{k})$, $\mathcal{G}_2(\bm{k})$ and $\overline{\kappa}^2+2k^2_{x}(pE\overline{\kappa}^{-1})^2$, can be simply approximated by $\overline{\kappa}^2$ when considering the small wavevector region of $k\overline{\kappa}^{-1}\ll 1$ at a moderate field strength of $pE\overline{\kappa}^{-1}\sim 1$.
This approximation allows eqn (\ref{appendix c answer}) to be reduced to
\begin{flalign}
&\frac{1}{(2\pi)^3}
\begin{pmatrix}
\mathcal{C}^{\mathrm{st}}_{\rho\rho}(\bm{k})\\
\mathcal{C}^{\mathrm{st}}_{qq}(\bm{k})
\end{pmatrix}
\nonumber\\
&\approx\frac{2\overline{n}\bm{k}^2}{\overline{\kappa}^2
\left\{
\bm{k}^2
+k^2_{x}(pE\overline{\kappa}^{-1})^2
\right\}}
\begin{pmatrix}
\overline{\kappa}^2\\
\bm{k}^2+2k^2_{x}(pE\overline{\kappa}^{-1})^2\\
\end{pmatrix},\nonumber\\
\end{flalign}
whose rearrangement leads to eqn (\ref{ans large}).

%%%%%%%%%%%%%%%%%
\section{Derivation of eqn (\ref{approx k sol1}) and (\ref{approx k sol2})}
The discriminant analysis of eqn (\ref{approx kirkwood eq1}) provides the determining equation for the Debye-H\"uckel length $\overline{\kappa}^{(*1)}$ on the Kirkwood crossover:
\begin{flalign}
&\left\{1-\alpha_1\left(\overline{\kappa}^{(*1)}\right)^2\sigma^2\right\} ^2\nonumber\\
&-4\alpha_2\left(\overline{\kappa}^{(*1)}\right)^2\sigma^2 \left\{
\left(\overline{\kappa}^{(*1)}\right)^2\sigma^2+\mathcal{E}^2
\right\}=0,
\label{appendix kirkwood eq1}%%%%%%%%%%%%%%%%%%%%
\end{flalign}
with $\mathcal{E}^2\equiv(pE)^2\sigma^2$.
Eqn (\ref{approx kirkwood eq1}) reads the following quadratic equation,
\begin{flalign}
\left(
\alpha_1^2-4\alpha_2
\right)X^2-2\left(
\alpha_1+2\alpha_2\mathcal{E}^2
\right)X+1=0,
\label{approx kirkwood eq2}%%%%%%%%%%%%%%%
\end{flalign}
for $X=\left(\overline{\kappa}^{(*1)}\right)^2\sigma^2$.
The solution $X$ to eqn (\ref{approx kirkwood eq2}) is
\begin{flalign}
X=\frac{1}{\alpha_1^2-4\alpha_2}\left[
\alpha_1+2\alpha_2\mathcal{E}^2
-2\sqrt{\alpha_2}(1+\alpha_1\mathcal{E}^2+\alpha_2\mathcal{E}^4)^{1/2}
\right].
\label{appendix quad general}%%%%%%%%%%%%%%%%%%%
\end{flalign}
At a low field strength of $\mathcal{E}\ll 1$, the numerator on the rhs of eqn (\ref{appendix quad general}) is approximated by
\begin{flalign}
&\alpha_1+2\alpha_2\mathcal{E}^2
-2\sqrt{\alpha_2}(1+\alpha_1\mathcal{E}^2+\alpha_2\mathcal{E}^4)^{1/2}\nonumber\\
&\qquad\qquad\approx
\alpha_1+2\alpha_2\mathcal{E}^2-2\sqrt{\alpha_2}\left(
1+\frac{\alpha_1\mathcal{E}^2}{2}
\right)\nonumber\\
&\qquad\qquad=\left(\alpha_1-2\sqrt{\alpha_2}\right)
(1-\sqrt{\alpha_2}\mathcal{E}^2),
\label{appendix low field}%%%%%%%%%%%%%%%%
\end{flalign}
whereas we have
\begin{flalign}
&\alpha_1+2\alpha_2\mathcal{E}^2
-2\sqrt{\alpha_2}(1+\alpha_1\mathcal{E}^2+\alpha_2\mathcal{E}^4)^{1/2}\nonumber\\
&\qquad\qquad\approx
2\alpha_2\mathcal{E}^2-2\sqrt{\alpha_2}(\alpha_2\mathcal{E}^4)^{1/2}=0,
\label{appendix high field}%%%%%%%%%%%%%%%%%%%
\end{flalign}
for the high field strength of $\mathcal{E}\gg 1$.
While the approximate form (\ref{appendix low field}) of the numerator results in the expression (\ref{approx k sol1}), the limiting result (\ref{approx k sol2}) is valid due to eqn (\ref{appendix high field}).

%%%%%%%%%%%%%%%%%%%%%%%%
%%%%%%%%%%%%%%%%%%%%%%%%
\section*{Acknowledgements}
The author thanks the anonymous referees for their valuable comments and suggestions.

\bibliographystyle{apsrev4-1}

\end{document}